\documentclass[twocolumn,preprintnumbers,amsmath,amssymb,superscriptaddress,nofootinbib,english]{revtex4-1}
\pdfoutput=1
\usepackage{times,amsmath,amsfonts,amssymb,epstopdf}
\usepackage{graphicx}
\usepackage{dcolumn}
\usepackage{bm}
\usepackage{enumerate}
\usepackage{epsfig}
\usepackage{graphicx}
\usepackage{hyperref}
\usepackage[usenames]{color}
\usepackage{url}
\usepackage[normalem]{ulem}
\usepackage[T1]{fontenc}
\usepackage{amsmath}

\usepackage[usenames]{color}

\def\be{\begin{equation}}
\def\ee{\end{equation}}
\def\ben{\begin{eqnarray}}
\def\een{\end{eqnarray}}
\def\ba{\begin{array}}
\def\ea{\end{array}}

\newcommand{\bq}{\begin{eqnarray}}
\newcommand{\eq}{\end{eqnarray}}
\newcommand{\bes}{\begin{subequations}}
\newcommand{\ees}{\end{subequations}}

\begin{document}
\newcommand{\half}{{\textstyle\frac{1}{2}}}
\allowdisplaybreaks[3]
\def\triangledown{\nabla}
\def\grad3{\hat{\nabla}}
\def\a{\alpha}
\def\b{\beta}
\def\g{\gamma}\def\G{\Gamma}
\def\d{\delta}\def\D{\Delta}
\def\ep{\epsilon}
\def\et{\eta}
\def\z{\zeta}
\def\t{\theta}\def\T{\Theta}
\def\l{\lambda}\def\L{\Lambda}
\def\m{\mu}
\def\f{\phi}\def\F{\Phi}
\def\n{\nu}
\def\r{\rho}
\def\s{\sigma}\def\S{\Sigma}
\def\ta{\tau}
\def\x{\chi}
\def\o{\omega}\def\O{\Omega}
\def\k{\kappa}
\def\pa {\partial}
\def\ov{\over}
\def\br{\\}
\def\ud{\underline}

\def\lcdm{\Lambda{\rm CDM}}
\def\qcdm{{\rm QCDM}}
\def\nloc{R\Box^{-2}R}
\def\msun{M_{\odot}/h}
\def\dw{f(X)}
\def\costhe{{\rm cos}\theta}
\def\sinthe{{\rm sin}\theta}
\def\cosphi{{\rm cos}\varphi}
\def\sinphi{{\rm sin}\varphi}
\def\sintwothe{{\rm sin}^2\theta}
\def\costwothe{{\rm cos}^2\theta}
\def\sintwophi{{\rm sin}^2\varphi}
\def\costwophi{{\rm cos}^2\varphi}
\def\hr{\rm HighRes}
\def\lr{\rm LowRes}

\def\pxpx{\left[\partial_x\partial_x\Phi\right]}
\def\pypy{\left[\partial_y\partial_y\Phi\right]}
\def\pzpz{\left[\partial_z\partial_z\Phi\right]}
\def\pxpy{\left[\partial_x\partial_y\Phi\right]}
\def\pxpz{\left[\partial_x\partial_z\Phi\right]}
\def\pypz{\left[\partial_y\partial_z\Phi\right]}

\newcommand\norm[1]{\left\lVert#1\right\rVert}
\newcommand\lsim{\mathrel{\rlap{\lower4pt\hbox{\hskip1pt$\sim$}}
    \raise1pt\hbox{$<$}}}
\newcommand\gsim{\mathrel{\rlap{\lower4pt\hbox{\hskip1pt$\sim$}}
    \raise1pt\hbox{$>$}}}
\newcommand\esim{\mathrel{\rlap{\raise2pt\hbox{\hskip0pt$\sim$}}
    \lower1pt\hbox{$-$}}}
\newcommand{\dpar}[2]{\frac{\partial #1}{\partial #2}}
\newcommand{\sdp}[2]{\frac{\partial ^2 #1}{\partial #2 ^2}}
\newcommand{\dtot}[2]{\frac{d #1}{d #2}}
\newcommand{\sdt}[2]{\frac{d ^2 #1}{d #2 ^2}}    

\title{{\tt RAY-RAMSES}: a code for ray tracing on the fly in N-body simulations}

\author{Alexandre Barreira}
\email[Electronic address: ]{barreira@mpa-garching.mpg.de}
\affiliation{Institute for Computational Cosmology, Department of Physics, Durham University, Durham DH1 3LE, U.K.}
\affiliation{Institute for Particle Physics Phenomenology, Department of Physics, Durham University, Durham DH1 3LE, U.K.}
\affiliation{Max-Planck-Institut f{\"u}r Astrophysik, Karl-Schwarzschild-Str. 1, 85748 Garching, Germany}

\author{Claudio Llinares}
\affiliation{Institute for Computational Cosmology, Department of Physics, Durham University, Durham DH1 3LE, U.K.}

\author{Sownak Bose}
\affiliation{Institute for Computational Cosmology, Department of Physics, Durham University, Durham DH1 3LE, U.K.}

\author{Baojiu Li}
\affiliation{Institute for Computational Cosmology, Department of Physics, Durham University, Durham DH1 3LE, U.K.}

\begin{abstract}
We present a ray tracing code to compute integrated cosmological observables on the fly in AMR N-body simulations. Unlike conventional ray tracing techniques, our code takes full advantage of the time and spatial resolution attained by the N-body simulation by computing the integrals along the line of sight on a cell-by-cell basis through the AMR simulation grid. Moroever, since it runs on the fly in the N-body run, our code can produce maps of the desired observables without storing large (or any) amounts of data for post-processing. We implemented our routines in the {\tt RAMSES} N-body code and tested the implementation using an example of weak lensing simulation. We analyse basic statistics of lensing convergence maps and find good agreement with semi-analytical methods. The ray tracing methodology presented here can be used in several cosmological analysis such as Sunyaev-Zel'dovich and integrated Sachs-Wolfe effect studies as well as modified gravity. Our code can also be used in cross-checks of the more conventional methods, which can be important in tests of theory systematics in preparation for upcoming large scale structure surveys. 

\end{abstract}

\maketitle

\section{Introduction}\label{sec:intro}

Observations of large scale structure in the Universe have been playing a crucial role in getting ever tighter constraints on competing theoretical cosmological models. Perhaps the most classical type of such observations consists in mapping the three-dimensional distribution of galaxies (which trace with some bias the total matter distribution) with spectroscopic surveys \cite{2000AJ....120.1579Y, 2003astro.ph..6581C, 2013AJ....145...10D}. These surveys measure the baryon acoustic oscillations (BAO) signal \cite{2005MNRAS.362..505C, 2005ApJ...633..560E} and clustering anisotropies due to redshift space distortions (RSD) \cite{2009MNRAS.393..297P, 2008Natur.451..541G, 2014MNRAS.440.2692S}, which allow to put constraints on the rate at which the Universe expands and the rate at which structure grows in it. A complementary approach to traditional galaxy surveys (and which is the focus of this paper) is to infer the large scale structure of matter by measuring its integrated effect on light that travels from background sources towards us. These include shifts in the temperature of cosmic microwave background (CMB) photons caused by inverse Compton scattering with high energy electrons inside galaxy clusters -- the so-called Sunyaev-Zel'dovich (SZ) effect \cite{1972CoASP...4..173S, 1980MNRAS.190..413S, carlstrom, 2012PhRvL.109d1101H}, shifts in the temperature of CMB photons as they cross time-evolving gravitational potentials -- the so-called integrated Sachs-Wolfe (ISW) effect \cite{1967ApJ...147...73S, 2002PhRvD..65j3510C, 2008PhRvD..78d3519H, 2008ApJ...683L..99G}, and magnification and distortions of background light sources as their emitted light bends due to strong and weak gravitational lensing effects \cite{2001PhR...340..291B, 2003ARA&A..41..645R, 2010CQGra..27w3001B, 2015RPPh...78h6901K}.

As observational surveys of large scale structure keep attaining higher precision, it is important that our theoretical understanding of the relevant physical processes keeps evolving as well. This helps in the interpretation of current data, as well as in the planning of future missions. In computing these theoretical predictions, theorists typically recourse to N-body simulation methods since these are currently the most accurate way to predict the clustering of matter on intermediate and small scales, where the density fluctuations have become nonlinear. N-body simulations allow also to include recipes to model the effects of baryonic physics and can be used in the generation of mock data sets to calibrate observational pipelines. Using N-body simulations to predict integrated effects along lines of sight that cover a redshift range is, in general, not as straightforward as getting predictions at fixed redshift values. For the latter, it often suffices to analyse the simulation output at a given snapshot of the particle distribution, whereas for the former it is required that the simulation results are analysed in a continuous range of redshift values. As a result of that, N-body methods for integrated observables are often subject to a number of approximations that are more or less valid depending on the exact observable studied. For instance, one of the most popular methods for cosmological weak lensing simulations consists of projecting the continuous matter distribution along the line of sight into a series of lens planes \cite{2000ApJ...530..547J, 2003ApJ...592..699V, 2008ApJ...682....1D, 2008MNRAS.391..435F, 2009A&A...497..335T, 2008MNRAS.388.1618C, Hilbert:2008kb, 2009ApJ...701..945S, 2012MNRAS.420..155K, 2013MNRAS.435..115B, 2014MNRAS.445.1942M, 2014MNRAS.445.1954P, 2015PhRvD..91f3507L, 2015arXiv151108211G}. This naturally erases the details of the time evolution of the fields along the line of sight. Furthermore, these projections assume that the superposition principle holds for the lensing effects of gravity, which is not necessarily true in theories beyond General Relativity that have nonlinear equations. Simulations of the ISW effect also make certain simplifying assumptions in the calculation of the time derivative of the gravitational potentials (see e.g.~\cite{2009MNRAS.396..772C}).

A great deal of effort is normally put into assessing the validity of the approximations made in these numerical methods, and in general, they seem to be robust enough. However, given the ever higher precision observations that lie ahead, it is desirable that the same observables can be computed with different methods, especially those which are subject to fewer approximations. This can allow for important checks of any residual theory systematics that could still be present. Moreover, current N-body methods to probe the clustering of matter along the line of sight sometimes require substantial amounts of data to be stored before it is post-processed to compute the desired signal. This provides extra motivation to develop new numerical techniques that are lighter in data storage, especially in light of the upcoming generation of surveys, which will require large simulations and mock data sets for calibration purposes.

This paper is precisely about a numerical method for integrated cosmological observables that goes beyond existing techniques in the number of approximations made and data storage concerns. Our method, which is based on the original idea of Refs.~\cite{whitehu2000, li2001}, is designed to trace rays from some source redshift to the observer, on the fly in the N-body simulation. Our algorithm is implemented in the adaptive mesh refinement (AMR) {\tt RAMSES} N-body code \cite{2002A&A...385..337T} and performs the integrations along the line of sight on a cell-by-cell basis, fully exploiting the spatial and time resolution provided by the N-body code. Moreover, since it runs on the fly in the N-body simulation, it can produce the desired maps (lensing, ISW, etc) without having to output the particle snapshots for post-processing analysis. The goal of this paper is to introduce the code implementation and illustrate its application in cosmological weak lensing simulations.

The outline of this paper is as follows. We start in Sec.~\ref{sec:mot} by describing in more detail the reasons that motivated us to develop the ray tracing code presented in this paper. Section \ref{sec:algo} explains the main aspects of the implementation of the algorithm in the {\tt RAMSES} code. In Sec.~\ref{sec:lens}, we explain the formalism to perform weak lensing studies with our code and test our implementation in Sec.~\ref{sec:gaussian} for a fixed Gaussian potential. Section \ref{sec:cosmo} is devoted to weak lensing cosmological simulations. We analyse our code results for one- and two-point statistics of the lensing convergence, where we assess the impact of N-body resolution and different integration methods. We also investigate the lensing signal around dark matter haloes in our simulations. Finally, we summarise and analyse the prospects for future developments and work in Sec.~\ref{sec:conc}.

\section{Motivating a new ray tracing code}\label{sec:mot}

In general, numerical studies of integrated observables employ the following three step strategy. First, one runs a N-body simulation for a given cosmology and stores the particle data at a specified number of redshift values. Second, the output from the simulations is used to generate a mock lightcone from some observer to a given source redshift\footnote{There are however ways to compute the lightcone on the fly in the simulation (e.g.~Refs.~\cite{2008ApJ...682....1D, 2009A&A...497..335T}).}. Here, one often needs to employ some interpolation scheme to construct a continuous matter distribution from the simulation results that are available only at a finite number of redshift values. Finally, rays are traced across the lightcone to probe the distribution of matter along the line of sight. This strategy has been employed most notably in weak lensing studies (see e.g.~Refs.~\cite{2000ApJ...530..547J, 2003ApJ...592..699V, 2008ApJ...682....1D, 2008MNRAS.391..435F, 2009A&A...497..335T, 2008MNRAS.388.1618C, Hilbert:2008kb, 2009ApJ...701..945S, 2012MNRAS.420..155K, 2013MNRAS.435..115B, 2014MNRAS.445.1942M, 2014MNRAS.445.1954P, 2015PhRvD..91f3507L, 2015arXiv151108211G} and references therein), but also in ISW \cite{2009MNRAS.396..772C, 2010MNRAS.407..201C, 2014MNRAS.438..412W} and SZ \cite{2000MNRAS.317...37D, 2001MNRAS.326..155D, 2001ApJ...549..681S, 2014MNRAS.440.3645M, 2015arXiv150905134D} related work.

One can identify, however, two less appealing aspects of this strategy. The first one is practical and relates to the large amounts of data that are needed to generate lightcones for post-processing. The second is related to the loss in resolution along the line of sight that follows from analysing a lightcone that has been constructed from a finite number of snapshots. To give a concrete example, conventional weak lensing studies usually employ the so-called multiple lens-plane approximation, in which the observables are calculated only on a series of planes, onto which the density field has been projected\footnote{See, however, the approach of Ref.~\cite{2012MNRAS.420..155K}, in which the lensing quantities are integrated using the three-dimensional distribution of the simulations (without projection onto planes), but which is still only available at a finite number of redshifts.}. Although one can always perform convergence tests on the number of planes used (e.g.~Ref.~\cite{2014MNRAS.445.1954P}), some of the detailed information on the time evolution along the line of sight is in general lost.

Our main motivation to develop the code presented in this paper was to overcome the two above-mentioned aspects. Namely, we aimed to implement a numerical method that (i) computes the integrated observables on the fly in the simulation, thereby avoiding the need to store large amounts of data; and (ii) takes full advantage of the spatial and time resolution of the N-body run to compute the integrals along the line of sight. Our numerical implementation is based on the original idea of Ref.~\cite{whitehu2000} for weak lensing simulations, which was later optimized in Ref.~\cite{li2001}. In particular, in the latter work, the authors realized that in particle-mesh (PM) N-body simulations, the integrated quantities can be computed analytically on a cell-by-cell basis as the simulation is running. These authors implemented their method in regular grid PM codes. In this paper, we follow a similar approach, but implement the algorithm in the publicly available {\tt RAMSES} code \cite{2002A&A...385..337T}, which can achieve a far greater resolution due to its AMR nature.

When designing this code, it was also our goal to make it general enough so that it could be used as a platform to perform studies of other types of integrated observables, and not just lensing. As a result, even though in this code presentation paper we illustrate the code operation for lensing, we stress that the algorithm is more general than that. In short, the code we present in this paper can calculate integrals of the form
\bq\label{eq:I1}
I = \int K(\chi) Q(x,y,z) {\rm d}\chi,
\eq
where $\chi$ is the comoving distance along some ray trajectory, $K$ is an integration kernel and $Q$ is any field that can be determined inside the simulation box at coordinates $x$, $y$ and $z$. The calculation of different observables corresponds to different expressions for $K$ and $Q$. For instance, for thermal and kinetic SZ studies, $Q$ is related to the density-weighted temperature and bulk velocity of electrons in clusters, respectively; for ISW studies, $Q$ is given by the time derivative of the lensing gravitational potential; and for lensing studies, $Q$ would be associated with second transverse derivatives of the lensing potential (cf.~Sec.~\ref{sec:lens}). {For the case of lensing, rays may also get their trajectories bent (although we anticipate here that this is not the case in this first version of the code)}.  As we commented above, there is already a substantial body of work available in the literature on these topics. The code we present here provides a different method to compute the same quantities which, amongst other things, can be used in important cross-checks of the traditional methods.

We note in passing that the ray-tracing machinery that we installed in {\tt RAMSES} may also serve as a starting point to develop a code that could be applied in radiative transfer studies (see e.g.~Refs.~\cite{2006MNRAS.371.1057I, 2009MNRAS.400.1283I, 2011MNRAS.414.3458W, 2013MNRAS.436.2188R, 2013MNRAS.434..748A} and references therein). This will, however, require some modifications to the code presented here, which is primarily oriented for integrated observables along lines of sight.


\section{Code description}\label{sec:algo}

In this section, we describe the main parts of the ray tracing code. We start by presenting a quick review of the default {\tt RAMSES} code, which is followed by an overview of the ray tracing algorithm and how it is implemented in {\tt RAMSES}. We then explain with more detail each of the main parts of the ray tracing code.

\subsection{Notation and basics of the {\tt RAMSES} code}

Our ray tracing modules are installed in the publicly available AMR {\tt RAMSES} N-body code \cite{2002A&A...385..337T}.  {\tt RAMSES} employs a number of simulation particles $N_p$ which act as discrete tracers of the underlying matter field. The simulation box is covered by a three dimensional mesh, on which the density values are calculated using the cloud-in-cell (CIC) interpolation scheme given the particle distribution at any time step. The code then solves for the gravitational potential field on the mesh, which can be finite-differenced to find the corresponding gravitational forces. The force at the particle positions is obtained by interpolating back from the mesh using the same CIC scheme to ensure momentum conservation. This is then used to update the particle's velocity and position at the next time step. The whole process is repeated from some initial time (typically redshift $z = 50-100$) to a later time (usually $z = 0$). The cubic cells of the 3D mesh can get refined if the effective number of particles contained in them exceeds some pre-specified threshold, $N_{\rm refine}$. Conversely, the cells are also de-refined if the number of particles drops below that threshold. This AMR nature of the code is useful in cosmological simulations, because it allows for high resolution in regions of high matter density, whilst saving computational resources in regions of lower density where the resolution can be lower. 

The term "domain level" refers to the coarsest mesh that regularly covers the whole simulation volume. In {\tt RAMSES}, the domain level contains $N_p^{1/3}$ cells in each direction\footnote{In this paper, we always assume three dimensional systems, although some times we shall use two-dimensional diagrams to facilitate the illustrations and explanations.}. If a cell of the domain level gets refined, then it is called a "father cell" with eight cubic "son cells". If the cell size of the father cell is $h$, then each of the eight sons has cell size $h/2$. The father cell, together with its son cells, form a so-called "grid" or "oct" of the refined level. If one of these eight sons gets further refined, then it will form a grid of the second refined level, i.e., its son cells will have cell size $h/4$. This series of grids accross refinement levels is organized in a tree structure. {\tt RAMSES} stores the grid and cell IDs in separate arrays and in a way that (i) given a son cell's relative position inside its grid and the ID of that grid, one can find the ID of the son cell, and vice versa; and (ii) given the parent cell's ID one can find the grid ID. Each level of refinement is labelled by $l$. The domain level $l = l_{\rm domain}$ is defined by $2^{l_{\rm domain}} = N_p^{1/3}$. For instance, if the simulation has $N_p = 512^3$, then $l_{\rm domain} = 9$. The first refinement level is labelled by $l_{\rm domain}+1$ and so on. Another characteristic of the {\tt RAMSES} code is that, at refinement boundaries, the coarse and fine sides differ only by one level of refinement. The size of the time steps is determined independently for each refinement level, with higher refinements taking smaller time steps. For example, one of the criteria to determine the size of the time steps is that the particles should move only by a fraction of the cell size they are currently in (cf.~Sec.~2.4 of Ref.~\cite{2002A&A...385..337T}).

{\tt RAMSES} is also efficiently parallelised using MPI. When run in parallel, each grid is "owned" by the same CPU that "owns" the parent cell of the grid. "To own" here means that the CPU knows all necessary information of a cell/grid, and is responsible for calculating all the relevant quantities inside the cell (density, potential, etc.), as well as the son cell position inside the grid. {As domain decomposition strategies, {\tt RAMSES} can employ Peano-Hilbert, angular and planar schemes. Of these, the Peano-Hilbert space filling curve is the most optimal for standard N-body runs. However, given the angular geometry of the ray tracing operations that we aim to perform, it is beneficial to consider the angular scheme since it distributes better the rays accross CPUs (specially when the number of rays becomes large).}

Our modifications to the {\tt RAMSES} code are mostly in the form of additional independent numerical modules, which keep the base code unchanged (except only for a few interfaces). We refer the reader to the {\tt RAMSES} code paper \cite{2002A&A...385..337T} for more details about its operation.

\subsection{Tiled simulation boxes}\label{sec:tile}

\begin{figure*}
	\centering
	\includegraphics[scale=0.365]{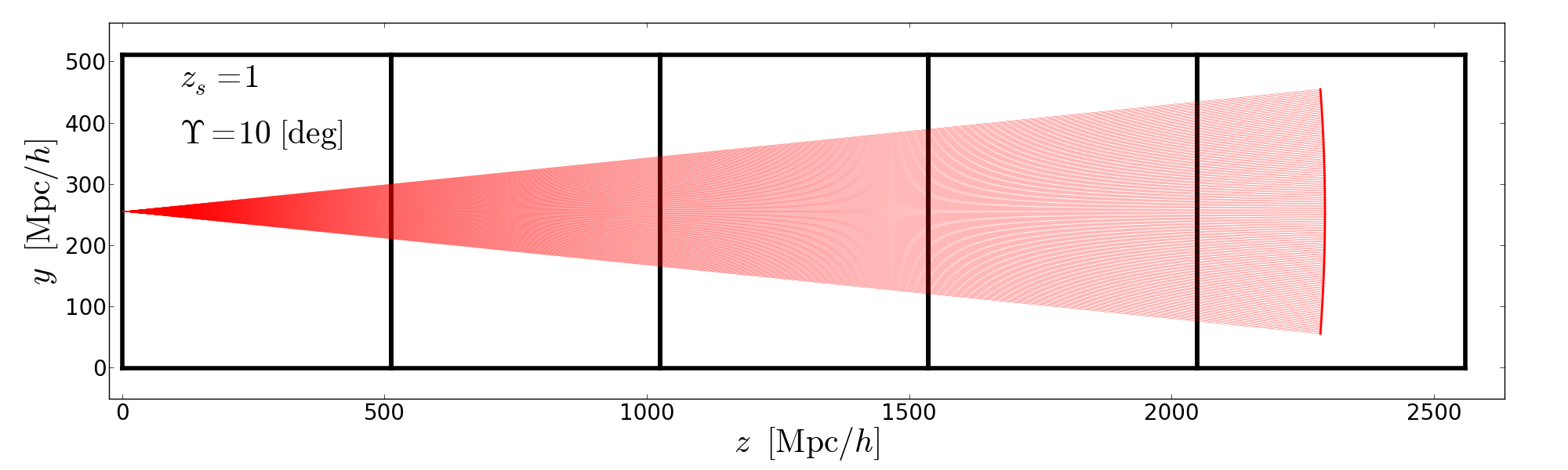}
	\caption{Example of a tiling scheme for ray tracing. The $x$-axis points into the plane of the figure. The boxes have size $L=512\ {\rm Mpc}/h$. The thin red lines illustrate the trajectory of the rays in a light bundle with opening angle $\Upsilon = 10\ {\rm deg}$ from $z_s = 1$. The different boxes should also simulate different realizations of the initial density field to avoid rays crossing the same structures at different times. If the ray trajectories are straight lines, then the boxes in the tile can be simulated simultaneously. In the case of bending rays, each box can start when the previous box (higher redshift) has finished the calculations. Although it is not the case in the figure, closer to the observer where the bundle covers a smaller volume, the simulation boxes can be made smaller to increase the particle resolution.}
\label{fig:tile}
\end{figure*}

High-resolution N-body simulations of boxes that are large enough to contain the distance travelled by photons from $z \gtrsim 1$ ($\chi \gtrsim 2\ {\rm Gpc}/h$) typically require massive computational resources (see e.g.~Refs.~\cite{2008MNRAS.391..435F, 2009A&A...497..335T}). To circumvent this, one can "tile" together a number of simulation boxes in order to fit the whole light bundle \cite{2000ApJ...530..547J, whitehu2000, li2001}. Figure~\ref{fig:tile} shows an example of a possible tiling scheme. The observer lies in the box that we refer to as the "last box", as opposed to the "first box", which contains the ray sources. The source redshift is $z_s = 1$ in this example. Each simulation box takes as input the position of the observer w.r.t.~its origin, $x_{\rm obs}, y_{\rm obs}, z_{\rm obs}$. For example, for the case illustrated in Fig.~\ref{fig:tile}, the observer is located at the center of $x$-$y$ face ($z = 0$) of the last box\footnote{We use the same letter, $z$, to denote redshift and one of the cartesian coordinates. The meaning of $z$ should be taken by the context.}, and so we have $x_{\rm obs} = y_{\rm obs} = 256\ {\rm Mpc}/h$, $z_{\rm obs} = 0$ for that box. For the first box, on the other hand, these would be $x_{\rm obs} = y_{\rm obs} = 256\ {\rm Mpc}/h$, $z_{\rm obs} = -2048\ {\rm Mpc}/h$. Given the geometry of the light bundle, the ray positions are more easily described using a spherical coordinate system with the observer at its origin
\bq\label{eq:raypos}
x_{\rm ray} &=& \chi\ \sinthe\ \cosphi, \nonumber \\
y_{\rm ray} &=& \chi\ \sinthe\ \sinphi, \nonumber \\
z_{\rm ray} &=& \chi\ \costhe,
\eq
where $\theta \in \left[0, \pi \right]$, $\varphi \in \left[0, 2\pi\right]$ are the two angular coordinates on the sky and $\chi$ is the radial coordinate. If the rays follow straight trajectories, then $\chi(z)$ is equal to the comoving distance $D_{\rm c}(z) = c\int^z_0 {\rm d}z/H(z)$, with $H(z)$ being the Hubble expansion rate, $z$ the redshift and $c$ the speed of light.

In the tiling scheme, a ray is only traced in a given box in the redshift interval during which the ray position lies within that box. For example, the integration of the rays in the first box would start at $z=1$ and it will last until $z \approx 0.86$, which is approximately when the rays "touch" the face of the box. Following the same reasoning, the second box would start the integrations at $z \approx 0.86$, which will continue until $z \approx 0.60$; and so on and so forth, until the rays reach the observer at $z=0$. Naturally, rays located in the outermost regions of the light bundle move from one box to the other before the more central rays. The conditions for the start and end of integration in each of the boxes are explained with more detail in Sec.~\ref{sec:boxtobox}. We note also that for straight ray cases, the boxes in the tile can be run simultaneously, since they all "know" {\it a priori} the position of the rays at all times. On the other hand, if rays bend, then boxes located closer to the observer can start tracing the rays after reading their positions from higher-redshift boxes.

We note that each simulation box should also start from different statistical realizations of the initial density field. This way, one ensures that the rays do not see the same structures throughout their trajectories (due to the periodic conditions of the simulation box). Finally, although not depicted in Fig.~\ref{fig:tile}, it is also worth mentioning that closer to the observer, where the spatial volume covered by the light bundle is smaller, the boxes in the tile can be made smaller to gain resolution without sacrificing computational efficiency (although not too small to still allow large enough structures to form).

\subsection{Outline of the code}\label{sec:outline}

\begin{figure*}
	\centering
	\includegraphics[scale=0.40]{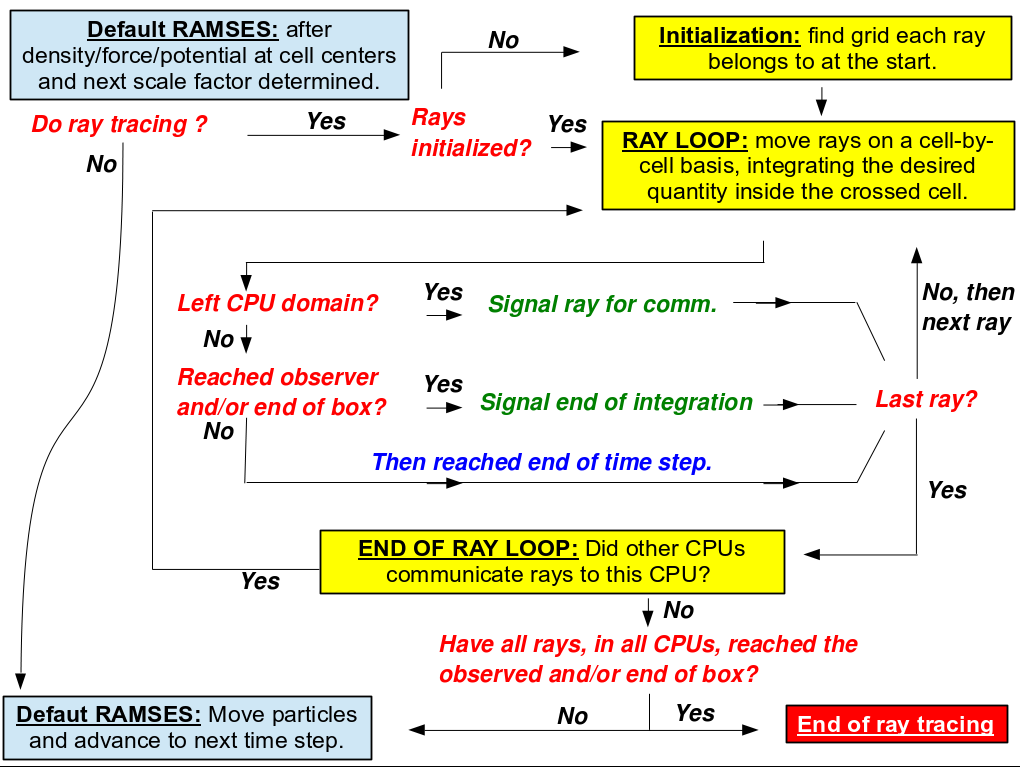}
	\caption{Sketch of the code flow. The ray tracing routines are called after default {\tt RAMSES} computes the field values and the next scale factor, but before the particles are moved. The ray tracing routines are initialized once each simulation box in the tile reaches the redshift at which the rays start their propagation there. In each time step, each CPU integrates each ray on a cell-by-cell basis (i) until it travels the maximum allowed distance light can travel, in which case the CPU goes directly to the next ray; (ii) until it reaches a CPU domain boundary, in which case the ray is marked for communication for another CPU to continue its integration; or (iii) until it reaches the observer and/or the end of the box, in which case the ray integration in the box is marked as finished for that ray.}
\label{fig:flow}
\end{figure*}

Figure \ref{fig:flow} shows a sketch of the flow of calculations in the code. The first operation of the ray tracing calculation consists of the initialization of the ray data structure (cf.~Sec.~\ref{sec:data}). The goal is to identify the ID of the grid that a given ray belongs to, i.e., determining the physical location of the ray within the grid structure (cf.~Sec.~\ref{sec:init}). This is performed only when the rays start the integration because, as the rays move through the mesh, it is possible to determine the ID of the next crossed grid, by searching for neighbouring grids. 

After the rays have been initialized, they are moved across the mesh on a cell-by-cell basis (cf.~Sec.~\ref{sec:move}), integrating a given quantity along the path inside each cell. As we explain in Sec.~\ref{sec:inte}, the integration can be done analytically by using the values of the desired quantity at each crossed cell centre or at its vertices. The latter have to be obtained by interpolation from the cell centres, which is where {\tt RAMSES} evaluates all fields (density, potential, etc.) by default (cf.~Sec.~\ref{sec:vertices}).

In a given time step, each CPU moves the rays that are currently within its spatial domain until one of the following possibilities happens:

\begin{enumerate}[i]
\item the rays travel the distance that light can travel in that time step;
\item the rays reach the observer/face of the box;
\item the rays reach the end of the CPU's spatial domain.
\end{enumerate}
If (i) is satisfied, then the CPU simply moves on to the next ray. If (ii) happens, then the ray's integration is marked as finished, and the CPU also proceeds to the next ray. Finally, if (iii) happens, then the ray is marked for communication and the CPU still carries on to its next ray. Once each CPU has dealt with its initial number of rays, it checks whether rays from other CPUs have been marked to enter its domain, and whether its own rays have been marked to leave. If there are rays entering and/or leaving the CPU's domain, then the relevant CPUs exchange ray data via MPI communication and repeat the above calculations for the incoming rays. This process is repeated until all rays satisfy (i) or (ii).

Our ray integration routines require as input the field values given by {\tt RAMSES} at a given time step, and hence, they are called after {\tt RAMSES} computes these quantities. Once the ray tracing calculations for this time step are finished, the code proceeds with the standard N-body part, until it is time to call the ray tracing routines again at the next time step. Our modifications consisted therefore in the development of independent modules that do not impact in any way the standard N-body part.

In the remainder of this section, we explain in more detail each of the steps and concepts involved in the propagation and integration of the rays across the mesh. The reader who wishes to skip these details can jump to Sec.~\ref{sec:lens}, from whereon we present tests and results from weak lensing ray tracing simulations.

\subsection{Ray data structure}\label{sec:data}

\begin{figure}
	\centering
	\includegraphics[scale=0.25]{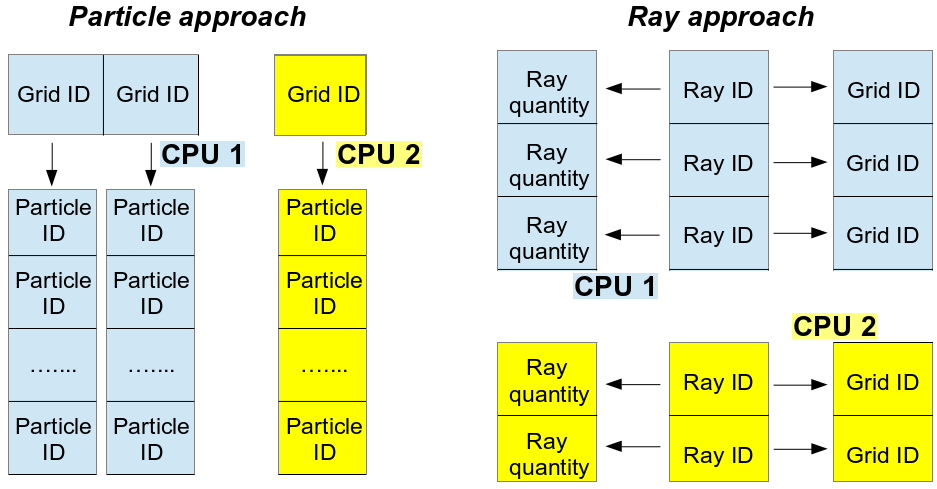}
	\caption{Sketch of two possible data structure schemes to link global ray IDs with local grid IDs. The {\it particle approach} treats rays as a different particle type in {\tt RAMSES} associating each grid with all the rays contained it. This approach, which is based on {\tt RAMSES}'s linked lists, is, however, computationally expensive because ray particles travel at the speed of light, which requires the linked lists to be update too many times. In the {\it ray approach}, two ordered lists link each ray to the grid it is currently in. If a ray leaves its current grid, then all there is to do is to update the entry of the grid list that corresponds to that ray.}
\label{fig:data}
\end{figure}

To implement our ray tracing algorithm in {\tt RAMSES} we need to establish a data structure that links ray and grid IDs \footnote{Once the link between grid and ray IDs is set, then the link between the ray and the relevant cell is made by checking in which octant of the grid the ray is in.}. One can think of at least two ways to do so. We call one the "particle approach" and the other the "ray approach". These two approaches are sketched in Fig.~\ref{fig:data}.

In crude terms, the particle approach determines {\it which rays are in each grid} (cf.~left-hand side of Fig.~\ref{fig:data}). The advantage of it is that it enables direct use of the existing {\tt RAMSES} structure for other types of "particles" (dark matter, stars, sinks, etc.), therefore making the coding easier. In this approach, a {\it linked list} data structure is used in {\tt RAMSES} to find the global IDs of the particles that lie within a grid, given the grid ID. The communication of rays between CPUs would also follow the strategy already set up for other types of particles in {\tt RAMSES}. However, in the code, one criterion for determining the size of the time step ensures that dark matter particles move only by a fraction of the current grid size. During this particle time step, photons, which travel at the speed of light, can cross many grids. This means that one has to either update the linked lists each time rays change grids, which involves a large numbers of operations and memory allocations/deallocations, or drastically reduce the particle time step so that rays do not cross more than one grid in a time step, which would make the code prohibitively slow.

The above drawbacks motivated us to implement the ray approach, in which one determines {\it which grid a ray is in} (cf.~right-hand side part of Fig.~\ref{fig:data}). In this case, the data structure consists of two ordered lists of global ray IDs and their corresponding local grid IDs. Ordered here is in the sense that the global ray ID that is in the $n$-th entry of the ray list physically resides inside the grid whose local grid ID is that of the $n$-th entry of the grid list. Note that, naturally, if a ray is inside a refined grid, then it also physically lies inside its parent grid. In the data structure, rays are always linked to the grid that is at the highest refinement level. With the ray approach, the integrated quantities are associated to each ray ID in the same way as grid IDs. For example, if the ray in the $n$-th entry of the ray list crosses one cell to enter another, then all the code needs to do is to (i) compute the integral for the crossed cell and accumulate it in the $n$-th entry of the quantity's list; and (ii) update the $n$-th entry of the grid list with the grid ID that contains the new cell (which can be the same grid). The ray approach is more efficient than the particle approach, even though it involved developing new strategies to move rays and communicate between CPUs (when rays move outside a CPU's spatial domain or when {\tt RAMSES} does {\it load balancing} to reassign domains to CPUs).

\subsection{Ray initialization}\label{sec:init}

The goal of the ray initialization is to set up the data structure described in the previous subsection. Just to guide the discussion in this section, we assume that the central ray of the bundle travels in the negative $z$ direction. The light bundle is specified by (i) its opening angle in the $x$ and $y$ directions, $\Upsilon_x$, $\Upsilon_y$,  respectively; and (ii) the number of rays in each of these two directions, $N_x$ and $N_y$. The angular positions of the rays in the field of view are assigned as
\bq\label{eq:txty}
\theta_a = \frac{\Upsilon_a}{N_a-1}\left(i_a - 1\right) - \frac{\Upsilon_a}{2}, \ \ \ \ \ \ \ \ \ a = x, y,
\eq
and we define the global ray ID number as\footnote{There is an abuse of notation with the subscripts $_x$ and $_y$. Here, they denote the two directions perpendicular to the line of sight and should not be confused with the 3D Cartesian coordinates.}
\bq\label{eq:rayid}
 {\tt rayid} = (i_y - 1)N_x + i_x,
\eq 
with $i_x \in \left[1, N_x\right]$ and $i_y \in \left[1, N_y\right]$. Once initialized, the global ray ID stays the same throughout the simulation, even when rays change CPUs. {At the end of the simulation, given a ray ID, one can reverse the above equations to find the ray's position in the 2D ray tracing map.}

In the {\tt RAMSES} structure, it is straightforward to retrieve the spatial location of a grid given its ID, but not the inverse operation: to find the grid ID given a certain (the ray's) spatial location. To achieve this, one can loop over all grids, and for each grid, loop over all rays to identify those that are inside it. We note that such a "brute force" search may not lead to huge overheads to the overall performance of the code since the initialization is performed only once per box. Nevertheless, we have developed a more efficient algorithm that loops over all grids, but for each, only loops over a smaller targeted range of ray IDs. The details of the algorithm are given in Appendix \ref{app:dandelin}, but in short, the idea is to compute the solid angle that subtends a sphere that contains the grid, with which one can determine a range of $\theta_x$ and $\theta_y$. This then enables us to significantly narrow down the ray IDs to loop for each grid using Eqs.~(\ref{eq:txty}) and (\ref{eq:rayid}).

For non-first boxes in the tile, the rays are initialized at the box face, and as a result, if the spatial location of the grid (which can be determined straightforwardly in {\tt RAMSES}) does not lie on the box face, then the loop over rays can be skipped. Note also that since rays are linked always to the finest grid (cf.~Sec.~\ref{sec:data}), one can also skip the loop over grids that are not at the highest refinement level.

As we commented above already, the initialization needs only to be performed once per box. During the course of the ray integrations, when a ray leaves a grid to enter another, there are other ways to efficiently determine the IDs of the new grid.

In this paper, we limit ourselves to analysing the code results for small opening angles, for which the flat-sky approximation holds. For full-sky ray tracing, one may benefit from using techniques such as {\tt HEALPix} \cite{2005ApJ...622..759G} to describe the ray distribution across the sphere. We leave the generalization to full-sky cases for future work.

\subsection{Moving rays}\label{sec:move}

In this section, we describe how the algorithm determines the path of a ray in a cell and how the rays are moved through the mesh across particle time steps and CPU domains.
 
\subsubsection{Trajectory inside cells}

\begin{figure}
	\centering
	\includegraphics[scale=0.155]{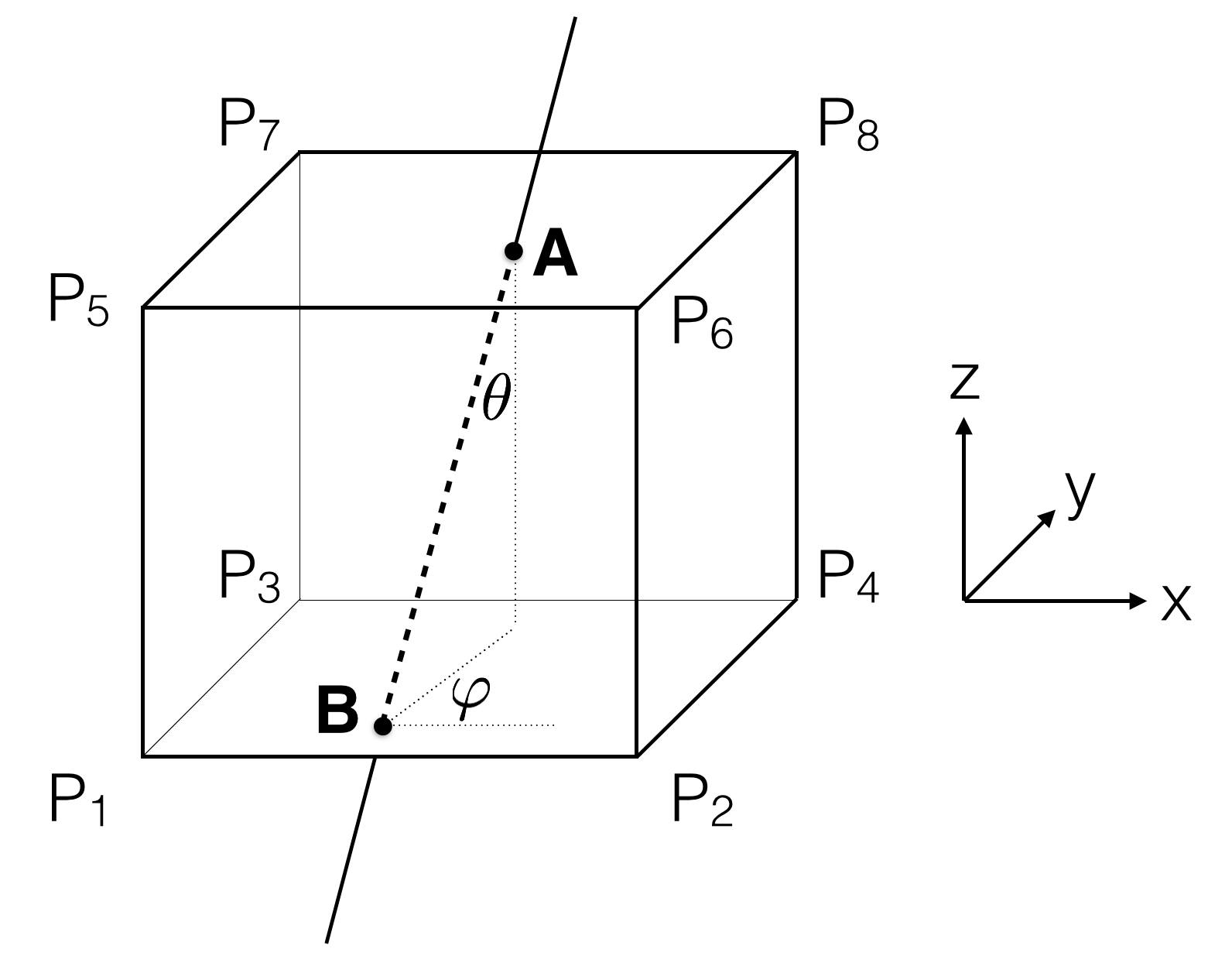}
	\caption{Example of a ray trajectory in a cell. Points $A$ and $B$ are, respectively, the ending and starting points of the trajectory. In this illustration, these two points are the intersection points of the ray trajectory and the cell faces, but in general points $A$ and $B$ can lie anywhere inside the cell.}
\label{fig:cell}
\end{figure}

Given the direction of a ray inside a cell, the calculation of its trajectory and the determination of the face from which the ray leaves the cell (which determines the next crossed cell) is mostly a geometrical exercise. Denoting by $\theta$ and $\varphi$ the two spherical coordinate angles that specify the direction of the ray, then the position of the ray can be parametrized by $\lambda$ in the equation \footnote{For straight rays, $\theta$ and $\varphi$ coincide with the angles that specify the spherical coordinates of the ray (cf.~Eq.~(\ref{eq:raypos})). We leave the generalization of our algorithm to follow non-straight trajectories for future work.}
\bq\label{eq:rayposition}
\vec{r} = \big(\chi_A + \lambda\big)\big(\sinthe\cosphi, \sinthe\sinphi, \costhe\big),
\eq
where $\chi_A$ is the comoving distance of the ray to the observer at the beginning of the trajectory in the cell (cf.~Fig.~\ref{fig:cell}). Hence,
\bq\label{eq:rayA}
\vec{r}_A = \chi_A\big(\sinthe\cosphi, \sinthe\sinphi, \costhe\big)
\eq
is the position vector of the ray at the beginning of its trajectory. Of all of the six cell faces the ray can cross, three can be ruled out by the signs of $\costhe$, $\sinphi$ and $\cosphi$. For instance, if $\costhe > 0$, then the ray is travelling in the negative $z$ direction ($\theta \in \left[0, \pi/2\right]$). This means that the ray cannot enter the neighbouring cell that lies in the positive $z$ direction. Similarly, if $\sinphi > 0$ ($\cosphi > 0$), then the ray cannot enter the neighbouring cell that lies in the positive $y$ ($x$) direction. More compactly, the faces from which the ray can leave the cell lie on one of three planes, each characterized by
\bq\label{eq:xtarget}
x =  x_{\rm target} &=& x_{\rm cell} - {\rm sign}(\cosphi) h/2, \nonumber \\
y =  y_{\rm target} &=& y_{\rm cell} - {\rm sign}(\sinphi) h/2, \nonumber \\
z =  z_{\rm target} &=& z_{\rm cell} - {\rm sign}(\costhe) h/2,
\eq
where $h$ is the cell size and $x_{\rm cell}$, $y_{\rm cell}$ and $z_{\rm cell}$ are the cell center coordinates. The trajectory of the ray, Eq.~(\ref{eq:rayposition}), intersects each of these three planes, respectively, at 
\bq\label{eq:rayB_tentative}
\vec{r}_{B, x} &=& x_{\rm target} \left(1, \frac{\sinphi}{\cosphi}, \frac{\costhe}{\sinthe\cosphi}\right) \nonumber \\
\vec{r}_{B, y} &=& y_{\rm target} \left(\frac{\cosphi}{\sinphi}, 1, \frac{\costhe}{\sinthe\sinphi}\right) \nonumber \\
\vec{r}_{B, z} &=& z_{\rm target} \left(\frac{\sinthe\cosphi}{\costhe}, \frac{\sinthe\sinphi}{\costhe}, 1\right),
\eq
where the subscript $_B$ denotes the end of the ray trajectory inside the cell (cf.~Fig.~\ref{fig:cell}). The face from which the ray leaves the cell is that to which the ray needs to travel the least. Therefore, one computes
\bq\label{eq:disttocross}
D_x &=& \norm{\vec{r}_A - \vec{r}_{B, x}}^2 = \chi_A - \frac{x_{\rm target}}{\sinthe\cosphi}, \nonumber \\
D_y &=& \norm{\vec{r}_A - \vec{r}_{B, y}}^2 = \chi_A - \frac{y_{\rm target}}{\sinthe\sinphi}, \nonumber \\
D_z &=& \norm{\vec{r}_A - \vec{r}_{B, z}}^2 = \chi_A - \frac{z_{\rm target}}{\costhe},
\eq
and the value of ${\rm min}\left\{D_x, D_y, D_z\right\}$ specifies the target face. For example, if  ${\rm min}\left\{D_x, D_y, D_z\right\} = D_x$, then the ray leaves from the face $x = x_{\rm target}$ (and analogously for $y$ and $z$).

Once the exiting face is found, then one can identify the next crossed cell by searching for neighbours using the default data structure in {\tt RAMSES}. If the new cell belongs to the same grid, then one needs only to update the cell information. If on the other hand the next crossed cell belongs to a different grid, then one updates the grid information as well. Note that the new grid can lie at a different refinement level and/or in a different CPU domain, and this information is also recorded in our ray data structure.

\subsubsection{From cell to cell across different time steps and CPU domains}\label{sec:acrosstime}

\begin{figure}
	\centering
	\includegraphics[scale=0.35]{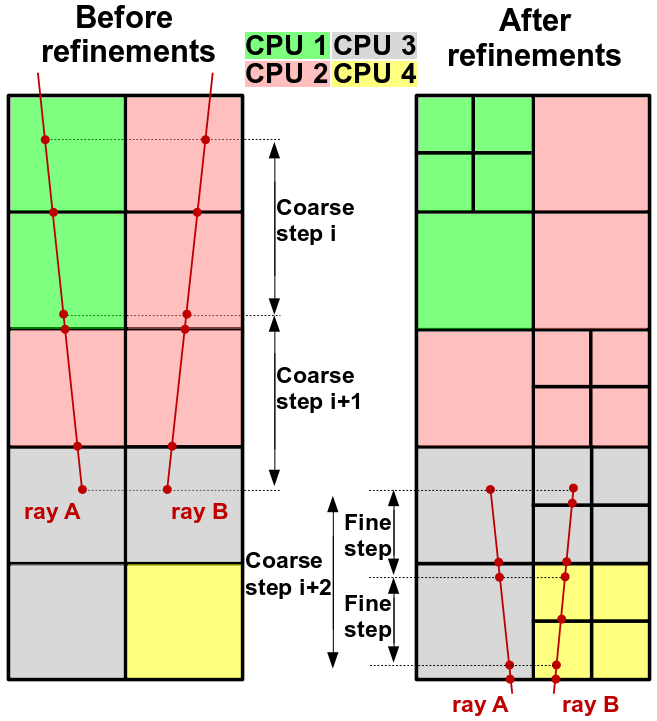}
	\caption{Two-dimensional sketch of two rays moving through the mesh. The colors indicate different CPU domains and the red dots denote the starting and ending points of a ray inside a cell. The rays move from top to bottom in the figure. There are three coarse time steps shown, and the mesh gets refined from the second to the third. Note that the distance that rays travel in a time step becomes smaller when there are refinements.}
\label{fig:move}
\end{figure}

Figure \ref{fig:move} shows a two-dimensional sketch of rays moving through the mesh. Shown are the trajectories of two rays during three coarse time steps. During the first two there are no refinements (left), and in the last coarse step some cells get refined (right). The different colors indicate different CPU domains. The red dots along the ray trajectories indicate where the rays cross cells or reach the end of a time step.

In the first coarse time step, ray A starts from the middle of a cell and is moved to the face of this cell. During this trajectory, the code integrates the desired quantity using the algorithm that shall be described in Sec.~\ref{sec:inte}. Ray A then continues its trajectory in the next crossed cell, but stops before reaching the new cell's face, at the point where the ray has travelled the maximally allowed distance in that time step. In the meantime, Ray B is moved similarly to ray A, but in the domain of CPU 2.

In the second coarse time step, CPU 1 moves ray A to the face of the cell and marks it for communication to CPU 2, which is found to own the next crossed cell (CPU 1 would then continue to move the other rays in its domain, if any). In the meantime, CPU 2 moves ray B until it is marked for communication to CPU 3. After CPU 2 has dealt with ray B (and all the other rays in its domain), it checks whether other CPUs have marked rays to be sent to it and moves these incoming rays as described above. Similarly, CPU 3 also checks for incoming rays. In the case sketched in Fig.\ref{fig:move}, CPU 2 receives ray A and integrates it until it is marked for communication to CPU 3, and CPU 3 receives ray B and integrates it until the end of the time step. These series of CPU communications occur until all rays have reached the end of the time step \footnote{We note in passing that, in addition to the communications that occur when rays leave CPU domains, there are other situations that require communicating the ray information, namely when the code performs load balancing (redistribution of the spatial domains across CPUs).}. In our particular example, CPU 3 receives ray A and integrates it until the end of the time step.

The way that rays move in the third coarse step is analogous to that of the other two, except that the distance that the rays travel before {\tt RAMSES} updates the field values is smaller, because of the smaller particle time step. In the current implementation of the code, all rays travel by the distance determined by the particle time step on the finest level, regardless of which level they belong to. In principle, this is not necessary since, if a ray only crosses unrefined cells, then it can be moved by the (larger) distance allowed in the coarse step, if the field values at the corresponding cell are not updated during the finer steps taken by the code. The implementation of this in {\tt RAMSES} is, however, slightly more involved and therefore we opted to have all rays moving by the distance determined by the finest cells of the mesh.

\subsubsection{From box to box in the tile}\label{sec:boxtobox}

As the bundle approaches the face of non-last boxes, some of the rays will cross the face earlier than other rays. For instance, the outermost rays in the bundle are the first to reach the box face, whereas the center rays are the last. To exemplify how the code deals with this transition, consider the trajectory of the outermost and center rays as they leave a given box (call it Box 2) to enter another (call it Box 1), and denote by $z_{\rm end, 2}^{\rm outermost}$ and $z_{\rm end, 2}^{\rm center}$ the redshift values at which the outermost and center rays cross the face of the box. In this case, Box 2 propagates the outermost ray until it reaches its face. At this point, the integration for this ray stops, but the calculation for the center ray is still ongoing. The simulation of Box 2 should therefore only be stopped for $z \leq z_{\rm end, 2}^{\rm center}$. As for Box 1, it initializes the ray data structure at $z = z_{\rm end, 2}^{\rm outermost}$ for all rays, including the center one whose position is known beforehand \footnote{As mentioned previously, the geometry is fixed for straight ray simulations and therefore Box 1 "knows" {\it a priori} the position of all rays at any redshift. In this case Box 1 and Box 2 can be run simultaneously.}. However, it starts by integrating only the outermost ray. The center ray remains "initialized" at the box face, and it only starts being integrated when Box 1 reaches $z =  z_{\rm end, 2}^{\rm center} < z_{\rm end, 2}^{\rm outermost}$.


\subsection{\bf Integration in cells}\label{sec:inte}

One of the key parts of our algorithm is the computation of the integral of a given field along the ray trajectories. Explicitly, we wish to evaluate integrals of the form of Eq.~(\ref{eq:I1}), which we repeat here for easy reference,
\bq\label{eq:I}
I = \int_0^{\chi_s} K(\chi) Q(x,y,z) {\rm d}\chi,
\eq
where $\chi_s = \chi(z_s)$ (with $z_s$ the source redshift), $K(\chi)$ is some weighting function or kernel that can be a polynomial of $\chi$ and $Q(x,y,z)$ is any quantity that can be evaluated in the cells. Compared with Eq.~(\ref{eq:I1}), Eq.~(\ref{eq:I}) is more specific as it encodes the information that we wish to follow rays from some distant source to an observer at redshift zero. The integral of Eq.~(\ref{eq:I}) can be split into the contribution from each cell crossed by a ray
\bq\label{eq:sumIc}
I = \sum_{\rm cells} I_c,
\eq
in which
\bq\label{eq:Ic}
I_c = \int_{\chi_B}^{\chi_A} K(\chi) Q(x,y,z) {\rm d}\chi,
\eq
with $\chi_A$ and $\chi_B$ being, respectively, the radial coordinate of the starting and ending points of the ray trajectory in a cell as the ray travels towards the observer ($\chi_A > \chi_B$) \footnote{The points $\chi_A$ and $\chi_B$ can lie anywhere inside the cell and not necessarily at the intersection of the line-of-sight with the cell faces. This is, for instance, the general case at the start and end of particle time steps.} (see Fig.~\ref{fig:cell}). To perform the integral we need to be able to evaluate the quantity $Q$ at any given point inside the cells. The simplest way to do so is to take the field to be constant inside the cell and equal to the value at its center $Q_c$, as given by default {\tt RAMSES}. In this case, $Q$ can be taken out of the integral in Eq.~(\ref{eq:Ic}) and the contribution of each cell to the total integral becomes
\bq\label{eq:Icngp}
I_c^{\rm NGP} = Q_c \int_{\chi_B}^{\chi_A} K(\chi) {\rm d}\chi,
\eq
where the superscript NGP stands for nearest grid point. We note that, in this way, the fields being integrated do not vary continuously when crossing cell boundaries.

Another, more involved, way to construct the field at an arbitrary point inside the cell is via trilinear interpolation of the values of $Q$ at each cell vertex:
\bq\label{eq:trilinear_int}
Q(x,y,z) &=& \alpha_1 + \alpha_2\Delta_x + \alpha_3\Delta_y + \alpha_4\Delta_z + \alpha_5\Delta_x\Delta_y \nonumber \\
               &+&  \alpha_6\Delta_y\Delta_z + \alpha_7\Delta_x\Delta_z + \alpha_8\Delta_x\Delta_y\Delta_z,
\eq
where the $\alpha_i$'s are given by
\bq\label{eq:trilinear_coef}
\alpha_1 &=& Q_1,\nonumber \\
\alpha_2 &=& Q_2 - Q_1,\nonumber \\
\alpha_3 &=& Q_3 - Q_1,\nonumber \\
\alpha_4 &=& Q_5 - Q_1,\nonumber \\
\alpha_5 &=& Q_4 - Q_3 - Q_2 + Q_1,\nonumber \\
\alpha_6 &=& Q_7 - Q_5 - Q_3 + Q_1,\nonumber \\
\alpha_7 &=& Q_6 - Q_5 - Q_2 + Q_1,\nonumber \\
\alpha_8 &=& Q_8 - Q_7 - Q_6 - Q_4 + Q_2 + Q_5 + Q_3 - Q_1,
\eq
in which $Q_i$ denote the values of $Q(x,y,z)$ at the cell vertices $P_i$, as indicated in Fig.~\ref{fig:cell}. In Eq.~(\ref{eq:trilinear_int}), $\Delta_x$, $\Delta_y$ and $\Delta_z$ are given by
\bq\label{eq:Deltas}
\Delta_x &=& \frac{x_{\rm ray} - x_1}{h} = \frac{a + \left(\chi - \chi_A\right){\rm sin}\theta{\rm cos}\varphi}{h}, \nonumber \\
\Delta_y &=& \frac{y_{\rm ray} - y_1}{h} = \frac{b + \left(\chi - \chi_A\right){\rm sin}\theta{\rm sin}\varphi}{h}, \nonumber \\
\Delta_z &=& \frac{z_{\rm ray} - z_1}{h} = \frac{c +  \left(\chi - \chi_A\right){\rm cos}\theta}{h},
\eq
where $h$ is the cell size, and $\left(x_{\rm ray},y_{\rm ray},z_{\rm ray}\right)$, $\left(x_1,y_1,z_1\right)$ and $(a,b,c)$ are, respectively, the coordinates of the given ray (cf.~Eq.~(\ref{eq:raypos})), the point $P_1$, and the point $A$ w.r.t.~point $P_1$.

Using Eqs.~(\ref{eq:trilinear_int}), (\ref{eq:trilinear_coef}) and (\ref{eq:Deltas}), it is possible to compactify the expression for $Q$ as
\bq\label{eq:Qcompact}
Q\left(\chi, \theta, \varphi\right) = \sum_{N=1}^4 d_N\left(\chi - \chi_A\right)^{N-1},
\eq
where the coefficients $d_N$ depend on $a,b,c,\alpha_i, \theta$ and $\varphi$, but not on $\chi$ (note also that we are now specifying $Q$ as a function of spherical coordinates). Their expression is given in Appendix \ref{app:dns}. Since the $d_N$'s do not depend on $\chi$, one can combine Eqs.~(\ref{eq:Ic}) and (\ref{eq:Qcompact}) to write
\bq\label{eq:Icfinal}
I_c = \sum_{N=1}^4 d_N \int_{\chi_B}^{\chi_A} K(\chi) \left(\chi - \chi_A\right)^{N-1}{\rm d}\chi.
\eq
The integrand in the above equation is a polynomial in $\chi$, and so the integral can be solved analytically \cite{li2001}. Therefore, the calculation of the integral in the cell reduces to the geometrical exercise of determining the position of points $A$ and $B$ and the evaluation of the quantity $Q$ at the cell vertices. In Secs.~\ref{sec:lenscode} and \ref{sec:lenscode_alt} we shall see a concrete example of the use of these formulae when we apply it to lensing.

Equations (\ref{eq:Icngp}) and (\ref{eq:Icfinal}) provide two possible ways to compute the final result. The latter has the advantage of allowing for the integration of a continuous field when crossing cell boundaries, by appropriately evaluating the fields at the vertices of the cells (see next section). On the other hand, the use of Eq.~(\ref{eq:Icngp}) does not involve evaluating the fields at the cell vertices (which do not exist in default {\tt RAMSES}), making it therefore more computationally efficient. In Sec.~\ref{sec:methods}, we compare results based on Eqs.~(\ref{eq:Icngp}) and (\ref{eq:Icfinal}).

\subsection{Field values at cell vertices}\label{sec:vertices}

\begin{figure}
	\centering
	\includegraphics[scale=0.35]{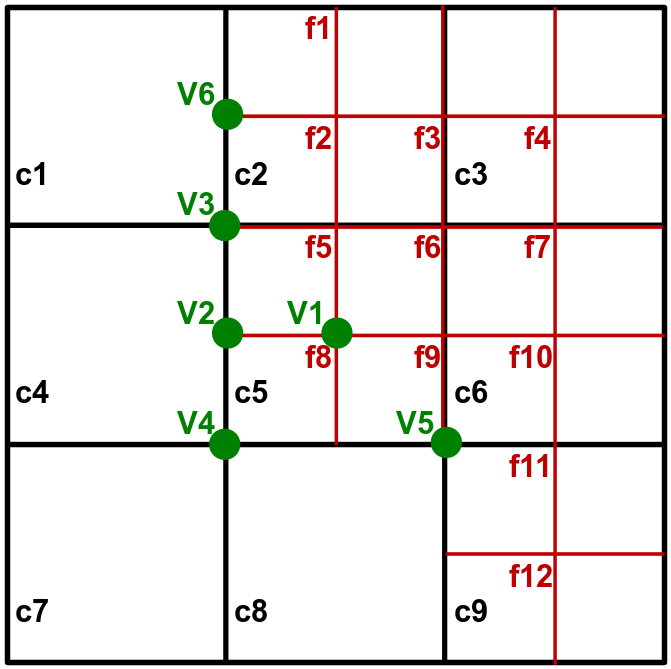}
	\caption{Two-dimensional illustration of a boundary region. The coarse cells are labelled from c1, c2, etc., and a few fine cells are labelled f1, f2, etc. The vertices V1 to V6 are illustrative vertices where the fields values are to be evaluated from interpolation from the cell centers, which is where {\tt RAMSES} evaluates the fields by default.}
\label{fig:2dgrid}
\end{figure}

In default {\tt RAMSES}, the field values are evaluated at cell centers, but the application of Eq.~(\ref{eq:Icfinal}) above requires them to be interpolated to cell vertices. When doing so, one can ensure that the fields reconstructed with trilinear interpolation (cf.~Eq.~(\ref{eq:trilinear_int})) vary continuously from cell to cell and also that the total mass \footnote{In this subsection, we use the word "mass" to denote the product of a surface or volume density with an area or volume.} is preserved. This is trivial on a regular mesh without AMR, but requires special care at the boundary of refined regions. We discuss these subtle issues in this subsection.

Figure \ref{fig:2dgrid} illustrates a refinement boundary in two dimensions (we shall later generalize the discussions to three dimensions). In the figure, the nine coarse level cells shown are labelled from c1 to c9 and a few fine level cells are labelled from f1 to f12. The vertices V1 to V6 represent different types of vertices at which one wishes the evaluate the fields. To be concrete, we shall take the example of the matter density field, $\rho$, but the interpolation scheme that we describe here is applied to any type of field.

For vertices of the type of vertex V1, which is shared by four cells of the same level, one can simply set $\rho_{\rm V1} = \big(\rho_{\rm f5} + \rho_{\rm f6} + \rho_{\rm f8} + \rho_{\rm f9}\big)/4$, where $\rho_{i}$ are the values at the relevant cell centers. This average ensures that $\rho$ varies continuously from one cell to another when computed with Eq.~(\ref{eq:trilinear_int}) (or its two-dimensional version). In regular mesh simulations, this is the only type of vertex. The vertex V2, which lies at the middle of a coarse cell edge and is shared by two fine cells, can be straightforwardly computed after the values in vertices V3 and V4 have been determined. Explicitly, $\rho_{\rm V2} = \big(\rho_{\rm V3} + \rho_{\rm V4}\big)/2$. This ensures, for instance, that two rays that are infinitesimally close to vertex V2, with one crossing cell c4 and another crossing fine cells f5 or f8, experience the same density field (as reconstructed with trilinear interpolation), i.e.,  there is no sharp discontinuity between the density experienced by one ray and the other.

The cases of vertices V3, V4 and V5 are more involved as they are shared by both coarse and fine cells. Let us consider vertex V3 as an example, which is shared by cells c1, c4, f2 and f5. It is natural to assume that the density at V3 depends on the density in each of these four cells. The simplest way to write this is as
\bq\label{eq:rhov3_1}
\rho_{\rm V3} = \alpha_{\rm c1}\rho_{\rm c1} + \alpha_{\rm c4}\rho_{\rm c4} + \alpha_{\rm f2}\rho_{\rm f2} + \alpha_{\rm f5}\rho_{\rm f5},
\eq
where the $\alpha$'s are some weights to be determined. Let us focus on $\alpha_{\rm c1}$. The mass associated with cell c1 is $\rho_{\rm c1}h^2$, where $h$ in this subsection is the coarse cell size. After the interpolation, we want the mass of this cell to be the sum of the masses associated with each of its vertices. Hence, the mass associated with vertex V3 due to cell c1 is $m_{\rm V3}^{\rm c1} = \rho_{\rm c1}h^2/4$. Due to mass conservation, the value of $m_{\rm V3}^{\rm c1}$ is {\it re-distributed} by vertex V3 to other cells at the boundary. In this sense, we can colloquially describe vertex V3 as a {\it mass reservoir that is collecting mass from c1 and redistributing it to neighbouring cells}. This mass distribution constraint can be written as
\bq\label{eq:massdist}
\frac{\rho_{\rm c1}h^2}{4} & =& \frac{\alpha_{\rm c1}\rho_{\rm c1}}{4} \Bigg(h^2 + h^2 + \left(\frac{h}{2}\right)^2 + \left(\frac{h}{2}\right)^2 \nonumber \\
&+& \frac{1}{2}\left(\frac{h}{2}\right)^2 +  \frac{1}{2}\left(\frac{h}{2}\right)^2 + \frac{1}{2}\left(\frac{h}{2}\right)^2 +  \frac{1}{2}\left(\frac{h}{2}\right)^2\Bigg). \nonumber \\
\eq
In this equation, the first two terms on the RHS represent the {\it mass from V3 due to cell c1 that is redistributed to cells c1 and c4} ($\alpha_{\rm c1}\rho_{\rm c1}/4$ is the density that V3 contributes to c1 and c4 and $h^2$ is the area of cells c1 and c4). The third and forth terms are the same as the first two, but for cells f2 and f5 (note that $(h/2)^2$ is the area of the fine cells). The last four terms on the RHS of Eq.~(\ref{eq:massdist}) must be included as vertex V3 also contributes to the masses in cells f1, f2, f5 and f8 via vertices V2 and V6. Since the contribution from V3 to $\rho_{\rm V2}$ and $\rho_{\rm V6}$ is only  $\rho_{\rm V3}/2$, each of these four terms gets a factor of $1/2$ compared to the third and forth terms. We can solve for $\alpha_{\rm c1}$ in Eq.~(\ref{eq:massdist}) and, to facilitate the discussions below, we write the result as
\bq\label{eq:alphaconst}
\alpha_{c1} = \left(N_c + \frac{1}{4}\sum_{i}^{N_f}N_{fv,i}\right)^{-1},
\eq
where $N_c$ is the number of coarse cells that share vertex V3, $N_{f}$ is the number of fine cells that share vertex V3 and $N_{fv,i}$ is the number of vertices that the fine cell $i$ that shares V3 has at the boundary ($i =$ f2, f5). Explicitly, for V3 we have $N_c = 2$, $N_{f} = 2$ and $N_{fv,\rm{f2}} = N_{fv,\rm{f5}} = 2$. The coefficient $\alpha_{\rm c4}$ in Eq.~(\ref{eq:rhov3_1}) is obtained in the same way: Eq.~(\ref{eq:massdist}) remains the same, just with $\rho_{\rm c1}$ replaced by $\rho_{\rm c4}$ (which appears on both sides of the equation and therefore cancels). Hence, $\alpha_{\rm c1}=\alpha_{\rm c4}$. 

The determination of the remaining coefficients, $\alpha_{\rm f2}$ and $\alpha_{\rm f5}$ differs from $\alpha_{\rm c1}$ and $\alpha_{\rm c4}$. To determine $\alpha_{\rm f2}$, the RHS of Eq.~(\ref{eq:massdist}) remains the same, just with $\rho_{\rm c1}$ and $\alpha_{\rm c1}$ replaced by $\rho_{\rm f2}$ and $\alpha_{\rm f2}$, respectively. From the reasoning that led to Eq.~\ref{eq:massdist}, one could naively think that the LHS would be simply given by {\it the mass that vertex V3 collects from f2}, $m_{V3}^{f2} = \rho_{f2}(h/2)^2/4$. However, recall that vertex V3 also contributes to the mass in cells f2 and f1 via V6 (this is why two of the last four terms in Eq.~(\ref{eq:massdist}) appear). As a result, the mass that {\it V6 collects from f2}, $m_{V6}^{f2} = \rho_{f2}(h/2)^2/4$ must also be included in the LHS of the mass distribution equation that determines $\alpha_{\rm f2}$. Explicitly,
\bq\label{eq:massdist2}
\frac{\rho_{\rm f2}}{2}\left(\frac{h}{2}\right)^2 &=&  \frac{\alpha_{\rm f2}\rho_{\rm f2}}{4} \Bigg(h^2 + h^2 + \left(\frac{h}{2}\right)^2 + \left(\frac{h}{2}\right)^2 \nonumber \\
&+& \frac{1}{2}\left(\frac{h}{2}\right)^2 +  \frac{1}{2}\left(\frac{h}{2}\right)^2 + \frac{1}{2}\left(\frac{h}{2}\right)^2 +  \frac{1}{2}\left(\frac{h}{2}\right)^2\Bigg). \nonumber \\
\eq
The equation for $\alpha_{\rm f5}$ is the same (apart from $_{\rm {f2}} \rightarrow _{\rm f5}$) and the equation for both can be written as
\bq\label{eq:alphaconst2}
\alpha_{\rm j} = \frac{N_{fv,j}}{4}\left(N_c + \frac{1}{4}\sum_{i}^{N_f}N_{fv,i}\right)^{-1},
\eq
where $i,j =$ f2, f5. The meaning of $N_c$, $N_f$ and $N_{fv, i}$ is the same as above and since $N_{fv,\rm f2}=N_{fv,\rm f5}$, then $\alpha_{\rm f2} = \alpha_{\rm f5}$

Equations (\ref{eq:alphaconst}) and (\ref{eq:alphaconst2}) were written in terms of $N_c$, $N_f$ and $N_{fv,i}$ because this way they also hold for the other vertices. By following the same steps for V4 and V5 we can write:
\bq
\label{eq:rhov4_1} \rho_{\rm V4} &=& \alpha_{\rm c4}\rho_{\rm c4} + \alpha_{\rm c7}\rho_{\rm c7} + \alpha_{\rm c8}\rho_{\rm c8} + \alpha_{\rm f8}\rho_{\rm f8}, \\
\label{eq:rhov5_1} \rho_{\rm V5} &=& \alpha_{\rm c8}\rho_{\rm c8} + \alpha_{\rm f9}\rho_{\rm f9} + \alpha_{\rm f10}\rho_{\rm f10} + \alpha_{\rm f11}\rho_{\rm f11},
\eq
where the coefficients associated with coarse and fine cells are obtained as in Eq.~(\ref{eq:alphaconst}) and (\ref{eq:alphaconst2}), respectively. For V4, $N_c = 3$, $N_f$ = 1 and $N_{fv,{\rm f8}} = 3$, whereas for V5 $N_c = 1$, $N_f = 3$, $N_{fv,{\rm f9}} = 2$, $N_{fv,{\rm f10}} = 1$ and $N_{fv,{\rm f11}} = 2$. Note that in Eqs.~(\ref{eq:rhov3_1}), (\ref{eq:rhov4_1}) and (\ref{eq:rhov5_1}), the summed value of the $\alpha$ weights adds up to unity, as it should.

\subsubsection{Generalization to three dimensions}

In three dimensions, the above derivation holds with only a few generalizations. When writing the RHS of mass distribution equations, Eqs.~(\ref{eq:massdist}) and (\ref{eq:massdist2}), for each vertex, in addition to considering the contribution from vertices that lie at the edge of coarse cell vertices (which get a factor of $1/2$), one must also consider the contribution to vertices that lie at the center of the coarse cell faces, which get a factor of $1/4$. Moreover, in two dimensions, the ratio of the area of a fine to coarse cell is $1/4$, whereas in three dimensions the ratio of their volumes is $1/8$. Bearing these two things in mind, it is possible to show that the weights associated with coarse cells are given by
\bq\label{eq:alphac3d}
\alpha_{\rm coarse} = \left(N_c + \frac{1}{8}\sum_{i}^{N_f}N_{fv,i}\right)^{-1},
\eq
and the weights associated with fine cells by
\bq\label{eq:alphaf3d}
\alpha_{j} = \frac{N_{fv,j}}{8}\left(N_c + \frac{1}{8}\sum_{i}^{N_f}N_{fv,i}\right)^{-1}.
\eq
These expressions differ from their two-dimensional counterparts by replacing the factors of $1/4$ by $1/8$. The meaning of $N_c$, $N_{f}$ and $N_{fv,i}$ is the same as in two dimensions. Just to give an example, consider a vertex V that is shared by seven coarse cells and one fine cell. In the scheme described above, the density at this vertex is
\bq\label{eq:rhoV_1}
\rho_{\rm V} = \alpha_{\rm coarse} \sum_{i = 1}^7 \rho_{i, {\rm coarse}} + \alpha_{\rm fine}\rho_{\rm fine},
\eq
where $\rho_{i, {\rm coarse}}$ is the density at the center of the $i$-th coarse cell and $\rho_{\rm fine}$ is the density at the center of the fine cell. For this case, $N_c = 7$, $N_{f} = 1$ and $N_{fv, {\rm fine}} = 7$ in Eqs.~(\ref{eq:alphac3d}) and (\ref{eq:alphaf3d}).

Analogously to the two dimensional case, once the density at the vertices that are shared by both coarse and fine cells is determined, then (i) the density at vertices that lie at the middle of a coarse cell edge is given by the average of the densities of the two coarse cell vertices of that edge; and (ii) the densities at vertices that lie at the center of a coarse cell face is given by the average of the densities at the four coarse cell vertices of that face. The density at vertices that are shared by eight cells of the same level (i.e.~not in a refinement boundary) is given by the average value of the density at those eight cell centers.

As a test of our interpolation scheme, we have measured the total mass inside simulation boxes by using the values at the cell centers and at the cell vertices. The agreement between the two ways of measuring the total mass was perfect for meshes with and without refinements, which confirms that the design and implementation of our interpolation is correct. We note also that these operations to interpolate the field values from cell centers to cell vertices naturally add some computational costs to the code, and hence, it is desirable to reduce the number of times these operations should be performed. For instance, since a single cell can contain a large number of rays, the interpolation needs to be performed only once to compute the integral for all rays. Morevoer, if the fields at the vertices of a given cell do not change from one time step to the other (e.g., if it is a coarse cell that is not at a refinement boundary and the time step taken was a fine one), then one can also store the interpolated values from the previous time step, thereby saving some computational time.

Before proceeding, we note that the interpolation scheme, as describe in this section, represents in practice a form of adaptive smoothing of the fields used in the ray integration \cite{Hilbert:2008kb}. This is in the sense that the field values at a given cell vertices (and hence the interpolated field inside that cell) depend on the field values on neighbouring cells. The reason why we dub this an adaptive smoothing is because the size of the {\it smoothing kernel} (roughly the volume occupied by all the cells that are used in the interpolation) depends on the sizes of the given cell and its neighbours, which in turn depends on the local matter density.

\section{Weak Lensing simulations: method}\label{sec:lens}

In this section, we explain how our algorithm can be applied to studies of weak gravitational lensing. 

\subsection{Lensing basics}\label{sec:basics}

We start with a brief recap of the basics of gravitational lensing (see e.g.~Refs.~\cite{2001PhR...340..291B, 2003ARA&A..41..645R, 2005PhRvD..72b3516C, 2010PhRvD..81h3002B, 2010CQGra..27w3001B, 2015RPPh...78h6901K}). In a perturbed Friedmann-Robertson-Walker (FRW) spacetime, the line element in the absence of anisotropic stress can be written as (considering only scalar perturbations)
\bq\label{eq:frw}
{\rm d}s^2 = \left(1 + \frac{2\Phi}{c^2}\right)c^2{\rm d}t^2 - a^2\left(1 - \frac{2\Phi}{c^2}\right){\rm d}s_{\rm space}^2,
\eq
where $a = 1/(1+z)$ is the scale factor. Photons travelling from distant sources towards an observer get their trajectories bent due to the intervening gravitational potential, $\Phi$. The (unobserved) angular position of the source on the source plane, $\vec{\beta}$, is related to the observed one, $\vec{\theta}$, by the lensing deflection angle $\vec{\alpha}$ as
\bq\label{eq:lenseq}
\beta^i &=& \theta^i + \alpha^i \nonumber \\
 &=& \theta^i - \frac{2}{c^2}\int_0^{\chi_s}\frac{\left(\chi_s - \chi\right)}{\chi_s}\nabla^{x^i}\Phi(\chi, \vec{\beta}(\chi)){\rm d}\chi \nonumber \\
&=& \theta^i - \frac{2}{c^2}\int_0^{\chi_s}\frac{\left(\chi_s - \chi\right)\chi}{\chi_s}\nabla^{\beta^i}\Phi(\chi, \vec{\beta}(\chi)){\rm d}\chi,
\eq
where $i = 1,2$ denotes the two perpendicular directions to the line-of-sight. The third line in Eq.~(\ref{eq:lenseq}) is obtained from the second one by defining the derivatives w.r.t.~the angular coordinate $ \nabla_{\beta^i} = \chi\nabla_{x^i}$, or equivalently, $\nabla^{\beta^i} = \nabla^{x^i}/\chi$. The Jacobian matrix, $A^i_{j}$, of this source-to-observer mapping is obtained by differentiating the above equation w.r.t. $\vec{\theta}$ as in
\bq\label{eq:jacobian}
A^i_{j} = \nabla_{\theta^j}\beta^i= \delta^i_{j} - \frac{2}{c^2}\int_0^{\chi_s}g\left(\chi_s, \chi\right)\nabla^{\beta^i}\nabla_{\theta^j}\Phi(\chi, \vec{\beta}(\chi)){\rm d}\chi, \nonumber \\
\eq
where $g\left(\chi_s, \chi\right) = \left(\chi_s - \chi\right)\chi/\chi_s$. Note that the integral is performed along the perturbed path of the photon, as indicated by the $\beta(\chi)$ dependence of the potential inside the integral. Note also that one of the derivatives of $\Phi$ is w.r.t. $\vec{\beta}$ and another w.r.t. $\vec{\theta}$. These two aspects add complication to the ray tracing, but they can be neglected to obtain approximate solutions. To first order, we can write
\bq\label{eq:jacobian2}
A^i_{j} = \delta^i_{j} - \frac{2}{c^2}\int_0^{\chi_s}g\left(\chi_s, \chi\right)\nabla^{\theta^i}\nabla_{\theta^j}\Phi(\chi, \vec{\theta}(\chi)){\rm d}\chi, \nonumber \\
\eq
in which the integral is now peformed along the unperturbed apparent direction of the photons, which is the so-called {\it Born approximation}, and the derivatives are now both w.r.t.~$\vec{\theta}$, which amounts to neglecting the so-called {\it lens-lens coupling}\footnote{Lens-lens coupling refers to the correlation between the distortions of the sources with the intervening sources that act as lenses, whose images and positions seen by the observer are also distorted.}). The lensing results that we present in this paper are obtained under these two approximations, which are generally found to be valid, at least in what concerns determinations of the power spectrum of lensing quantities \cite{Hilbert:2008kb}. The generalization of our ray tracing calculations to follow the rays in their perturbed paths, as well as calculations that take lens-lens coupling into account is the subject of ongoing work (see e.g.~the Appendix of Ref.~\cite{li2001} for a discussion).

Equation~(\ref{eq:jacobian2}) can be written in matrix form
\begin{equation}\everymath{\displaystyle}\hat{A}\label{eq:jacobian3}
=
\begin{bmatrix}
1 - \kappa - \gamma_1 & -\gamma_2  \\
-\gamma_2 & 1 - \kappa + \gamma_1
\end{bmatrix},
\end{equation}
to define the {\it lensing convergence}, $\kappa$,
\bq\label{eq:kappa}
\kappa &=& 1  - \left(A^1_1 + A^2_2\right)/2 \nonumber \\
 &=& \frac{1}{c^2}\int_0^{\chi_s}g\left(\chi_s, \chi \right)\left[\nabla^{1}\nabla_1\Phi + \nabla^{2}\nabla_2\Phi\right]{\rm d}\chi,
\eq
and the two components of the lensing shear $\vec{\gamma} = \left(\gamma_1, \gamma_2\right)$,
\bq
\label{eq:gamma1} \gamma_1 &=& -\left(A^1_1 - A^2_2\right)/2 \nonumber \\ 
&=& \frac{1}{c^2}\int_0^{\chi_s}g\left(\chi_s, \chi \right)\left[\nabla^1\nabla_1\Phi - \nabla^2\nabla_2\Phi\right]{\rm d}\chi, \\
\label{eq:gamma2} \gamma_2 &=& -A^1_2 = -A^2_1\nonumber \\
&=& \frac{2}{c^2}\int_0^{\chi_s}g\left(\chi_s, \chi \right)\nabla^1\nabla_2\Phi{\rm d}\chi,
\eq
where we have now denoted $\nabla_i \equiv \nabla_{\theta^i}$ for compactness of notation.

\subsection{Lensing integration in the code}\label{sec:lenscode}

The integrals of Eqs.~(\ref{eq:kappa}), (\ref{eq:gamma1}) and (\ref{eq:gamma2}) can be found by using the algorithm outlined in Sec.~\ref{sec:inte}. In the case of lensing, the integration kernel in Eq.~(\ref{eq:Ic}) is given by (up to factors $\propto 1/c^2$) $K \equiv g(\chi_s, \chi)$, and the quantity $Q$ that one needs to evaluate at cell vertices is $Q = \nabla^1\nabla_1\Phi + \nabla^2\nabla_2\Phi$ for $\kappa$, $Q = \nabla^1\nabla_1\Phi - \nabla^2\nabla_2\Phi$ for $\gamma_1$, and $Q = \nabla^1\nabla_2\Phi$, for $\gamma_2$. Consequently, the contribution of each crossed cell to the integral according to method NGP (cf.~Eq.~(\ref{eq:Icngp})) is given by
\bq\label{eq:Iclensing_ngp}
I_c &=& Q_c \int_{\chi_B}^{\chi_A} \frac{\chi_s - \chi}{\chi_s} \chi {\rm d}\chi \nonumber \\
&=& \frac{Q_c}{\chi_s}\left[\frac{\chi_s}{2}\left(\chi_A^2 - \chi_B^2\right) - \frac{1}{3}\left(\chi_A^3 -\chi_B^3\right)\right].
\eq
Alternatively, according to Eq.~(\ref{eq:Icfinal}), the contribution is
\bq\label{eq:Iclensing}
I_c &=& \frac{1}{\chi_s}\sum_{N=1}^4 d_N \int_{\chi_B}^{\chi_A} \chi\left(\chi_s-\chi\right) \left(\chi - \chi_A\right)^{N-1}{\rm d}\chi \nonumber \\
&=&\frac{1}{\chi_s}\sum_{N=1}^4 d_N \Bigg[\frac{R^{N}}{N}\left(\chi_A - \chi_s\right)\chi_A + \frac{R^{N+1}}{N+1}\left(2\chi_A - \chi_s\right) \nonumber \\ 
&&\ \ \ \ \ \ \ \ \ \ \ \ \ \ \ \ \ \ \ \ \ \ \ \ \ \ \ \ \ \ \ \ \ \ \ \ \ \ \ \ \ \ \ \ \ \ \ \ \ \ \ \ \  + \frac{R^{N+2}}{N+2}\Bigg],
\eq
where $R = \chi_B - \chi_A$.

What is left to specify is the relation between the quantities $\nabla^1\nabla_1\Phi$, $\nabla^2\nabla_2\Phi$ and $\nabla^1\nabla_2\Phi$, with the values of $\partial_a\partial_b\Phi$ ($a,b = x,y,z$) that are actually computed on the simulation mesh (see the subsection below). This relation is
\begin{widetext}
\bq
\label{eq:n1n1}\nabla^1\nabla_1 \Phi & = & \sintwophi\pxpx + \costwophi\pypy - {\rm sin}2\varphi\pxpy, \\
&& \nonumber \\
\label{eq:n2n2}\nabla^2\nabla_2 \Phi & = &   \costwophi\costwothe\pxpx + \sintwophi\costwothe\pypy + \sintwothe\pzpz + {\rm sin}2\varphi\costwothe\pxpy - \sinphi{\rm sin}2\theta \pypz \nonumber \\ 
&& - \cosphi{\rm sin}2\theta \pxpz ,\\
\label{eq:n1n2}\nabla^1\nabla_2 \Phi & = & \frac{\costhe}{\sinthe}\Bigg[\cosphi\sinphi\Big(\pypy - \pxpx\Big) + \Big(\costwophi - \sintwophi\Big)\pxpy\Bigg]  + \sinphi\pxpz  - \cosphi\pypz.
\eq
\end{widetext}

The above equations are derived by associating the two spherical coordinates $\varphi$ and $\theta$ that specify the incoming direction of the rays with $\theta_1$ and $\theta_2$. Then, the expressions follow straightforwardly from applying $\nabla_i\nabla_j\Phi = \partial_i\partial_j\Phi - \Gamma^{k}_{ij}\partial_k\Phi$, with $\Gamma^{k}_{ij}$ being the Christoffel symbols of the line element\footnote{We note in passing that by using this line element one takes into account the curvature of the sky. However, Eqs.~(\ref{eq:n1n1}), (\ref{eq:n2n2}) and (\ref{eq:n1n2}) remain the same if $\nabla_1$ and $\nabla_2$ are interpreted as being derivatives w.r.t.~the coordinates $\left(x_1, x_2\right) = \chi(\theta_1, \theta_2)$, i.e., by taking the {\it flat sky} approximation. This coordinate system is essentially a Cartesian system rotated such that the direction of the incoming ray is perpendicular to the $x_1-x_2$ plane. A simple argument for this equivalence is that the sphere is locally flat, which means that the curvature can in practice be neglected when one takes derivatives.} ${\rm d}s_{\rm space}^2 = {\rm d}\chi^2 + \chi^2{\rm d}\theta^2 + \chi^2{\rm sin}^2\theta{\rm d}\varphi^2$, and then writing $\partial_\varphi$ and $\partial_\theta$ in terms of $\partial_x$, $\partial_y$, $\partial_z$ according to Eq.~(\ref{eq:raypos}).

\subsection{Calculation of $\partial_a\partial_b\Phi$}\label{sec:tidalincode}

The values of $\partial_a\partial_b\Phi$ ($a,b = x,y,z$) can be computed at the center of a given cell by finite differencing the values of $\Phi$ on neighbouring cells. If a cell has all its neighbours at the same refinement level, this calculation is straightforward. For instance, if $\Phi_{i,j,k}$ is the gravitational potential on the cell labelled by $\left\{i,j,k\right\}$, then we have
\bq\label{eq:tidalexample}
\partial_x\partial_x\Phi &=& \frac{\Phi_{i-1,j,k} - 2\Phi_{i,j,k} + \Phi_{i+1,j,k}}{h^2} \nonumber \\
\partial_x\partial_y\Phi &=& \frac{\Phi_{i+1,j+1,k} + \Phi_{i-1,j-1,k} - \Phi_{i+1,j-1,k} - \Phi_{i-1,j+1,k}}{4h^2}, \nonumber \\
\eq
as two representative examples. The other components of $\partial_a\partial_b\Phi$ are obtained similarly. However, some complications arise at boundary regions of refinements. As an example, consider that we wish to compute $\partial_x\partial_x\Phi$ on cell f5 in Fig.~\ref{fig:2dgrid}, where $x$ and $y$ are, respectively, the horizontal and vertical directions on the figure. The fine cell f5 is missing the neighbour that would exist if coarse cell c4 had been refined. One can think of two ways to compute the missing values that are needed for the finite difference. One option is to interpolate the values of $\Phi$ obtained from the coarse level to the point in cell c4 where the center of the relevant son cell would be if it existed. This value could then be used in a fine-level finite difference to compute $\partial_x\partial_x\Phi$ in cell f5. Another option is interpolating directly the coarse values of $\partial_x\partial_x\Phi$ in cells c1, c2, c4 and c5 to the center of cell f5, without finite differencing. 

To test these two approaches, we have set up a grid with more that one refinement level in the code and used the cell centers to define a Gaussian potential on the mesh. We then compared the analytical result of $\partial_a\partial_b\Phi$ with the result given by the code. We have found that the second approach agrees very well with the analytical result, but the first option showed larger discrepancies at the refinement boundaries. This is because by taking the finite difference using interpolated values, one amplifies the interpolation error in $\Phi$ by the factor of $h^{-2}$, which enters in the finite difference. In the results that follow, we have therefore implemented the second approach, which is also more computationally efficient\footnote{As a technical point, imagine that there is a CPU domain along the line that contains vertices V6, V2, V3 and V4 in Fig.~\ref{fig:2dgrid}.  In {\tt RAMSES}, there are "communication buffers" at the CPU domain boundaries, i.e., regions in the next CPU's domain that are available to the present CPU. In order to compute the value of $\partial_a\partial_b\Phi$ in cell f5, then its CPU needs to access the value of $\partial_a\partial_b\Phi$ at cell C4, whose calculation involves $\Phi$ on cells further left of C4 (not shown in Fig.~\ref{fig:2dgrid}). These latter cells are outside of the communication buffer of standard {\tt RAMSES}, which means that we had to increase its size. This is one of the few changes made to the main code.}. Once $\partial_a\partial_b\Phi$ is evaluted at the center of the cells, then its interpolation to cell vertices is as described in Sec.~\ref{sec:vertices}.

\subsection{Alternative lensing integration in the code}\label{sec:lenscode_alt}

The lensing methodology described above involves the calculation of $\partial_a\partial_b\Phi$ ($a,b = x,y,z$) on the mesh, which inevitably adds some computational overheads. However, as described in this section, it is possible to compute $\kappa$ and $\gamma$ by integrating only quantities that are computed by default in {\tt RAMSES}.

Equation (\ref{eq:kappa}) can be written as
\begin{widetext}
\bq\label{eq:kappa2}
c^2\kappa &=& \int_0^{\chi_s} g\left(\chi_s, \chi\right)\left[\nabla^2\Phi - \nabla^2_\chi\Phi\right]{\rm d}\chi \nonumber \\
&=& \frac{3}{2}\Omega_{m0}H_0^2 \int_0^{\chi_s} g\left(\chi_s, \chi\right) \frac{\delta}{a}{\rm d}\chi \ \ \ + \ \ \ \int_0^{\chi_s}\nabla_\chi\Phi\ \partial_\chi g{\rm d}\chi \ \ \ + \ \ \ \frac{1}{c}\int_0^{\chi_s}g\ \partial_t\left(\nabla_\chi\Phi\right){\rm d}\chi \ \ \ - \ \ \   g\nabla_\chi\Phi\Big|_0^{\chi_s},
\eq
\end{widetext}
where in the second equality we have used the Poisson equation to relate the comoving three-dimensional Laplacian to the matter density contrast, $\delta$, and the term $g\nabla^2_\chi\Phi$ was integrated by parts using $\nabla_\chi = \partial_t/c + \partial_\chi$, where $t$ is the physical time. The integration of the first term on the RHS of Eq.~(\ref{eq:kappa2}) is the same as that of Eqs.~(\ref{eq:kappa}), (\ref{eq:gamma1}) and (\ref{eq:gamma2}), but with $\delta/a$ as the quantity $Q$\footnote{In the current implementation of the code, $a$ is taken to be constant during the time step integration. This should not lead to big errors in high-resolution simulations if the time steps are sufficiently small. There are however ways to go beyond this by, for instance, implementing the relation $a(\chi)$ in the integration.}. The second term is obtained analogously, but with $Q = \nabla_\chi\Phi = \sinthe\cosphi\partial_x\Phi + \sinthe\sinphi\partial_y\Phi + \costhe\partial_z\Phi$ (where $\partial_x\Phi$, $\partial_y\Phi$ and $\partial_z\Phi$ are the negative of the three components of the gravitational force) and the kernel $K = \partial_\chi g = 1 - 2\chi/\chi_s$. The third term involves the time derivative of the potential, which is not calculated in {\tt RAMSES} by default. In the results presented in this paper, we neglect the contribution from this term, which is expected to be small anyway (see e.g.~Ref.~\cite{2000ApJ...530..547J}).

The last term on the RHS of Eq.~(\ref{eq:kappa2}), which is a surface term, is exactly zero in theory, since the lensing kernel $g$ vanishes at $\chi = 0$ and $\chi = \chi_s$. However, if one breaks down the calculation of this term into the contributions coming from each time step
\bq\label{eq:2nterm}
g\nabla_\chi\Phi\Big|_0^{\chi_s} = \sum_{\rm time\ steps}g\nabla_\chi\Phi\Big|_{\chi_{f}}^{\chi_{i}},
\eq
where $\chi_{i}$ and $\chi_{f}$ are, respectively, the values of $\chi$ at the start and end of a particle time step, then the result is not zero due to discontinuities between particle time steps and at the boundary of boxes in the tile. This is because, if {\tt RAMSES} updates the values of the potential from one time step to the next, then its value at the end of the current time step (when the ray is at $\chi = \chi_{f}^{t_{\rm current}}$) is not the same as at the start of the next time step (when the ray is at $\chi = \chi_{i}^{t_{\rm next}}$), where $\chi_{f}^{t_{\rm current}} = \chi_{i}^{t_{\rm next}}$ \footnote{That is, the ending point at the current time step is the starting point at the next time step integration.}. As a result, 
\bq\label{eq:2dterm2}
g\nabla_\chi\Phi\Big|_0^{\chi_s} &=&  g\nabla_\chi\Phi\Big|_{\chi = \chi_s} - g\nabla_\chi\Phi\Big|_{\chi = 0} + \epsilon_{\rm disc} \nonumber \\
&=& \epsilon_{\rm disc},
\eq
where $\epsilon_{\rm disc}$ denotes the cumulative error that arises due to the discontinuities at each time step. The latter are unavoidable since they are linked to the intrinsic discreteness of N-body simulations. Note however that the nonzero value of $\epsilon_{\rm disc}$, which comes from integrating the second term in the bracket of the first line of Eq.~(\ref{eq:kappa2}), means that the same discreteness also affects the integration of the first term there. Therefore, having $\epsilon_{\rm disc}$ included in the calculation for each time step can reduce the error in the density integral of Eq.~(\ref{eq:kappa2}) that comes from the same discreteness. We have checked explicitly that including $\epsilon_{\rm disc}$ brings the $\kappa$ map obtained using Eq.~(\ref{eq:kappa2}) into closer agreement with the result from Eq.~(\ref{eq:kappa}).

Note that the discreteness in time also introduces another source of error. For example, when computing the second integral in Eq.~(\ref{eq:kappa2}), the values of $\nabla_\chi\Phi$ are assumed to be constant in a time step, and therefore the integral misses the contribution that comes from the time evolution of fields within the time step. This also affects the integral of Eq.~(\ref{eq:kappa}). However, we expect this error to be small in cosmological simulations where the fine time step is typically of order $\Delta a \sim 10^{-4}$ or smaller. In principle, our algorithm can be straightforwardly extended to interpolate the fields between neighbouring time steps so that they are continuous in time; however, given the small error that this discontinuity causes, we will leave this for future implementations. 

In the end, we have two ways to compute the lensing convergence. One is that of Eq.~(\ref{eq:kappa}), which we call Method B and it involves only one integral term. The other, which we call Method A\footnote{This naming is the same as in Ref.~\cite{li2001}.}, uses Eq.~(\ref{eq:kappa2}) and in the current implementation of the code involves two integral terms and the inclusion of the surface term at each time step to cancel some of the errors of the integral terms. These two methods therefore respond differently to the discontinuities between time steps. {Another numerical difference between these two methods is that Method B involves the calculation of $\partial_a\partial_b\Phi$, which for each cell requires using a different number of field values on neighbouring cells, compared to the evaluation of $\delta$ or $\nabla_\chi\Phi$ in Method A. In Sec.~\ref{sec:methods}, we shall see that the two methods give consistent results, which is telling that these numerical differences are not critical.}

As we mentioned above, Method A is more convenient than Method B in the sense that it avoids the computationally expensive calculation of $\partial_a\partial_b\Phi$. On the other hand, it does not yield directly the lensing shear, which needs to be obtained indirectly from the $\kappa$ maps. This can be done by (i) Fourier transforming $\kappa(\vec{\theta})$ to obtain $\tilde{\kappa}\big(\vec{\ell}\big)$; (ii) computing the Fourier transform of the shear as \cite{1993ApJ...404..441K}
\bq\label{eq:gammaind}
\left(\tilde{\gamma}_1, \tilde{\gamma}_2\right) = \left(\frac{\ell_1^2 - \ell_2^2}{\ell_1^2 + \ell_2^2}\tilde{\kappa}, \frac{2\ell_1\ell_2}{\ell_1^2 + \ell_2^2}\tilde{\kappa}\right),
\eq
where $\ell_1 = \pi/\theta_1$, $\ell_2 = \pi/\theta_2$; and finally, (iii) inverse Fourier transforming $\left(\tilde{\gamma}_1, \tilde{\gamma}_2\right)$ to get $\left({\gamma}_1(\vec{\theta}), {\gamma}_2(\vec{\theta})\right)$ \footnote{Equation (\ref{eq:gammaind}) is valid only in the flat-sky approximation, in which Fourier modes and spherical harmonic multipoles are equivalent.}.

\section{A code test: lensing by a Gaussian potential}\label{sec:gaussian}

\begin{figure*}
	\centering
	\includegraphics[scale=0.39]{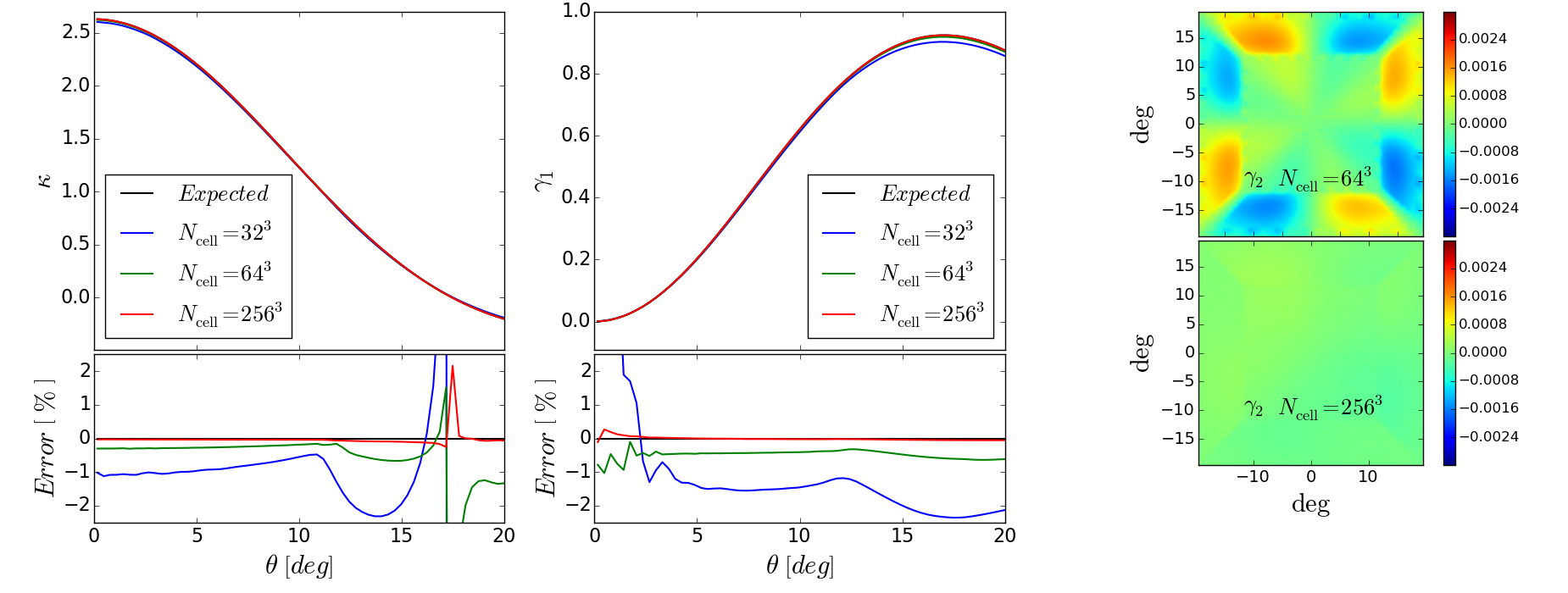}
	\caption{Code lensing tests with a fixed Gaussian potential. From left to right, the panels show the profiles of $\kappa$, $\gamma_1$ and the maps of $\gamma_2$, for different domain grid resolutions, as labelled. The black curve shows the expected result obtained using the same potential and ray geometry, but with the equations solved with an independent Runge-Kutta numerical integrator. The panels below the $\kappa$ and $\gamma_1$ profiles show the percentage error relative to the expected result. In the $\gamma_2$ panels, the expected result is $\gamma_2 = 0$.}
\label{fig:gaussian}
\end{figure*}

As a check of our ray tracing implementation for lensing, we tested the code results in a controlled setup for which we can obtain solutions with other integration methods. Specifically, we compute the lensing signal associated with a fixed Gaussian potential 
\bq\label{eq:gaussian}
\tilde{\Phi}(r) = A{\rm exp}\left[-\frac{r^2}{2\sigma^2}\right],
\eq
where the tilde denotes that the potential is written in code units. We set the parameters as $A = 10^4$ and $\sigma = 2\ {\rm Mpc}/h$ (the exact values are not critical for our tests). The center of the potential is located at the center of a box of size $L = 10\ {\rm Mpc}/h$ and the potential is defined on cell centers. The observer lies at the center of one of the box faces and we integrate $128\times128$ rays covering a $40\times40\ {\rm deg}^2$ field of view. The source redshift for this test is $z_s = 0.95$, but the rays are only integrated in the box that contains the observer and the Gaussian potential. We consider three domain mesh resolutions with $32$, $64$ and $256$ cells per dimension. For each domain mesh resolution, we hierarchically refine the mesh towards the inner region of the box using two cubic-shaped refined levels. That is, if $l_d$ is the domain level, then a ray moving away from the center of the potential will go through cells on level $l_d+2$, then on level $l_d+1$ and finally $l_d$. We create the refinement levels by appropriately distributing particles inside the box. However, these particles are only used to define the AMR structure and play no other role in this test, e.g., the potential felt by the rays is that of Eq.~(\ref{eq:gaussian}) and not that associated with the particle distribution. We let the code run as if it was a normal N-body simulation, but at each time step the particles are kept from moving to ensure that the AMR structure remains fixed. We ran this test on 8 CPUs using integration method B with Eq.~(\ref{eq:Iclensing}) (cf.~Sec.~\ref{sec:lenscode}), and compared our code results with the integral solutions obtained with an adaptive 1D numerical integrator from the GSL library \cite{Gough:2009:GSL:1538674} for the same ray settings.

The outcome of this test is shown in Fig.~\ref{fig:gaussian}, which displays $\kappa$ profiles (left), $\gamma_1$ profiles (middle) and $\gamma_2$ maps (right), for the tested resolutions (the $\kappa$ and $\gamma_1$ profiles correspond to a given radial slice of the corresponding maps). Figure \ref{fig:gaussian} illustrates the very good agreement between the expected result and that from our ray tracing code. In the $\kappa$ and $\gamma_1$ panels, the error (the difference between the expected result and the ray tracing one) for the highest mesh resolution is kept well within $1\%$. The error is also smaller than $1\%$ for the intermediate resolution, and even for the poorest resolution case it never exceeds $\approx 3\%$. When quoting these figures, we do not consider the radial scales where $\kappa$ and $\gamma_1$ cross zero, since this artificially amplifies the relative error there. We have checked that these small errors are mostly caused by errors in the interpolation of the potential values from cell centers to cell vertices\footnote{We checked this by comparing the reconstructed potential values at cell vertices (cf.~Sec.~\ref{sec:vertices}) with the values determined by Eq.~(\ref{eq:gaussian}).}, and not due to the integration routines. This is why the agreement with the expected result becomes noticeably better when the cells become finer, and hence the interpolation more accurate \footnote{Although not explicit in Eq.~(\ref{eq:gaussian}), when written in code units, $\Phi$ acquires a factor of $a^2$, which we take to be constant in each time step. Since the time steps get smaller with increased resolution, this also contributes to the better accuracy seen in the higher resolution setting.}. For a spherically symmetric potential, $\gamma_2 = 0$, which makes it harder to quantify the code error. However, the $\gamma_2$ panels do show that its absolute value is close to zero (up to some weak noisy pattern) and that the agreement with the expected result improves with increased resolution. Note also that our $\kappa$, $\gamma_1$ and $\gamma_2$ results show no evidence of any inaccuracies caused by the interpolation at refinement boundaries.

This test represents an important validation of not only the integration algorithm, but also of the calculation of the tidal tensor, $\partial_a\partial_b\Phi$ and the interpolation scheme from cell centers to cell vertices. These are all routines that do not exist in default {\tt RAMSES} and this successful test demonstrates that they have been implemented correctly. Finally, we note that this test serves also as an important check of the code infrastructure, namely the fact that the ray propagation through the AMR structure and the communications between CPUs are done correctly -- otherwise there would be errors coming from artifacts at the boundaries of CPU domains or not all rays would reach the observer at redshift $z = 0$.

\section{Cosmological weak lensing simulations}\label{sec:cosmo}

In this section, we show and discuss some code results for cosmological weak lensing simulations. We start by summarising our simulation settings and ray geometry, and then show our code results for one- and two-point statistics, as well as for the lensing signal around dark matter haloes.

\subsection{Summary of the simulation and tile settings}\label{sec:sumsim}

Our lensing maps were obtained by tracing $N_{\rm ray} = 2048\times 2048$ rays from $z_s=1$, covering a field of view with $10\times10\ {\rm deg}^2$. This yields an angular resolution of $0.005\ {\rm deg} = 0.3'$. To encompass this lightcone geometry, we employed the tiling scheme depicted in Fig.~\ref{fig:tile} with five $L = 512\ {\rm Mpc}/h$ boxes, in which we simulate a flat $\lcdm$ cosmology using $N_p = 512^3$ and $N_p = 1024^3$ dark matter tracer particles. We refer to these two resolutions as $\lr$ and $\hr$, respectively. Our $\lcdm$ parameters are $\Omega_{b0} = 0.049$, $\Omega_{c0} = 0.267$, $\Omega_{\Lambda0} = 1 - \Omega_{b0} - \Omega_{c0}$, $h = 0.671$, $n_s = 0.9624$, $\sigma_8 = 0.834$, in accordance with the recent results from the Planck satellite \cite{2014A&A...571A..16P, 2015arXiv150201589P} (but assuming that all neutrino species are massless). The simulation in each box in the tile stops after all rays have reached the end of the integration there (cf.~Secs.~\ref{sec:tile}, \ref{sec:outline}) and the observer is located at the center of the face of the last box (that which is furthest away from the sources). In particular, from the first to the last box, the ray integrations are performed in the redshift intervals, $\left[1, 0.86\right]$, $\left[0.86, 0.6\right]$, $\left[0.6, 0.38\right]$, $\left[0.38, 0.18\right]$ and $\left[0.18, 0\right]$, respectively.

For each particle resolution, we considered five realizations of the particle initial conditions (generated with the {\tt MPGRAFIC} code \cite{2008ApJS..178..179P} at $z =49$) for each of the five boxes in the tile. This allows to construct $5^5 = 3125$ lensing maps by combining the integration results from each box realization. These maps are not all independent from one another \footnote{For instance, consider the lensing map obtained by one realization of the lensing tile. Then, the lensing map constructed from this map by replacing the result of first box with another realization of the first box still gets the same contribution from all the other boxes.}, but they are equally likely realizations of the lightcone.  This "shuffling" of different realizations of the boxes in the tile allows for a measure of the uncertainties associated with cosmic variance, specially for those boxes in the tile whose volume is only very partially covered by the light bundle. In these lensing map constructions, we do not mix the results from $\lr$ and $\hr$ boxes. If one requires all tiles to be completely independent from one another then we can construct 5 tiles for each resolution \footnote{This is in the sense that using the available box realizations, one can only construct 5 tiles using each box realization only once.}.

All these lensing maps are obtained using integration Method A with Eq.~(\ref{eq:Iclensing}) (cf.~Sec.~\ref{sec:lenscode_alt}). However, for testing purposes, we have selected one combination of initial conditions for a $\hr$ tile to construct lensing maps with method B, using both Eq.~(\ref{eq:Iclensing_ngp}) and Eq.~(\ref{eq:Iclensing}). We refer to the results obtained using Eq.~(\ref{eq:Iclensing_ngp}) as Method B (NGP). If not specified, when we refer to Method A and Method B, we mean the result obtained using Eq.~(\ref{eq:Iclensing}). We compare the different integration methods in Sec.~\ref{sec:methods}.

\subsection{Convergence probability distribution function}\label{sec:pdf}

\begin{figure*}
	\centering
	\includegraphics[scale=0.42]{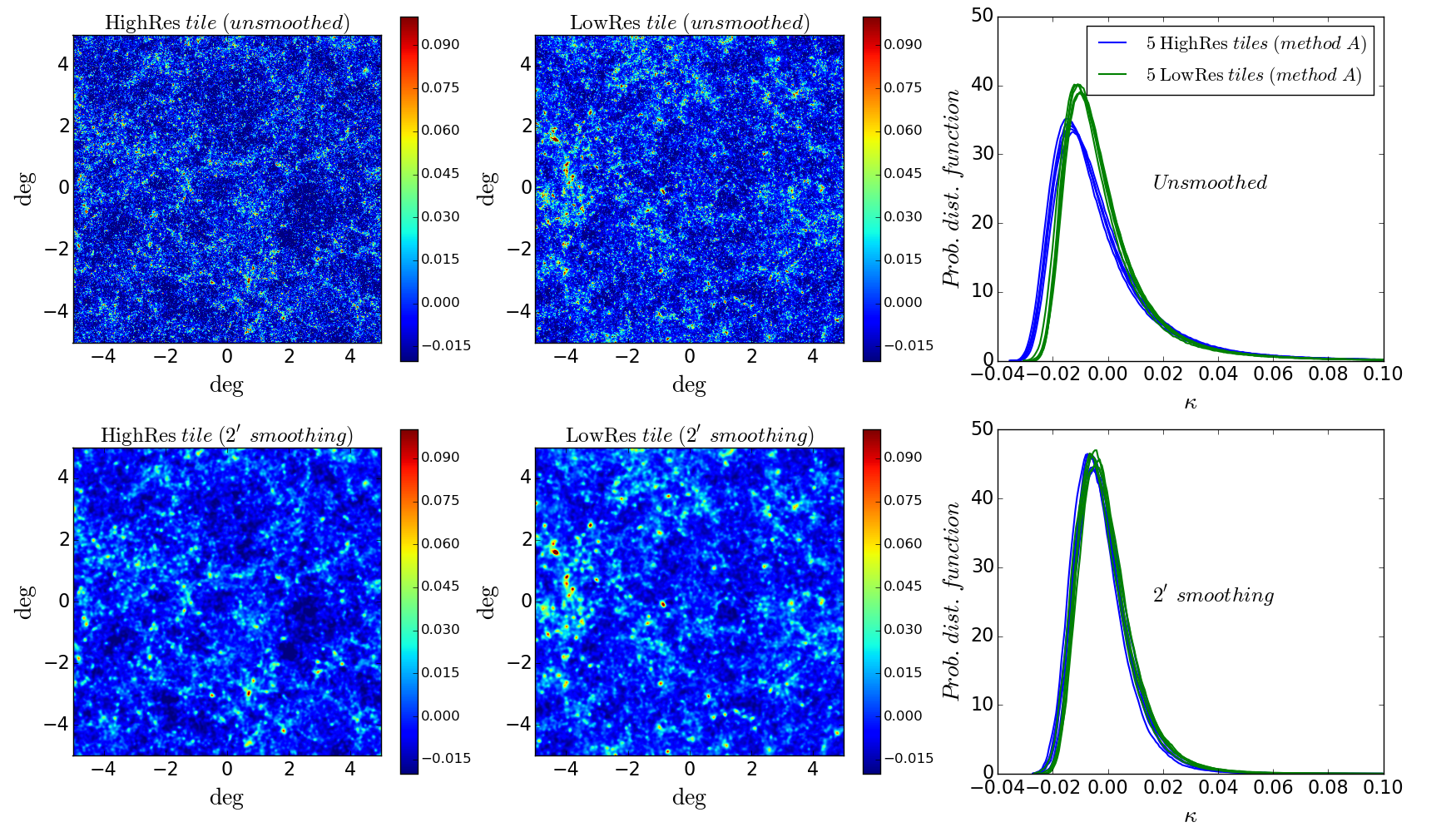}
	\caption{Lensing convergence maps and $\kappa$ probability distribution functions (PDFs). The color maps show the $\kappa$ field obtained for a realization of the $\hr$ and $\lr$ tiles without any smoothing and with a Gaussian smoothing with width $2' \approx 0.03\ {\rm deg}$, as indicated in the panel titles. The right panels show the PDFs of the $\kappa$ field displayed in the color maps and for four other independent tile realizations with and without Gaussian smoothing, as labelled. These results correspond to integration Method A with Eq.~(\ref{eq:Iclensing}).}
\label{fig:pdf}
\end{figure*}

The color maps in Fig.~\ref{fig:pdf} show the $\kappa$ fields obtained from one particular tile realization of the $\hr$ and $\lr$ ray tracing simulations. The upper panels show the maps as computed by our code and the lower panels show the maps smoothed by a Gaussian filter with size $2' \approx 0.03\ {\rm deg}$. The corresponding rightmost panels show the probability distribution function (PDF) of the $\kappa$ fields shown in the color maps, but also for the other 4 independent tile realizations that we can construct from our simulations (cf.~Sec.~\ref{sec:sumsim}).

The upper right panel shows that the PDFs of the $\hr$ and $\lr$ tiles are in good agreement for $\kappa \gtrsim 0.02$, but show some discrepancies at smaller values. In particular, the distribution of the $\hr$ realizations is shifted towards lower values of $\kappa$, relative to the $\lr$ ones. This result can be attributed to the differences in resolution. In particular, the particle CIC clouds in the $\lr$ case are larger than in the $\hr$ case and the particle mass is distributed to a larger volume. Hence, the $\lr$ simulations do not develop density troughs that are as deep as in the $\hr$ case, which pushes the low-$\kappa$ tail of the distribution in the $\lr$ tiles to larger values of $\kappa$, as seen in the upper right panel of Fig.~\ref{fig:pdf}. By the same reasoning, the PDF of the $\hr$ simulations should be higher for larger values of $\kappa$ because of the better resolved high density peaks. However, the PDFs for large values of $\kappa$ becomes noteciably suppressed, which makes that assessment more difficult.

The lower right panel of Fig.~\ref{fig:pdf} shows the same as the upper right panel, but for the smoothed maps. The size of the filter ($2'$) corresponds roughly to cluster size scales at $z \approx 0.5$ and to typical smoothing scales employed on real lensing maps \cite{2013MNRAS.433.3373V, 2015PhRvD..92b2006V}. For the smoothed maps, the two resolutions now agree quite well for all values of $\kappa$. Compared to the unsmoothed cases, the smoothing suppresses the PDF for $\kappa \gtrsim 0.02$, which indicates that these values of $\kappa$ were due to peaks with size smaller than $2'$. The smoothing, however, does not noticeably suppress the amplitude of the PDF for $\kappa \lesssim 0$.

\subsection{Convergence power spectrum}\label{sec:ps}

\begin{figure}
	\centering
	\includegraphics[scale=0.48]{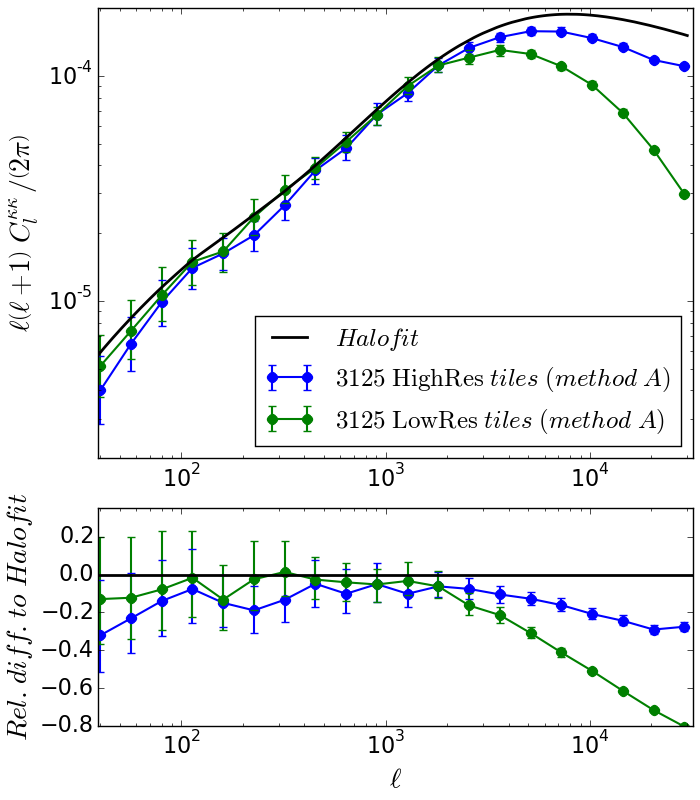}
	\caption{Lensing convergence power spectrum from the maps constructed with integration method A (with Eq.~(\ref{eq:Iclensing})) for the $\hr$ (blue) and $\lr$ (green) tiles. The dots indicate the median and the errorbars indicate the $25\%$ and $75\%$ percentiles of the distribution of the 3125 tile realizations for each particle resolution. The black line indicates the semi-analytical prediction based on the Halofit formalism (Eq.~(\ref{eq:halofit})).}
\label{fig:ps}
\end{figure}

\begin{figure}
	\centering
	\includegraphics[scale=0.48]{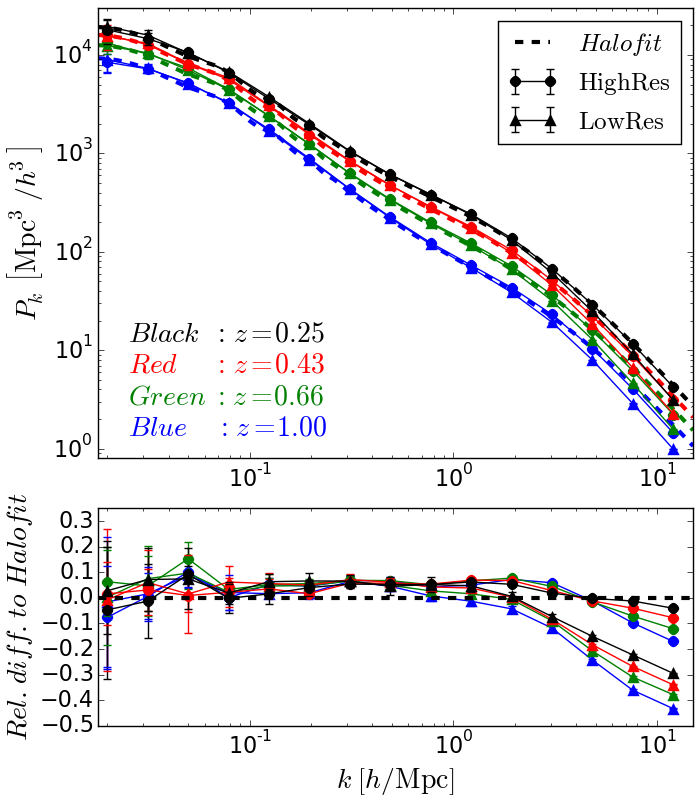}
	\caption{Matter power spectrum measured from the $\hr$ (dots) and $\lr$ (triangles) simulations. For each redshift shown (different colors, as labelled), the symbols indicate the mean and the errorbars the standard deviation of the five realizations of the boxes that will trace the rays right after those redshifts. For example, the result for $z = 0.25$ was measured in the last boxes (those that have the observer), $z = 0.43$ in the second-to-last boxes, and so on. The dashed lines indicate the prediction using the Halofit formalism. }
\label{fig:3dps}
\end{figure}

Figure \ref{fig:ps} shows our convergence power spectrum results for the tiles constructed with the $\hr$ (blue) and $\lr$ (green) simulation boxes. The solid lines indicate the median of the 3125 power spectra that we can construct for each resolution and the errorbars indicate the $25\%$ and $75\%$ percentiles. The black solid line shows the result given by the formula\footnote{We have used the publicly available {\tt CAMB Sources} software (http://camb.info/sources/) to compute this integral. Note that we are assuming the Limber approximation, which is valid for small fields of view like ours (compared to full sky). Morevoer, for small fields of view, the assumption of a flat sky is a good approximation, which in practice means that Fourier and spherical harmonic transforms give equivalent results.}
\bq\label{eq:halofit}
C_{\ell}^{\kappa\kappa} = \frac{9\Omega_m^2}{4}\left(\frac{H_0}{c}\right)^4 \int_0^{\chi_s} g^2\left(\chi_s, \chi\right)\frac{P_{\delta}(k = \ell/\chi, \chi)}{a^2} {\rm d}\chi, \nonumber \\
\eq
{with $P_{\delta}$ being the nonlinear matter power spectrum computed in the Halofit formalism presented in Ref.~\cite{2003MNRAS.341.1311S}, and later tuned by Ref.~\cite{2012ApJ...761..152T} (we use the latter). The spectra from the $\hr$ and $\lr$ tiles are within each other's errorbars and agree also with the Halofit prediction for $\ell \lesssim 2.0\times10^3$. The small differences in between the two resolutions and with the Halofit prediction can be attributed to cosmic variance. The latter can be particularly important in weak lensing studies with small opening angles (compared to full sky), as observed for instance in the full-sky analysis of Ref.~\cite{2008MNRAS.391..435F}.}

For $\ell \gtrsim 2\times10^3$, the Halofit prediction is above our code predictions. This result can be associated with at least two factors. The first one relates to the adaptive smoothing effect that is caused by our interpolation scheme from cell centers to cell vertices (to use Eq.~(\ref{eq:Iclensing})), and which works to suppress the power below a given angular scale (recall the discussion at the end of Sec.~\ref{sec:vertices}). The second factor relates to the accuracy with which Halofit describes the three-dimensional clustering in our simulations. Figure \ref{fig:3dps} compares the Halofit prediction for $P_\delta$ with the nonlinear matter power spectrum measured from our simulations at a number of epochs, as labelled. The figure shows that the clustering power in our $\hr$ and $\lr$ simulations is lower than what Halofit predicts for $\k \gtrsim 4\ h/{\rm Mpc}$ and $\k \gtrsim 1\ h/{\rm Mpc}$, respectively. These differences naturally propagate into the two-dimensional convergence power spectrum on small angular scales, which helps to explain why our simulations underpredict the Halofit result in Fig.\ref{fig:ps}. This also explains why the suppression in power is more pronounced in the $\lr$ case. Although we do not explicitly test for that, we note that the agreement between our code results and Eq.~(\ref{eq:halofit}) is expected to improve if instead of using Halofit to compute $P_\delta$, one uses directly the nonlinear matter power spectrum measured from our simulations (see e.g.~Refs.~\cite{2003ApJ...592..699V, 2009A&A...499...31H}).

\subsection{Comparison of different integration methods}\label{sec:methods}

\begin{figure*}
	\centering
	\includegraphics[scale=0.41]{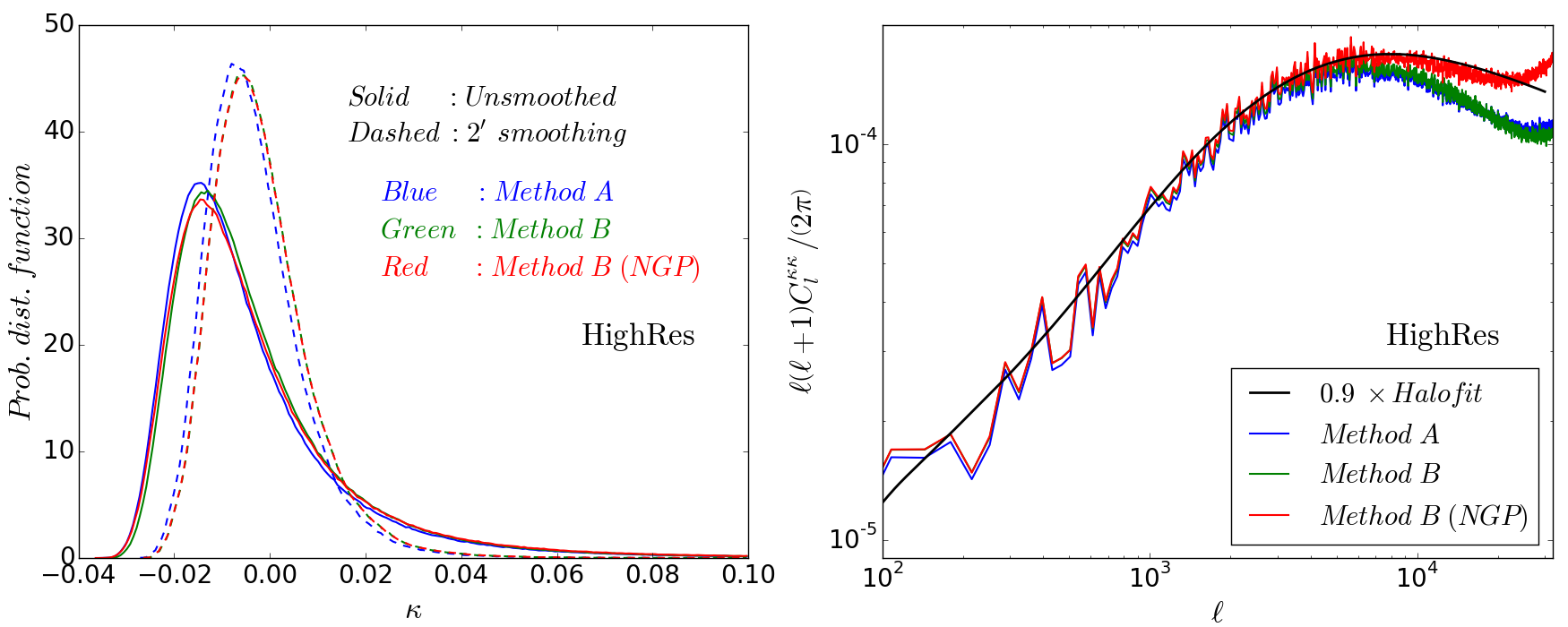}
	\caption{Comparison of integration methods A, B and B (NGP). Method A and Method B make use of Eq.~(\ref{eq:Iclensing}), whereas Method B (NGP) uses Eq.~(\ref{eq:Iclensing_ngp}). The left panel shows the PDF obtained from unsmoothed and smoothed maps for the three methods, as labelled. The right panel shows the convergence power spectrum of (unsmoothed) maps obtained with the three integrations methods. In this panel, the Halofit curve was scaled down to facilitate comparisons between the shape of the curves at high-$\ell$. All these results correspond to the same realization of a $\hr$ tile.}
\label{fig:vs}
\end{figure*}

Figure \ref{fig:vs} compares the results from different ray integration methods for the PDF of $\kappa$ (left panel) and its power spectrum (right panel). The three methods shown are method A, method B and method B (NGP), as labelled. Recall that method A and method B make use of Eq.~(\ref{eq:Iclensing}) and method B (NGP) uses Eq.~(\ref{eq:Iclensing_ngp}). The three integration methods are overall in good agreement in their PDF results, for both the unsmoothed and the smoothed cases. One notes that the PDFs of the two maps from method B are slightly shifted to higher values of $\kappa$, compared to the method A result. This could be attributed to differences in the detailed implementation of methods A and B (cf.~Sec.~\ref{sec:lenscode_alt}). Nevertheless, we stress that these are only small differences, which are in fact smaller than spread of the five method A realizations shown in Fig.~\ref{fig:pdf}. This agreement between the different integration methods is a reassuring result that can be regarded as a check of the consistency of our ray tracing implementation for lensing.

For the case of the convergence power spectrum, integration methods A and B are also in very good agreement for all the scales shown, which once again demonstrates the robustness of our ray tracing modules {(any observed difference is much smaller than the spread due to cosmic variance shown in Fig.~\ref{fig:ps})}. The integration method B (NGP) agrees also very well with method A and B for $\ell \lesssim 5\times10^3$, but for larger values of $\ell$ the shape of its power spectrum agrees better with the Halofit prediction. This is due to the adaptive smoothing effects of our interpolation scheme from cell centers to cell vertices that affects methods A and B, but does not affect method B (NGP). In particular, the shape of the curve from method B (NGP) agrees with Halofit up to $\ell \sim 2\times10^4$. A more detailed assessment of the behavior of the convergence power spectrum on small scales would benefit from ray tracing simulations with higher resolution than those used for this paper, and hence we defer such investigations for future work.

On scales $\ell \gtrsim 2.0\times10^4$, the three curves exhibit an upturn that is caused by ray shot noise. We have checked that decreasing the number of rays traced makes the spurious effects of shot noise more noticeable at lower $\ell$ (not shown, but see e.g.~Ref.~\cite{2000ApJ...530..547J}).

\subsection{Halo lensing}\label{sec:halos}

\begin{figure*}
	\centering
	\includegraphics[scale=0.40]{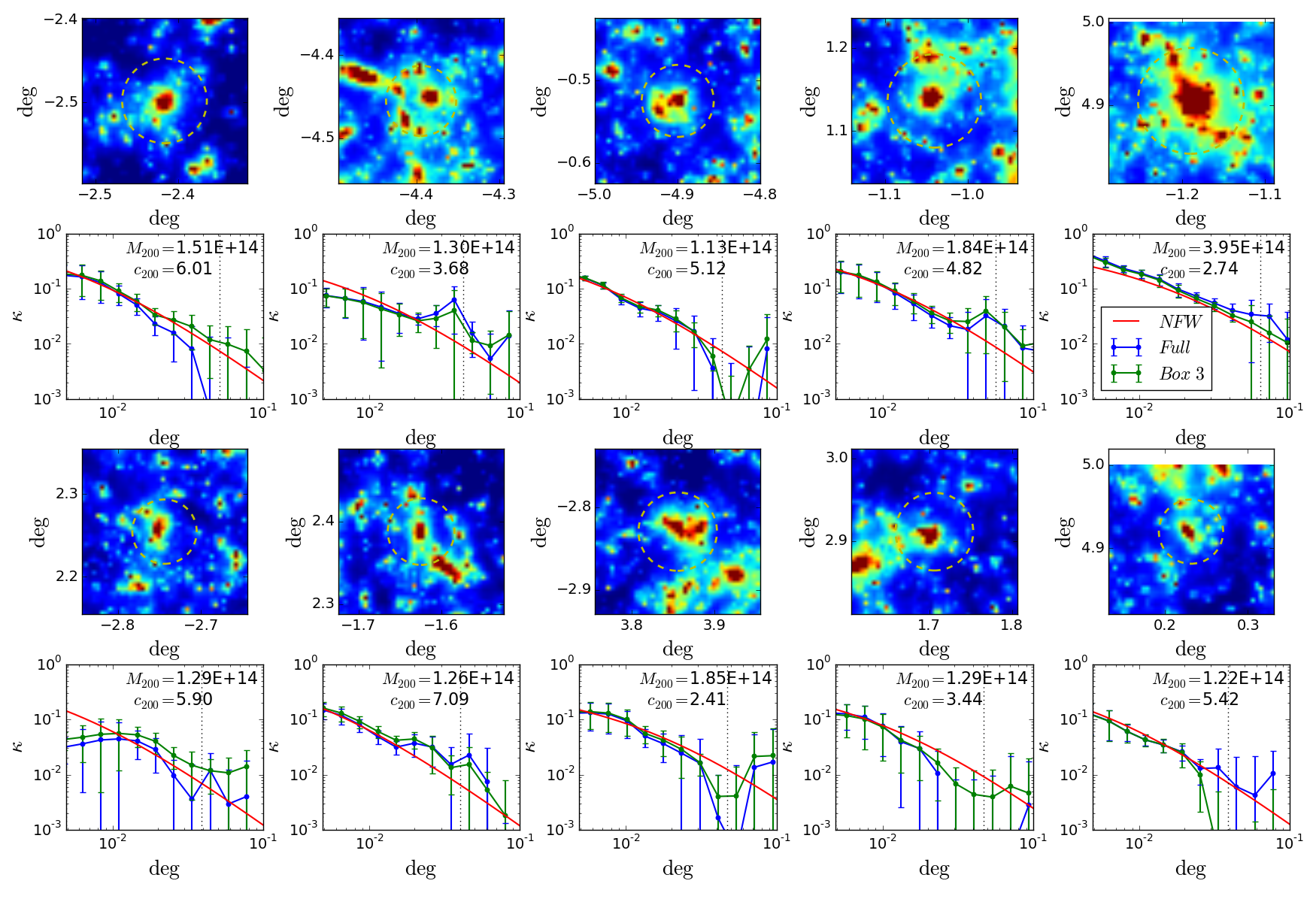}
	\caption{Spherically averaged lensing signal around individual dark matter haloes. The color maps show the $\kappa$ field zoomed into a $0.2\times0.2\ \rm{deg}^2$ region around the center of mass of haloes found at $z = 0.5$ in the middle box of a $\hr$ tile. This box, called Box 3, is the one in which the redshift interval of the ray integration brackets $z = 0.5$. The dashed yellow lines depict the angular size of the haloes, $R_{200}$. The ten haloes shown are those which lie inside the lightcone geometry and have $M_{200} > 10^{14}\ M_{\odot}/h$, $c/a > 0.55$, $b/a > 0.75$ and $f_{\rm sub} < 0.1$. The panels below each color map show the corresponding spherically averaged $\kappa$ profiles computed around the halo's center of mass taking into account the full lensing signal of the tile (blue) or just the contribution from Box 3 (green), where the haloes were found (the color maps show the full lensing signal, and not just the contribution from Box 3). The points are obtained by taking the mean value of $\kappa$ at a number of points that sample a circle of a given angular scale around the cluster center. The errorbars show the standard deviation around that mean. The solid red lines show the analytical NFW prediction computed using the halo mass and concentration values (upper right part of each panel) determined by the {\tt Rockstar} halo finder. The vertical dashed lines indicate the angular size of the haloes.}
\label{fig:halo}
\end{figure*}

Figure~\ref{fig:halo} shows the lensing signal around dark matter haloes found in the middle box of a $\hr$ tile (we call this Box 3) at $z = 0.5$. The halo catalogue was built using the {\tt Rockstar} code \cite{2013ApJ...762..109B}. We applied a mass\footnote{Our mass definition is $M_{200} = \left(4\pi/3\right)\rho_c(z)200R_{200}^3$, where $R_{200}$ is the halo radius defined as the radial distance to the halo center within which the mean density is equal to $200$ times the critical density in the Universe $\rho_c(z)$.} cut $M_{200} > 10^{14}\ M_{\odot}/h$ and considered only haloes with shape parameters $c/a > 0.55$ and $b/a > 0.75$, where $a>b>c$ are the three ellipsoidal axis. We have also kept only haloes with $f_{\rm sub} < 0.1$, where $f_{\rm sub}$ is the fraction of halo mass associated with resolved substructure. The ten haloes shown are those which lie within the field of view. The color maps of Fig.~\ref{fig:halo} show the $\kappa$ field zoomed into a $0.2\times0.2\ \rm{deg}^2$ region around the halo center determined by {\tt Rockstar}. The profile panels below each color map show the corresponding spherically averaged convergence profile around the halo center \footnote{The points at each angular scale show the mean $\kappa$ at a number of points that sample a circle around the halo center with that angular size. The errorbars show the standard deviation around this mean. We evaluate the values of $\kappa$ from the map via bilinear interpolation given the values of $\kappa$ at each pixel.}. The blue points correspond to the profiles obtained using the maps from the full tile, whereas the green ones show the profiles obtained from the signal computed by Box 3 alone. This helps to separate the contribution from the halo itself and from foreground and background structures. The solid red lines show the analytical Navarro-Frenk-White (NFW) \cite{Navarro:1996gj} result (see e.g.~Refs.~\cite{1999astro.ph..8213O, 1996A&A...313..697B, 2010arXiv1002.3952U}) computed with the mass and concentration ($M_{200}, c_{200}$) values given by {\tt Rockstar}. The angular size associated with $R_{200}$ is depicted by the yellow dashed lines in the color maps and by the vertical dashed lines in the profile panels.

The fact that $\kappa$ peaks in our maps coincide with the halo angular positions determined by another independent code constitutes another consistency check of our ray tracing implementation for lensing. Furthermore, for all ten haloes shown and despite some expected differences, Fig.~\ref{fig:halo} shows that there is good agreement between our code results and the NFW analytical prediction for the amplitude and shape of the $\kappa$ profiles. The observed differences can be caused by a variety of effects. For instance, despite our attempt to select the haloes that are the "most spherical" ($c/a > 0.55$ and $b/a > 0.75$) and devoid from substructure ($f_{\rm sub} < 0.1$), they are naturally still not perfectly spherical nor have perfectly smooth mass distributions. The nonsphericity and irregular shape of haloes leaves room for projection effects that are known to induce a bias between concentration and mass values estimated from lensing and those estimated from the 3D mass distribution in the simulations (see e.g.~Refs.~\cite{2014ApJ...797...34M, 2015ApJ...806....4M, 2015ApJ...814..120D}. Moreover, owing also to projection effects and substructure, there can be offsets between the angular position of the halo center (around which we compute our lensing profiles) and of the $\kappa$ peaks, which can lead to some differences to the NFW analytical prediction. Some of these differences are particularly noticeable for the second halo from the left in the upper part of the figure, whose lensing map reveals the presence of several $\kappa$ peaks inside $R_{200}$. Similarly, the first halo from the left in the lower part of the figure shows an offset between its concentrated lensing peak and the center of the halo (center of the dashed yellow circle). 

Figure~\ref{fig:halo} shows also that the contribution from foreground and background structures barely modifies the lensing profiles in the inner regions, where the $\kappa$ values are larger. This is indicative that the rays that crossed the inner regions of the ten haloes did not cross the inner regions of other haloes along their trajectories, which would have otherwise induced a difference between the blue and green lines at small radii \cite{2011PhRvD..84d3529Y, 2004MNRAS.350..893H, 2010ApJ...719.1408F}. At larger angular scales, the amplitude of the convergence decreases, which makes the lensing signal there more sensitive to the influence of intervening matter. Illustrative of this situation is the first halo from the left in the upper part of the figure. For this halo, rays that propagated through its outskirts picked up the lensing signal from matter that surrounds the halo, but seemed to have travelled through mostly underdense regions ($\kappa < 0$) in the rest of their trajectory from the source to the observer (which is why the blue dots are below the green ones). 

A more thorough study of halo lensing using our code could involve applying more strict criteria to select haloes (based on their relaxed state, shape, substructure, etc), choosing different points around which to evaluate profiles (halo center, halo density peaks, $\kappa$ peaks) or studying the average lensing profiles of a stack of haloes \cite{2012MNRAS.420.3213O, 2011MNRAS.414.1851O}. Our several tile realizations can also be used in studies of the contamination to the lensing signal along the line of sight. The latter may not be a critical source of systematics in cluster lensing related work, but that may not necessarily be the case for voids \cite{2014MNRAS.440.2922M, 2015MNRAS.454.3357C, 2013ApJ...762L..20K, 2013MNRAS.432.1021H}, which have an intrinsically smaller lensing amplitude. We leave these investigations for future work.

\section{Summary and Outlook}\label{sec:conc}

\subsection{Summary}

We have presented a ray tracing code to compute integrated cosmological observables (cf.~Eq.~\ref{eq:I1}) that runs on the fly in AMR N-body simulations. Our algorithm is based on the original ideas of Refs.~\cite{whitehu2000, li2001}, but we implemented it here in the efficiently parallelised {\tt RAMSES} AMR N-body code, which makes it possible to reach the resolution levels that are required by current and future observational surveys. The routines we described in this paper move the rays on a cell-by-cell basis, from some source redshift until an observer at redshift zero. The ray initialization routines (cf.~Sec.~\ref{sec:init}) can self-consistently handle cases where a light bundle is initialized inside the simulation box or at its faces. This ensures that one does not need to simulate a box that is large enough to encompass the whole lightcone but can, instead, "tile" several smaller boxes and let the rays move from one to the other (cf.~Sec.~\ref{sec:tile}). The integral along the whole line of sight is obtained by summing up the contribution from each crossed cell (cf.~Eq.~(\ref{eq:sumIc})). The latter can be performed analytically either by treating the field as constant inside each cell (called NGP field, Eq.~(\ref{eq:Icngp})) or by reconstructing it via trilinear interpolation from the field values at cell vertices (cf.~Eq.~(\ref{eq:Icfinal})). The default {\tt RAMSES} code does not evaluate the fields at cell vertices, but at cell centers. For ease of our integration routines (namely those which employ Eq.~(\ref{eq:Icfinal})), we designed an interpolation scheme from cell centers to cell vertices, which we used to ensure that the reconstructed fields from trilinear interpolation vary continuously when crossing cell boundaries and are mass conserving (cf.~Sec.~\ref{sec:vertices}).

Since it runs on the fly in the N-body simulation, our code can produce maps of the integrated observables without requiring large amounts of data (or even any) to be stored and further post-processed for ray tracing. Furthermore, our code takes full advantage of the time and spatial resolution available in the N-body run, which is not the case in standard ray tracing numerical studies. {These constitute two of the main improvements of our code over more conventional ray tracing methods.}

We have tested our ray tracing implementation by applying it to gravitational lensing. We have explained how the lensing convergence, $\kappa$, in our code can be computed in two different ways. The first one involves direct integration of the two-dimensional transverse Laplacian of the lensing potential (which we call Method B, Eq.~(\ref{eq:kappa})), whereas the second makes use of the three-dimensional Poisson equation to relate the transverse derivatives of the potential to the density field and a series of other integral and surface terms that depend on the radial gradient of the potential (called Method A, Eq.~(\ref{eq:kappa2})). These two methods make use of Eq.~(\ref{eq:Icfinal}) (or Eq.~(\ref{eq:Iclensing})). We have also computed the lensing signal with a method we called Method B (NGP), which is the same as Method B but using Eq.~(\ref{eq:Icngp}) (or Eq.~(\ref{eq:Iclensing_ngp})) instead. The lensing shear field can be obtained by integrating directly the corresponding combination of second derivatives of the potential  (cf.~Eqs.~(\ref{eq:gamma1}) and (\ref{eq:gamma2})) or indirecly via the $\kappa$ result from Method A using Eq.~(\ref{eq:gammaind}). We have tested the numerical solutions of our code by comparing them with the solutions from an independent numerical integrator for lensing through a fixed Gaussian potential (cf.~Sec.~\ref{sec:gaussian}).

As an illustration of the application of our code, we have used it to perform cosmological simulations of weak gravitational lensing. We traced $N_{\rm ray} = 2048^2$ rays in a $10\times10\ {\rm deg}^2$ field of view from $z_s = 1$. We simulated a $\lcdm$ model on five boxes with size $L = 512\ {\rm Mpc}/h$ to encompass that lightcone. We considered two particle resolutions: $N_p = 1024^3$ ($\hr$) and $N_p = 512^3$ ($\lr$). For each resolution setting, we simulated each of the five boxes in the tile using five realizations of the initial conditions. In total, this allows us to build $5^5 = 3125$ different lensing maps. We have analysed 1- and 2-point statistics of the $\kappa$ maps as well as the lensing profiles around dark matter haloes. Our weak lensing results can be briefly summarised as follows:

\hspace{0.2 cm} $\bullet$ The sets of $\kappa$ maps constructed with Method A from both particle resolutions agree very well in their PDF results (Fig.~\ref{fig:pdf}). There are some expected differences induced by the difference in resolution. These disappear after the maps are smoothed on a few arcmin scales, which is what is typically done in observational studies (cf.~Sec.~\ref{sec:pdf}). The PDF results obtained with method B and method B (NGP) are also in very good agreement with method A (cf.~Fig.~\ref{fig:vs}), which constitutes a validity check of the code implementation.

\hspace{0.2 cm} $\bullet$ The $\kappa$ power spectrum results obtained from the $\hr$ and $\lr$ tiles with method A are in good agreement with each other and with the semi-analytical prediction computed in the Halofit formalism for $\ell \lesssim 2\times10^3$ (cf.~Fig.~\ref{fig:ps}). For larger values of $\ell$ (small angular scales), our method A results exhibit a suppression of power relative to the Halofit result. This can be attributed to the resolution limit of our simulations and to the adaptive smoothing effects that follow from interpolating the field values from cell centers to cell vertices (cf.~the discussion at the end of Sec.~\ref{sec:vertices}). The results from method B and method B (NGP) are also in good agreement with power spectrum obtained with method A. The lack of small scale power in method B (NGP) is not as pronounced as in the other two cases, which is related to the fact that method B (NGP) is not affected by the adaptive smoothing caused by the reconstruction of the fields at cell vertices.

\hspace{0.2 cm} $\bullet$ Our $\kappa$ maps show amplitude peaks at the angular positions of haloes found in the box that is at the middle of the tile. This constitutes a trivial (but successful) validation test of our results (cf.~Fig.~\ref{fig:halo}). Furthermore, the spherically averaged $\kappa$ profiles in our maps exhibit the expected level of agreement with the analytic prediction for NFW haloes computed using the mass and concentration inferred from the three-dimensional particle distribution.

\subsection{Future code developments}

We comment below on a list of possible ways to go beyond the current implementation of the code.

\underline{\it Beyond tiling of simulation boxes}: {The cosmological weak lensing results presented in this paper were obtained by running one N-body simulation for each box that makes up a tile (although not all these simulations have to reach $z=0$).} We note, however, that this is not necessary. For instance, when the light bundle leaves a simulation box, instead of terminating that simulation and starting another box which is closer to the observer, one can keep using the same box, but initializing another bundle of rays for an observer at a different location. This process of changing the position of the observer can continue until all rays reach $z = 0$. The consecutive positions of the observer can be chosen to minimize the repetition of structures, e.g., by having the rays crossing regions of the box that had not been previously crossed or regions that had been crossed but in different directions. {Hence, if one needs $N_{\rm box}$ to encompass a given lightcone, then for $N_{\rm sim}$ sets of initial conditions one has $N_{\rm box} \times N_{\rm sim}$ map {\it portions}\footnote{We call a map portion the contribution coming from each single box crossing.} that can be combined to build map realizations. If one uses each portion only once, then one can build $N_{\rm sim}$ fully independent maps. This number increases if different map realizations (no longer fully independent) can share some map portions. The shared portions can be appropriately chosen to minimize the effects of structure repetition (e.g.~repeat the portions close to the observer or to the sources, where the lensing kernel is small).}

\underline{\it Beyond first order approximations for lensing}: The ray tracing implementation for lensing that we presented in this paper neglected the effects from second and higher order perturbations, which include the well known Born correction and lens-lens coupling (see Ref.~\cite{2010PhRvD..81h3002B} for a thorough account of higher order corrections in lensing). As a first step to go beyond this, the effects from these higher order terms can be added to our code by following the strategy outlined in Appendices A1 and A2 of Ref.~\cite{li2001}. {In particular, the effects of lens-lens coupling (and other nonlinear couplings) can be included in a perturbative manner along straight trajectories, and in a way that still allows the relevant integrals to be computed analytically. The high line-of-sight resolution of our code can prove useful in clarifying the importance of first-order approximations in lensing studies \cite{2014arXiv1410.8452H, 2015JCAP...03..049C}. To go beyond the Born approximation one can appropriately deflect the rays every time they cross a cell. The integral that determines the deflection angle $\vec{\alpha}$ inside a cell is evaluated analogously to any other integral in the code (e.g.~that of $\kappa$).} For these higher-order studies (specially beyond the Born approximation), some difficulties may arise as it is no longer certain that the rays all reach the observer at $z = 0$. The significance of these issues and possible ways around it (e.g.~construction of lensing maps using only rays that are sufficiently close to the observer or running the simulation backwards in time) is the subject of ongoing work.

\underline{\it Ray-Ramses also as a post-processing tool}: {The application of our code is not restricted to ray-tracing simulations on the fly. Given a number of particle snapshots from some N-body simulation, {\tt RAMSES} can construct the appropriate AMR grid structure, on which our routines can compute the integrals. Note that this does not involve projections into planes and the calculation will remain a three-dimensional one. The ray-tracing algorithm remains essentially unchanged, except for the fact that the particle distribution is now static, and hence, the ray trajectories are not constrained by the size of code time steps (cf.~Sec.~\ref{sec:acrosstime}). This can also facilitate the implementation of non-straight ray trajectories, since rays can be traced away from the observer location. These developments are currently ongoing.}

\underline{\it All-sky lensing simulations}: In this paper, we have restricted our analysis for fields of view that are small enough for the flat-sky approximation to hold. However, in studies where the light source is the CMB (ISW, SZ, CMB lensing), the field of view is normally the full celestial sphere (up to masked areas), in which case the flat-sky approximation naturally breaks down. Other surveys such as LSST \cite{2012arXiv1211.0310L} and Euclid \cite{2011arXiv1110.3193L} are also expected to cover ever larger fractions of the sky for lensing, which motivates extending our code for larger fields of view. This can be done with a few steps. Perhaps the most involved one is related to the initialization of the rays, which would benefit from using the {\tt HEALPix} \cite{2005ApJ...622..759G} pixelation on the sphere. The initial directions and positions can be generated elsewhere and stored in a file that is read by our code when it is time to initialize the rays. Other steps should involve a careful determination of the box sizes used in the tiles and the location of the observer. We note that the routines that control the ray trajectories and integration would remain as presented in this paper.

\underline{\it Source redshift distribution for lensing}: {Although this was not tested in this paper, we note that it is possible to initialize the different rays at different source redshifts. For instance, it is straightforward to sample values of $z_s$ across the rays using the source distribution for some observational survey and only let each ray start the integration once the simulations have reached the value of $z_s$ assigned to it. We note also that it is possible to design ways to compute the lensing signal for different source redshift distributions without having to re-run the code. Consider, for instance, two source redshifts, $z_s^a < z_s^b$, with comoving distance $\chi_s^a < \chi_s^b$. Taking the case of lensing applications as an example, the signal associated with $z_s^b$ is given by the integral}
\bq\label{eq:Isources}
I &=& \frac{1}{c^2}\int_0^{\chi_s^b} \chi Q {\rm d}\chi - \frac{1}{c^2\chi_s^b}\int_0^{\chi_s^b} \chi^2 Q {\rm d}\chi \nonumber \\
&=& \frac{1}{c^2}\mathcal{A} - \frac{1}{c^2\chi_s^b}\mathcal{B},
\eq
{where $Q$ is the desired combination of second angular derivatives of the lensing potential and the second equality serves to define the integrals $\mathcal{A}$ and $\mathcal{B}$. The above equation is the same as the equations solved in this paper, but written in this way it becomes clearer that $\chi_s^b$ appears as a multiplicative term and as the limit of integration, but not inside the integrand. The integrals $\mathcal{A}$ and $\mathcal{B}$ are in the form of Eq.~(\ref{eq:I}) and so they can be solved using our routines. Focusing on the case of $\mathcal{A}$, we can decompose its calculation as}
\bq\label{eq:Isources1}
\mathcal{A} = - \int_{\chi_s^b}^{\chi_s^a} \chi Q {\rm d}\chi -  \int_{\chi_s^a}^0 \chi Q {\rm d}\chi,
\eq
{where we have flipped the integration ranges, just to emphasize that in our code, the integrations are done from the source towards to observer. Hence, given the value of $\mathcal{A}$ computed for $\chi_s^b$, we can get its value for $\chi_s^a$, by subtracting the first term on the right-hand side of Eq.~(\ref{eq:Isources1}). This term can be obtained by letting the code output the convergence calculation accumulated for each ray from $z_s^b$ to $z_s^a$. The same considerations hold for the integral $\mathcal{B}$. The final step to get the value of $I$ in Eq.~(\ref{eq:Isources}) for $z_s^a$ is to replace $\chi_s^b$ by $\chi_s^a$ in the second term on the right-hand side. This reasoning can be generalized to more source redshift values. This involves having to output the accumulated integrals at a number of redshifts within some range where one expects source galaxies to exist. Note, however, that this is not very demanding from a data storage point of view since the outputted lensing maps are relatively light (compared with particle snapshots from simulations, for instance). A scheme such as this can have interesting applications is assessing fairly quickly the impact of different source redshift distributions on the lensing signal.}

\subsection{Code applications}

To conclude our discussion, we comment briefly on a number of possible applications of our code.

\underline{\it Baryonic effects}: The inclusion of baryon physics is relatively straightforward as it involves simply turning on any hydrodynamical modules that are already existing or that can be added to the {\tt RAMSES} code. The degree of complexity of such baryon physics recipes would depend on the exact application in mind. For instance, Refs.~\cite{2011MNRAS.417.2020S, 2014arXiv1410.6826M, 2015ApJ...806..186O} investigate the effects of gas cooling, stellar feedback, AGN feedback, and others on weak lensing observables. Still in the context of hydrodynamical simulations, our code can also be used in studies of the SZ effect. In practice, this would amount to associating the quantity $Q$ in Eq.~(\ref{eq:I1}) with $n_eT$ for thermal SZ and $n_ev_b$ for kinetic SZ effects, where $n_e$ is the electron gas density, $T$ its temperature and $v_b$ its bulk velocity.

\underline{\it ISW simulations}: One of the possible applications of our code is in studies of the ISW effect. This can be done by setting $Q$ in Eq.~(\ref{eq:I1}) to the physical time derivative of the lensing potential $\dot{\Phi}_{\rm len}$. The latter can be computed implicitly in each cell via finite-differencing using values of the potential at the current ($\Phi_{\rm curr}$) and previous ($\Phi_{\rm prev}$) time steps: $\dot{\Phi} = \left(\Phi_{\rm curr} - \Phi_{\rm prev}\right)/\Delta t$, where $\Delta t$ is the time step interval. Our numerical implementation is particularly suited for ISW studies since it allows to directly compute the time-derivative of the potential on all scales, rather than making use of the continuity equation to relate the velocity field in the simulations to $\dot{\Phi}$. The latter approach is what is done in conventional ISW studies \cite{2009MNRAS.396..772C, 2010MNRAS.407..201C}. In a future work, we plan to use our ray tracing code to study the ISW effect, particularly its impact on nonlinear scales (also known as the Rees-Sciama effect \cite{1968Natur.217..511R}).

\underline{\it Modified gravity}: An interesting application of our ray tracing code can be in the context of theories of modified gravity (see e.g.~\cite{2012PhR...513....1C, Joyce:2014kja, 2015arXiv150404623K} for reviews). This is straightforward in practice, as it amounts to installing the routines presented in this paper for {\tt RAMSES} into codes such as {\tt ECOSMOG} \cite{2012JCAP...01..051L, baojiudgp, 2013JCAP...11..012L, 2015arXiv151108200B} and {\tt ISIS} \cite{2014AA...562A..78L}, which are themselves also modified versions of {\tt RAMSES}. For those theories that modify directly the lensing potential (e.g.~Galileon \cite{2009PhRvD..79f4036N, 2009PhRvD..79h4003D, 2015PhRvD..91f4012P, 2015MNRAS.454.4085B, 2015JCAP...08..028B, 2014JCAP...08..059B, 2013JCAP...10..027B, 2012PhRvD..86l4016B}, Nonlocal \cite{2007PhRvL..99k1301D, 2014IJMPA..2950116F, 2014JCAP...09..031B}, K-mouflage \cite{2015PhRvD..91f3528B, 2014PhRvD..90b3508B} gravity, etc.) the time and spatial resolutions along the line of sight have particular importance because of the time evolution of the modified gravity effects. Moreover, appealing theories of modified gravity usually have screening mechanisms to suppress the modifications to gravity on small scales (like in the Solar System), which follow from nonlinearities in the equations that govern the potential. The nonlinearity implies that the superposition principle does not hold, which means that it is not straightforward to use the multiple lens-plane approximation to study lensing in these theories (in the recent work of Ref.~\cite{2015JCAP...10..036T}, the authors use the multiple lens-plane approximation, but focus on models that do not modify the lensing potential directly). Our code can therefore also be seen as a platform where the lensing effects of these theories of gravity can be studied self-consistently.

\underline{\it Other applications}: In the context of lensing, other applications of our code may include the study of cosmic flexions \cite{2005ApJ...619..741G, 2006MNRAS.365..414B, 2007ApJ...660..995O, 2008A&A...485..363S, 2013MNRAS.435..822R}, which are lensing effects sensitive to third spatial derivatives of the lensing potential, as well as the lensing effects associated with vector and tensor perturbations of the metric \cite{2014PhRvD..89d4010B, 2015JCAP...09..021T}. The ray-tracing machinery that we have installed in {\tt RAMSES} may also serve as a backbone to develop a code for radiative transfer and cosmic reionization studies \cite{2006MNRAS.371.1057I, 2009MNRAS.400.1283I, 2011MNRAS.414.3458W, 2013MNRAS.436.2188R, 2013MNRAS.434..748A}.

\bigskip

\bigskip

In conclusion, the code we presented in this paper provides a different way to compute integrated observables that is, in general, subject to fewer approximations compared to conventional ray tracing methods. One can also argue that it is more practical to use, in the sense that the calculations are done on the fly in the simulation and not at post-processing stages. We believe that works performed with this code can provide a valuable set of results that could complement those obtained with other methods. This should yield a more robust theoretical understanding of the physical processes that determine a number of integrated cosmological observables, which could help to plan better current and future observational missions. {In the future, we plan to widen up the range of applications of {\tt Ray-Ramses} and make the code publicly available to the research community.}

\begin{acknowledgments}

We thank Romain Teyssier for useful clarifications about {\tt RAMSES} and Xiangkun Liu for helpful numerical tests. We also thank Carlton Baugh, Martin Feix, Lindsay King, Richard Massey, Fabian Schmidt, Gongbo Zhao and Hongsheng Zhao for encouraging this project and useful comments and discussions, and Lydia Heck for invaluable numerical support. This work was supported by the Science and Technology Facilities Council [grant number ST/L00075X/1]. This work used the DiRAC Data Centric system at Durham University, operated by the Institute for Computational Cosmology on behalf of the STFC DiRAC HPC Facility (www.dirac.ac.uk). This equipment was funded by BIS National E-infrastructure capital grant ST/K00042X/1, STFC capital grant ST/H008519/1, and STFC DiRAC Operations grant ST/K003267/1 and Durham University. DiRAC is part of the National E-Infrastructure. AB thanks the support from FCT-Portugal through grant SFRH/BD/75791/2011 during part of this work. CLL acknowledges support from STFC consolidated grant ST/L00075X/1. SB is supported by STFC through grant ST/K501979/1. BL acknowledges support by the UK STFC Consolidated Grant No. ST/L00075X/1 and No. RF040335.

\end{acknowledgments}

\appendix

\section{The Dandelin sphere algorithm in the ray initialization}\label{app:dandelin}

\begin{figure*}
	\centering
	\includegraphics[scale=0.45]{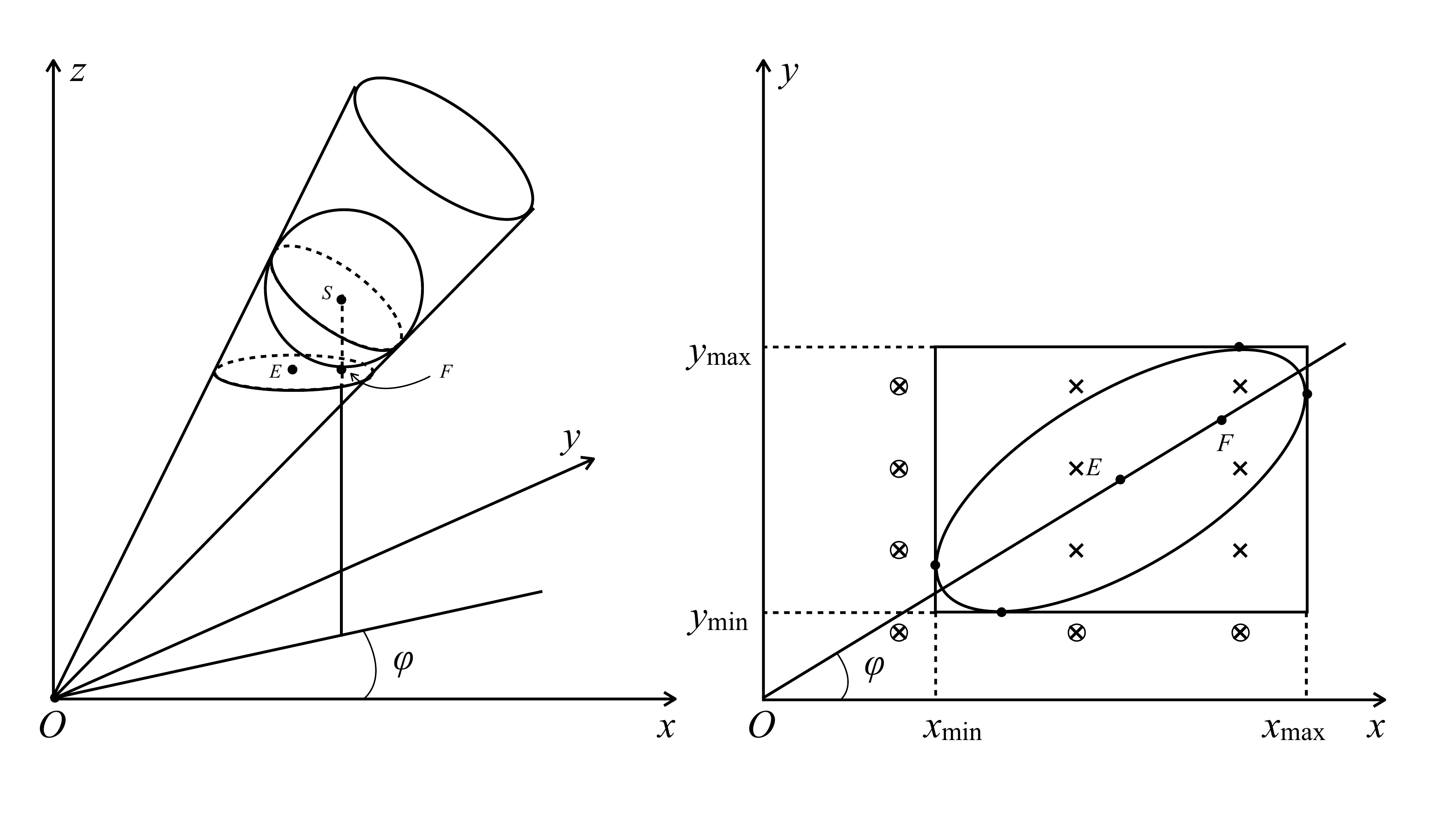}
	\caption{Dandelin sphere construction used in the ray initialization routines. The left diagram shows the narrowest cone from the observer that contains a sphere, which is the smallest sphere that contains a given cubic grid. The point $S$ is the center of the sphere and of the grid (the grid is not shown to keep the plot less busy). In the left diagram, poins $E$ and $F$ are, respectively, the center and one of the foci of the ellipse that results from {\it cutting} the cone perpendicularly to the $z$ axis and through the point of lowest $z$-coordinate of the sphere (which is point $F$). The right diagram shows this ellipse projected onto the $x$-$y$ plane. The crosses illustrate the positions of rays crossing the $x$-$y$ plane. Those that do not fall inside the rectangle that encompasses the ellipse (circled crosses) can be left out from the loops in the initialization routine.}
\label{fig:dandelin}
\end{figure*}

In this appendix, we describe with more detail the algorithm that is used in the initialization routines to narrow down which ray IDs can be physically inside a given grid. Recall that the initialization can always be performed by looping over all grids, and for each, looping over all rays and checking whether they lie in the volume covered by the grid. The goal of this algorithm is to filter out the number of rays to loop for each grid. This is a nontrivial exercise because the mesh structure is cubic and the ray positions are described in spherical coordinates.

The first step consists of determining the smallest sphere that contains the cubic grid. This is shown in the left panel of Fig.~\ref{fig:dandelin}, where the sphere is centered at point $S$, which is also the center of the grid (not shown). Then, one identifies the narrowest cone from the observer $O$ that contains this sphere, as illustrated also in the figure. Naturally, those rays whose angular positions do not fall within the opening angles of the cone cannot be in that grid. The problem is then reduced to finding the rays that are inside the cone. To do so, we choose to work within a plane that is perpendicular to the $z$-axis. 

Consider the elliptical cross section obtained by {\it cutting} through the cone with a plane perpendicular to the $z$-axis and tangential to the lowest $z$-coordinate of the sphere (point $F$ in Fig.~\ref{fig:dandelin}). In this case, the sphere is called a Dandelin sphere and has the property that point $F$ is one of the two foci of the ellipse, whose center is point $E$. Then the question reduces to determining which rays lie inside the ellipse, which is parallel to the $x$-$y$ plane. In the right panel of Fig.~\ref{fig:dandelin}, we show this ellipse projected onto the $x$-$y$ plane. Points $E$, $F$, and $O$ lie on a line that makes an angle $\varphi$ with the $x$-axis. In our algorithm, one finds the smallest rectangle that contains the ellipse and whose sides are parallel to the $x$- and $y$-axes. As shown in the figure, the rectangle is tangential to the ellipse at four points, which are, repectively, the maximum and minimum $x$-coordinates ($x_{\rm max}$ and $x_{\rm min}$) and $y$-coordinates ($y_{\rm max}$ and $y_{\rm min}$) of the ellipse. The rays that lie inside the sphere must have $x_{\rm ray}\in[x_{\rm min},x_{\rm max}]$ and $y_{\rm ray}\in[y_{\rm min},y_{\rm max}]$. Given these ranges, it is then straightforward to use Eqs.~(\ref{eq:txty}) and (\ref{eq:rayid}) to determine which {\tt rayid} values need to be checked (note that $x,y$ are related to $\theta_x,\theta_y$ in Sec.~\ref{sec:init}).

Using straightforward geometry, one can determine the boundaries of the rectangle as
\begin{eqnarray}
x_{\rm min} &=& x_F - \frac{a(1-e^2)\cos\xi}{1-e\cos(\varphi-\xi)},\nonumber\\
x_{\rm max} &=& x_F + \frac{a(1-e^2)\cos\xi}{1+e\cos(\varphi-\xi)},\nonumber\\
y_{\rm min} &=& y_F - \frac{a(1-e^2)\sin\zeta}{1-e\cos(\varphi-\zeta)},\nonumber\\
y_{\rm max} &=& y_F + \frac{a(1-e^2)\sin\zeta}{1+e\cos(\varphi-\zeta)},
\end{eqnarray}
where $(x_F, y_F)$ is the coordinate of the focus of the ellipse in the (projected) $x$-$y$ plane, $a$, $e$ are respectively the length of the semi-long axis and the eccentricity of the ellipse, and 
\begin{eqnarray}
\xi &=& \sin^{-1}(e\sin\varphi),\nonumber\\
\zeta &=& \cos^{-1}(e\cos\varphi).
\end{eqnarray}

To exemplify, we show in the right panel of Fig.~\ref{fig:dandelin} a number of rays (crosses) that cross the plane containing the ellipse (note that all rays are equally spaced in $x$ and $y$). The rays that fall outside the rectangles are circled, which means that they certainly do not cross the grid, and as a result, do not need to be looped over. For most applications, for each grid, the majority of the rays lies outside of the rectangle, which is why this algorithm speeds up substantially the initialization of the ray data structure. 

\section{$d_N$ expressions in Eq.~(\ref{eq:Qcompact})}\label{app:dns}

The $d_N$ coefficients in Eq.~(\ref{eq:Qcompact}) that are used in the trilinear interpolation inside cells are given by

\begin{widetext}
\bq\label{eq:dns}
d_1 &=& \alpha_1 + \frac{1}{h}\Big(\alpha_2 a + \alpha_3b+ \alpha_4c\Big) + \frac{1}{h^2}\Big(\alpha_5ab + \alpha_6bc + \alpha_7ac\Big) + \frac{1}{h^3}\alpha_8abc \\
d_2 &=& \frac{1}{h}\Big(\alpha_2\sinthe\cosphi + \alpha_3\sinthe\sinphi + \alpha_4\costhe \Big),  \nonumber \\ 
&+&\frac{1}{h^2}\Big(\left[\alpha_6c + \alpha_5a\right]\sinthe\sinphi + \left[\alpha_7c + \alpha_5b\right]\sinthe\cosphi + \left[\alpha_7a + \alpha_6b\right]\costhe\Big) \nonumber \\
&+& \frac{1}{h^3}\Big(\alpha_8bc\sinthe\cosphi + \alpha_8ab\costhe + \alpha_8ac\sinthe\sinphi\Big), \\
d_3 &=& \frac{1}{h^2}\Big(\alpha_5\sintwothe\cosphi\sinphi + \alpha_7\sinthe\cosphi\costhe + \alpha_6\sinthe\sinphi\costhe\Big) \nonumber \\
&+& \frac{1}{h^3}\Big(\alpha_8a\sinthe\sinphi\costhe + \alpha_8b\sinthe\cosphi\costhe + \alpha_8c\sintwothe\cosphi\sinphi\Big),  \\
d_4 &=& \frac{\alpha_8}{h^3}\Big(\sintwothe\costhe\cosphi\sinphi\Big),
\eq
\end{widetext}
where $h$ is the cell size.

\bibliography{ramsesray.bib}

\begin{thebibliography}{100}%
\makeatletter
\providecommand \@ifxundefined [1]{%
 \ifx #1\undefined \expandafter \@firstoftwo
 \else \expandafter \@secondoftwo
\fi
}%
\providecommand \@ifnum [1]{%
 \ifnum #1\expandafter \@firstoftwo
 \else \expandafter \@secondoftwo
\fi
}%
\providecommand \enquote [1]{``#1''}%
\providecommand \bibnamefont  [1]{#1}%
\providecommand \bibfnamefont [1]{#1}%
\providecommand \citenamefont [1]{#1}%
\providecommand\href[0]{\@sanitize\@href}%
\providecommand\@href[1]{\endgroup\@@startlink{#1}\endgroup\@@href}%
\providecommand\@@href[1]{#1\@@endlink}%
\providecommand \@sanitize [0]{\begingroup\catcode`\&12\catcode`\#12\relax}%
\@ifxundefined \pdfoutput {\@firstoftwo}{%
 \@ifnum{\z@=\pdfoutput}{\@firstoftwo}{\@secondoftwo}%
}{%
 \providecommand\@@startlink[1]{\leavevmode\special{html:<a href="#1">}}%
 \providecommand\@@endlink[0]{\special{html:</a>}}%
}{%
 \providecommand\@@startlink[1]{%
  \leavevmode
  \pdfstartlink
   attr{/Border[0 0 1 ]/H/I/C[0 1 1]}%
   user{/Subtype/Link/A<</Type/Action/S/URI/URI(#1)>>}%
  \relax
 }%
 \providecommand\@@endlink[0]{\pdfendlink}%
}%
\providecommand \url  [0]{\begingroup\@sanitize \@url }%
\providecommand \@url [1]{\endgroup\@href {#1}{\urlprefix}}%
\providecommand \urlprefix [0]{URL }%
\providecommand \Eprint[0]{\href }%
\@ifxundefined \urlstyle {%
  \providecommand \doi [1]{doi:\discretionary{}{}{}#1}%
}{%
  \providecommand \doi [0]{doi:\discretionary{}{}{}\begingroup
  \urlstyle{rm}\Url }%
}%
\providecommand \doibase [0]{http://dx.doi.org/}%
\providecommand \Doi[1]{\href{\doibase#1}}%
\providecommand \bibAnnote [3]{%
  \BibitemShut{#1}%
  \begin{quotation}\noindent
    \textsc{Key:}\ #2\\\textsc{Annotation:}\ #3%
  \end{quotation}%
}%
\providecommand \bibAnnoteFile [2]{%
  \IfFileExists{#2}{\bibAnnote {#1} {#2} {\input{#2}}}{}%
}%
\providecommand \typeout [0]{\immediate \write \m@ne }%
\providecommand \selectlanguage [0]{\@gobble}%
\providecommand \bibinfo [0]{\@secondoftwo}%
\providecommand \bibfield [0]{\@secondoftwo}%
\providecommand \translation [1]{[#1]}%
\providecommand \BibitemOpen[0]{}%
\providecommand \bibitemStop [0]{}%
\providecommand \bibitemNoStop [0]{.\EOS\space}%
\providecommand \EOS [0]{\spacefactor3000\relax}%
\providecommand \BibitemShut [1]{\csname bibitem#1\endcsname}%
\bibitem{2000AJ....120.1579Y}%
  \BibitemOpen
  \bibfield{author}{%
  \bibinfo {author} {\bibfnamefont{D.~G.}\ \bibnamefont{{York et al}}},\ }%
  \bibfield{journal}{%
  \Doi{10.1086/301513}{\bibinfo {journal} {AJ}}\ }%
  \textbf{\bibinfo {volume} {120}},\ \bibinfo {pages} {1579} (\bibinfo {month}
  {Sep.}\ \bibinfo {year} {2000}),\
  \Eprint{http://arxiv.org/abs/astro-ph/0006396}{astro-ph/0006396}%
  \bibAnnoteFile{NoStop}{2000AJ....120.1579Y}%
\bibitem{2003astro.ph..6581C}%
  \BibitemOpen
  \bibfield{author}{%
  \bibinfo {author} {\bibfnamefont{M.}~\bibnamefont{{Colless et al}}},\ }%
  \bibfield{journal}{%
  \bibinfo {journal} {ArXiv Astrophysics e-prints}}%
   (\bibinfo {month} {Jun.}\ \bibinfo {year} {2003}),\
  \Eprint{http://arxiv.org/abs/astro-ph/0306581}{astro-ph/0306581}%
  \bibAnnoteFile{NoStop}{2003astro.ph..6581C}%
\bibitem{2013AJ....145...10D}%
  \BibitemOpen
  \bibfield{author}{%
  \bibinfo {author} {\bibfnamefont{K.~S.}\ \bibnamefont{{Dawson et al}}},\ }%
  \bibfield{journal}{%
  \Doi{10.1088/0004-6256/145/1/10}{\bibinfo {journal} {AJ}}\ }%
  \textbf{\bibinfo {volume} {145}},\ \bibinfo {eid} {10} (\bibinfo {month}
  {Jan.}\ \bibinfo {year} {2013}),\
  \Eprint{http://arxiv.org/abs/1208.0022}{arXiv:1208.0022}%
  \bibAnnoteFile{NoStop}{2013AJ....145...10D}%
\bibitem{2005MNRAS.362..505C}%
  \BibitemOpen
  \bibfield{author}{%
  \bibinfo {author} {\bibfnamefont{S.}~\bibnamefont{{Cole et al}}},\ }%
  \bibfield{journal}{%
  \Doi{10.1111/j.1365-2966.2005.09318.x}{\bibinfo {journal} {MNRAS}}\ }%
  \textbf{\bibinfo {volume} {362}},\ \bibinfo {pages} {505} (\bibinfo {month}
  {Sep.}\ \bibinfo {year} {2005}),\
  \Eprint{http://arxiv.org/abs/astro-ph/0501174}{astro-ph/0501174}%
  \bibAnnoteFile{NoStop}{2005MNRAS.362..505C}%
\bibitem{2005ApJ...633..560E}%
  \BibitemOpen
  \bibfield{author}{%
  \bibinfo {author} {\bibfnamefont{D.~J.}\ \bibnamefont{{Eisenstein et al}}},\
  }%
  \bibfield{journal}{%
  \Doi{10.1086/466512}{\bibinfo {journal} {APJ}}\ }%
  \textbf{\bibinfo {volume} {633}},\ \bibinfo {pages} {560} (\bibinfo {month}
  {Nov.}\ \bibinfo {year} {2005}),\
  \Eprint{http://arxiv.org/abs/astro-ph/0501171}{astro-ph/0501171}%
  \bibAnnoteFile{NoStop}{2005ApJ...633..560E}%
\bibitem{2009MNRAS.393..297P}%
  \BibitemOpen
  \bibfield{author}{%
  \bibinfo {author} {\bibfnamefont{W.~J.}\ \bibnamefont{{Percival}}}\ and\
  \bibinfo {author} {\bibfnamefont{M.}~\bibnamefont{{White}}},\ }%
  \bibfield{journal}{%
  \Doi{10.1111/j.1365-2966.2008.14211.x}{\bibinfo {journal} {MNRAS}}\ }%
  \textbf{\bibinfo {volume} {393}},\ \bibinfo {pages} {297} (\bibinfo {month}
  {Feb.}\ \bibinfo {year} {2009}),\
  \Eprint{http://arxiv.org/abs/0808.0003}{arXiv:0808.0003}%
  \bibAnnoteFile{NoStop}{2009MNRAS.393..297P}%
\bibitem{2008Natur.451..541G}%
  \BibitemOpen
  \bibfield{author}{%
  \bibinfo {author} {\bibfnamefont{L.}~\bibnamefont{{Guzzo et al}}},\ }%
  \bibfield{journal}{%
  \Doi{10.1038/nature06555}{\bibinfo {journal} {NAT}}\ }%
  \textbf{\bibinfo {volume} {451}},\ \bibinfo {pages} {541} (\bibinfo {month}
  {Jan.}\ \bibinfo {year} {2008}),\
  \Eprint{http://arxiv.org/abs/0802.1944}{arXiv:0802.1944}%
  \bibAnnoteFile{NoStop}{2008Natur.451..541G}%
\bibitem{2014MNRAS.440.2692S}%
  \BibitemOpen
  \bibfield{author}{%
  \bibinfo {author} {\bibfnamefont{A.~G.}\ \bibnamefont{{S{\'a}nchez et al}}},\
  }%
  \bibfield{journal}{%
  \Doi{10.1093/mnras/stu342}{\bibinfo {journal} {MNRAS}}\ }%
  \textbf{\bibinfo {volume} {440}},\ \bibinfo {pages} {2692} (\bibinfo {month}
  {May}\ \bibinfo {year} {2014}),\
  \Eprint{http://arxiv.org/abs/1312.4854}{arXiv:1312.4854}%
  \bibAnnoteFile{NoStop}{2014MNRAS.440.2692S}%
\bibitem{1972CoASP...4..173S}%
  \BibitemOpen
  \bibfield{author}{%
  \bibinfo {author} {\bibfnamefont{R.~A.}\ \bibnamefont{{Sunyaev}}}\ and\
  \bibinfo {author} {\bibfnamefont{Y.~B.}\ \bibnamefont{{Zeldovich}}},\ }%
  \bibfield{journal}{%
  \bibinfo {journal} {Comments on Astrophysics and Space Physics}\ }%
  \textbf{\bibinfo {volume} {4}},\ \bibinfo {pages} {173} (\bibinfo {month}
  {Nov.}\ \bibinfo {year} {1972})%
  \bibAnnoteFile{NoStop}{1972CoASP...4..173S}%
\bibitem{1980MNRAS.190..413S}%
  \BibitemOpen
  \bibfield{author}{%
  \bibinfo {author} {\bibfnamefont{R.~A.}\ \bibnamefont{{Sunyaev}}}\ and\
  \bibinfo {author} {\bibfnamefont{I.~B.}\ \bibnamefont{{Zeldovich}}},\ }%
  \bibfield{journal}{%
  \bibinfo {journal} {MNRAS}\ }%
  \textbf{\bibinfo {volume} {190}},\ \bibinfo {pages} {413} (\bibinfo {month}
  {Feb.}\ \bibinfo {year} {1980})%
  \bibAnnoteFile{NoStop}{1980MNRAS.190..413S}%
\bibitem{carlstrom}%
  \BibitemOpen
  \bibfield{author}{%
  \bibinfo {author} {\bibfnamefont{J.~E.}\ \bibnamefont{{Carlstrom}}}, \bibinfo
  {author} {\bibfnamefont{G.~P.}\ \bibnamefont{{Holder}}},\ and\ \bibinfo
  {author} {\bibfnamefont{E.~D.}\ \bibnamefont{{Reese}}},\ }%
  \bibfield{journal}{%
  \Doi{10.1146/annurev.astro.40.060401.093803}{\bibinfo {journal} {ARAA}}\ }%
  \textbf{\bibinfo {volume} {40}},\ \bibinfo {pages} {643} (\bibinfo {year}
  {2002}),\ \Eprint{http://arxiv.org/abs/astro-ph/0208192}{astro-ph/0208192}%
  \bibAnnoteFile{NoStop}{carlstrom}%
\bibitem{2012PhRvL.109d1101H}%
  \BibitemOpen
  \bibfield{author}{%
  \bibinfo {author} {\bibfnamefont{N.}~\bibnamefont{{Hand et al}}},\ }%
  \bibfield{journal}{%
  \Doi{10.1103/PhysRevLett.109.041101}{\bibinfo {journal} {Physical Review
  Letters}}\ }%
  \textbf{\bibinfo {volume} {109}},\ \bibinfo {eid} {041101} (\bibinfo {month}
  {Jul.}\ \bibinfo {year} {2012}),\
  \Eprint{http://arxiv.org/abs/1203.4219}{arXiv:1203.4219}%
  \bibAnnoteFile{NoStop}{2012PhRvL.109d1101H}%
\bibitem{1967ApJ...147...73S}%
  \BibitemOpen
  \bibfield{author}{%
  \bibinfo {author} {\bibfnamefont{R.~K.}\ \bibnamefont{{Sachs}}}\ and\
  \bibinfo {author} {\bibfnamefont{A.~M.}\ \bibnamefont{{Wolfe}}},\ }%
  \bibfield{journal}{%
  \Doi{10.1086/148982}{\bibinfo {journal} {APJ}}\ }%
  \textbf{\bibinfo {volume} {147}},\ \bibinfo {pages} {73} (\bibinfo {month}
  {Jan.}\ \bibinfo {year} {1967})%
  \bibAnnoteFile{NoStop}{1967ApJ...147...73S}%
\bibitem{2002PhRvD..65j3510C}%
  \BibitemOpen
  \bibfield{author}{%
  \bibinfo {author} {\bibfnamefont{A.}~\bibnamefont{{Cooray}}},\ }%
  \bibfield{journal}{%
  \Doi{10.1103/PhysRevD.65.103510}{\bibinfo {journal} {\prd}}\ }%
  \textbf{\bibinfo {volume} {65}},\ \bibinfo {eid} {103510} (\bibinfo {month}
  {May}\ \bibinfo {year} {2002}),\
  \Eprint{http://arxiv.org/abs/astro-ph/0112408}{astro-ph/0112408}%
  \bibAnnoteFile{NoStop}{2002PhRvD..65j3510C}%
\bibitem{2008PhRvD..78d3519H}%
  \BibitemOpen
  \bibfield{author}{%
  \bibinfo {author} {\bibfnamefont{S.}~\bibnamefont{{Ho}}}, \bibinfo {author}
  {\bibfnamefont{C.}~\bibnamefont{{Hirata}}}, \bibinfo {author}
  {\bibfnamefont{N.}~\bibnamefont{{Padmanabhan}}}, \bibinfo {author}
  {\bibfnamefont{U.}~\bibnamefont{{Seljak}}},\ and\ \bibinfo {author}
  {\bibfnamefont{N.}~\bibnamefont{{Bahcall}}},\ }%
  \bibfield{journal}{%
  \Doi{10.1103/PhysRevD.78.043519}{\bibinfo {journal} {PRD}}\ }%
  \textbf{\bibinfo {volume} {78}},\ \bibinfo {eid} {043519} (\bibinfo {month}
  {Aug.}\ \bibinfo {year} {2008}),\
  \Eprint{http://arxiv.org/abs/0801.0642}{arXiv:0801.0642}%
  \bibAnnoteFile{NoStop}{2008PhRvD..78d3519H}%
\bibitem{2008ApJ...683L..99G}%
  \BibitemOpen
  \bibfield{author}{%
  \bibinfo {author} {\bibfnamefont{B.~R.}\ \bibnamefont{{Granett}}}, \bibinfo
  {author} {\bibfnamefont{M.~C.}\ \bibnamefont{{Neyrinck}}},\ and\ \bibinfo
  {author} {\bibfnamefont{I.}~\bibnamefont{{Szapudi}}},\ }%
  \bibfield{journal}{%
  \Doi{10.1086/591670}{\bibinfo {journal} {APJL}}\ }%
  \textbf{\bibinfo {volume} {683}},\ \bibinfo {pages} {L99} (\bibinfo {month}
  {Aug.}\ \bibinfo {year} {2008}),\
  \Eprint{http://arxiv.org/abs/0805.3695}{arXiv:0805.3695}%
  \bibAnnoteFile{NoStop}{2008ApJ...683L..99G}%
\bibitem{2001PhR...340..291B}%
  \BibitemOpen
  \bibfield{author}{%
  \bibinfo {author} {\bibfnamefont{M.}~\bibnamefont{{Bartelmann}}}\ and\
  \bibinfo {author} {\bibfnamefont{P.}~\bibnamefont{{Schneider}}},\ }%
  \bibfield{journal}{%
  \Doi{10.1016/S0370-1573(00)00082-X}{\bibinfo {journal} {PhysRep}}\ }%
  \textbf{\bibinfo {volume} {340}},\ \bibinfo {pages} {291} (\bibinfo {month}
  {Jan.}\ \bibinfo {year} {2001}),\
  \Eprint{http://arxiv.org/abs/astro-ph/9912508}{astro-ph/9912508}%
  \bibAnnoteFile{NoStop}{2001PhR...340..291B}%
\bibitem{2003ARA&A..41..645R}%
  \BibitemOpen
  \bibfield{author}{%
  \bibinfo {author} {\bibfnamefont{A.}~\bibnamefont{{Refregier}}},\ }%
  \bibfield{journal}{%
  \Doi{10.1146/annurev.astro.41.111302.102207}{\bibinfo {journal} {ARAA}}\ }%
  \textbf{\bibinfo {volume} {41}},\ \bibinfo {pages} {645} (\bibinfo {year}
  {2003}),\ \Eprint{http://arxiv.org/abs/astro-ph/0307212}{astro-ph/0307212}%
  \bibAnnoteFile{NoStop}{2003ARA&A..41..645R}%
\bibitem{2010CQGra..27w3001B}%
  \BibitemOpen
  \bibfield{author}{%
  \bibinfo {author} {\bibfnamefont{M.}~\bibnamefont{{Bartelmann}}},\ }%
  \bibfield{journal}{%
  \Doi{10.1088/0264-9381/27/23/233001}{\bibinfo {journal} {Classical and
  Quantum Gravity}}\ }%
  \textbf{\bibinfo {volume} {27}},\ \bibinfo {eid} {233001} (\bibinfo {month}
  {Dec.}\ \bibinfo {year} {2010}),\
  \Eprint{http://arxiv.org/abs/1010.3829}{arXiv:1010.3829}%
  \bibAnnoteFile{NoStop}{2010CQGra..27w3001B}%
\bibitem{2015RPPh...78h6901K}%
  \BibitemOpen
  \bibfield{author}{%
  \bibinfo {author} {\bibfnamefont{M.}~\bibnamefont{{Kilbinger}}},\ }%
  \bibfield{journal}{%
  \Doi{10.1088/0034-4885/78/8/086901}{\bibinfo {journal} {Reports on Progress
  in Physics}}\ }%
  \textbf{\bibinfo {volume} {78}},\ \bibinfo {eid} {086901} (\bibinfo {month}
  {Jul.}\ \bibinfo {year} {2015}),\
  \Eprint{http://arxiv.org/abs/1411.0115}{arXiv:1411.0115}%
  \bibAnnoteFile{NoStop}{2015RPPh...78h6901K}%
\bibitem{2000ApJ...530..547J}%
  \BibitemOpen
  \bibfield{author}{%
  \bibinfo {author} {\bibfnamefont{B.}~\bibnamefont{{Jain}}}, \bibinfo {author}
  {\bibfnamefont{U.}~\bibnamefont{{Seljak}}},\ and\ \bibinfo {author}
  {\bibfnamefont{S.}~\bibnamefont{{White}}},\ }%
  \bibfield{journal}{%
  \Doi{10.1086/308384}{\bibinfo {journal} {APJ}}\ }%
  \textbf{\bibinfo {volume} {530}},\ \bibinfo {pages} {547} (\bibinfo {month}
  {Feb.}\ \bibinfo {year} {2000}),\
  \Eprint{http://arxiv.org/abs/astro-ph/9901191}{astro-ph/9901191}%
  \bibAnnoteFile{NoStop}{2000ApJ...530..547J}%
\bibitem{2003ApJ...592..699V}%
  \BibitemOpen
  \bibfield{author}{%
  \bibinfo {author} {\bibfnamefont{C.}~\bibnamefont{{Vale}}}\ and\ \bibinfo
  {author} {\bibfnamefont{M.}~\bibnamefont{{White}}},\ }%
  \bibfield{journal}{%
  \Doi{10.1086/375867}{\bibinfo {journal} {APJ}}\ }%
  \textbf{\bibinfo {volume} {592}},\ \bibinfo {pages} {699} (\bibinfo {month}
  {Aug.}\ \bibinfo {year} {2003}),\
  \Eprint{http://arxiv.org/abs/astro-ph/0303555}{astro-ph/0303555}%
  \bibAnnoteFile{NoStop}{2003ApJ...592..699V}%
\bibitem{2008ApJ...682....1D}%
  \BibitemOpen
  \bibfield{author}{%
  \bibinfo {author} {\bibfnamefont{S.}~\bibnamefont{{Das}}}\ and\ \bibinfo
  {author} {\bibfnamefont{P.}~\bibnamefont{{Bode}}},\ }%
  \bibfield{journal}{%
  \Doi{10.1086/589638}{\bibinfo {journal} {APJ}}\ }%
  \textbf{\bibinfo {volume} {682}},\ \bibinfo {pages} {1} (\bibinfo {month}
  {Jul.}\ \bibinfo {year} {2008}),\
  \Eprint{http://arxiv.org/abs/0711.3793}{arXiv:0711.3793}%
  \bibAnnoteFile{NoStop}{2008ApJ...682....1D}%
\bibitem{2008MNRAS.391..435F}%
  \BibitemOpen
  \bibfield{author}{%
  \bibinfo {author} {\bibfnamefont{P.}~\bibnamefont{{Fosalba}}}, \bibinfo
  {author} {\bibfnamefont{E.}~\bibnamefont{{Gazta{\~n}aga}}}, \bibinfo {author}
  {\bibfnamefont{F.~J.}\ \bibnamefont{{Castander}}},\ and\ \bibinfo {author}
  {\bibfnamefont{M.}~\bibnamefont{{Manera}}},\ }%
  \bibfield{journal}{%
  \Doi{10.1111/j.1365-2966.2008.13910.x}{\bibinfo {journal} {MNRAS}}\ }%
  \textbf{\bibinfo {volume} {391}},\ \bibinfo {pages} {435} (\bibinfo {month}
  {Nov.}\ \bibinfo {year} {2008}),\
  \Eprint{http://arxiv.org/abs/0711.1540}{arXiv:0711.1540}%
  \bibAnnoteFile{NoStop}{2008MNRAS.391..435F}%
\bibitem{2009A&A...497..335T}%
  \BibitemOpen
  \bibfield{author}{%
  \bibinfo {author} {\bibfnamefont{R.}~\bibnamefont{{Teyssier}}}, \bibinfo
  {author} {\bibfnamefont{S.}~\bibnamefont{{Pires}}}, \bibinfo {author}
  {\bibfnamefont{S.}~\bibnamefont{{Prunet}}}, \bibinfo {author}
  {\bibfnamefont{D.}~\bibnamefont{{Aubert}}}, \bibinfo {author}
  {\bibfnamefont{C.}~\bibnamefont{{Pichon}}}, \bibinfo {author}
  {\bibfnamefont{A.}~\bibnamefont{{Amara}}}, \bibinfo {author}
  {\bibfnamefont{K.}~\bibnamefont{{Benabed}}}, \bibinfo {author}
  {\bibfnamefont{S.}~\bibnamefont{{Colombi}}}, \bibinfo {author}
  {\bibfnamefont{A.}~\bibnamefont{{Refregier}}},\ and\ \bibinfo {author}
  {\bibfnamefont{J.-L.}\ \bibnamefont{{Starck}}},\ }%
  \bibfield{journal}{%
  \Doi{10.1051/0004-6361/200810657}{\bibinfo {journal} {AAP}}\ }%
  \textbf{\bibinfo {volume} {497}},\ \bibinfo {pages} {335} (\bibinfo {month}
  {Apr.}\ \bibinfo {year} {2009}),\
  \Eprint{http://arxiv.org/abs/0807.3651}{arXiv:0807.3651}%
  \bibAnnoteFile{NoStop}{2009A&A...497..335T}%
\bibitem{2008MNRAS.388.1618C}%
  \BibitemOpen
  \bibfield{author}{%
  \bibinfo {author} {\bibfnamefont{C.}~\bibnamefont{{Carbone}}}, \bibinfo
  {author} {\bibfnamefont{V.}~\bibnamefont{{Springel}}}, \bibinfo {author}
  {\bibfnamefont{C.}~\bibnamefont{{Baccigalupi}}}, \bibinfo {author}
  {\bibfnamefont{M.}~\bibnamefont{{Bartelmann}}},\ and\ \bibinfo {author}
  {\bibfnamefont{S.}~\bibnamefont{{Matarrese}}},\ }%
  \bibfield{journal}{%
  \Doi{10.1111/j.1365-2966.2008.13544.x}{\bibinfo {journal} {MNRAS}}\ }%
  \textbf{\bibinfo {volume} {388}},\ \bibinfo {pages} {1618} (\bibinfo {month}
  {Aug.}\ \bibinfo {year} {2008}),\
  \Eprint{http://arxiv.org/abs/0711.2655}{arXiv:0711.2655}%
  \bibAnnoteFile{NoStop}{2008MNRAS.388.1618C}%
\bibitem{Hilbert:2008kb}%
  \BibitemOpen
  \bibfield{author}{%
  \bibinfo {author} {\bibfnamefont{S.}~\bibnamefont{Hilbert}}, \bibinfo
  {author} {\bibfnamefont{J.}~\bibnamefont{Hartlap}}, \bibinfo {author}
  {\bibfnamefont{S.~D.~M.}\ \bibnamefont{White}},\ and\ \bibinfo {author}
  {\bibfnamefont{P.}~\bibnamefont{Schneider}},\ }%
  \bibfield{journal}{%
  \Doi{10.1051/0004-6361/200811054}{\bibinfo {journal} {Astron. Astrophys.}}\
  }%
  \textbf{\bibinfo {volume} {499}},\ \bibinfo {pages} {31} (\bibinfo {year}
  {2009}),\ \Eprint{http://arxiv.org/abs/0809.5035}{arXiv:0809.5035
  [astro-ph]}%
  \bibAnnoteFile{NoStop}{Hilbert:2008kb}%
\bibitem{2009ApJ...701..945S}%
  \BibitemOpen
  \bibfield{author}{%
  \bibinfo {author} {\bibfnamefont{M.}~\bibnamefont{{Sato}}}, \bibinfo {author}
  {\bibfnamefont{T.}~\bibnamefont{{Hamana}}}, \bibinfo {author}
  {\bibfnamefont{R.}~\bibnamefont{{Takahashi}}}, \bibinfo {author}
  {\bibfnamefont{M.}~\bibnamefont{{Takada}}}, \bibinfo {author}
  {\bibfnamefont{N.}~\bibnamefont{{Yoshida}}}, \bibinfo {author}
  {\bibfnamefont{T.}~\bibnamefont{{Matsubara}}},\ and\ \bibinfo {author}
  {\bibfnamefont{N.}~\bibnamefont{{Sugiyama}}},\ }%
  \bibfield{journal}{%
  \Doi{10.1088/0004-637X/701/2/945}{\bibinfo {journal} {APJ}}\ }%
  \textbf{\bibinfo {volume} {701}},\ \bibinfo {pages} {945} (\bibinfo {month}
  {Aug.}\ \bibinfo {year} {2009}),\
  \Eprint{http://arxiv.org/abs/0906.2237}{arXiv:0906.2237 [astro-ph.CO]}%
  \bibAnnoteFile{NoStop}{2009ApJ...701..945S}%
\bibitem{2012MNRAS.420..155K}%
  \BibitemOpen
  \bibfield{author}{%
  \bibinfo {author} {\bibfnamefont{M.}~\bibnamefont{{Killedar}}}, \bibinfo
  {author} {\bibfnamefont{P.~D.}\ \bibnamefont{{Lasky}}}, \bibinfo {author}
  {\bibfnamefont{G.~F.}\ \bibnamefont{{Lewis}}},\ and\ \bibinfo {author}
  {\bibfnamefont{C.~J.}\ \bibnamefont{{Fluke}}},\ }%
  \bibfield{journal}{%
  \Doi{10.1111/j.1365-2966.2011.20023.x}{\bibinfo {journal} {MNRAS}}\ }%
  \textbf{\bibinfo {volume} {420}},\ \bibinfo {pages} {155} (\bibinfo {month}
  {Feb.}\ \bibinfo {year} {2012}),\
  \Eprint{http://arxiv.org/abs/1110.4894}{arXiv:1110.4894}%
  \bibAnnoteFile{NoStop}{2012MNRAS.420..155K}%
\bibitem{2013MNRAS.435..115B}%
  \BibitemOpen
  \bibfield{author}{%
  \bibinfo {author} {\bibfnamefont{M.~R.}\ \bibnamefont{{Becker}}},\ }%
  \bibfield{journal}{%
  \Doi{10.1093/mnras/stt1352}{\bibinfo {journal} {MNRAS}}\ }%
  \textbf{\bibinfo {volume} {435}},\ \bibinfo {pages} {115} (\bibinfo {month}
  {Oct.}\ \bibinfo {year} {2013})%
  \bibAnnoteFile{NoStop}{2013MNRAS.435..115B}%
\bibitem{2014MNRAS.445.1942M}%
  \BibitemOpen
  \bibfield{author}{%
  \bibinfo {author} {\bibfnamefont{R.~B.}\ \bibnamefont{{Metcalf}}}\ and\
  \bibinfo {author} {\bibfnamefont{M.}~\bibnamefont{{Petkova}}},\ }%
  \bibfield{journal}{%
  \Doi{10.1093/mnras/stu1859}{\bibinfo {journal} {MNRAS}}\ }%
  \textbf{\bibinfo {volume} {445}},\ \bibinfo {pages} {1942} (\bibinfo {month}
  {Dec.}\ \bibinfo {year} {2014}),\
  \Eprint{http://arxiv.org/abs/1312.1128}{arXiv:1312.1128}%
  \bibAnnoteFile{NoStop}{2014MNRAS.445.1942M}%
\bibitem{2014MNRAS.445.1954P}%
  \BibitemOpen
  \bibfield{author}{%
  \bibinfo {author} {\bibfnamefont{M.}~\bibnamefont{{Petkova}}}, \bibinfo
  {author} {\bibfnamefont{R.~B.}\ \bibnamefont{{Metcalf}}},\ and\ \bibinfo
  {author} {\bibfnamefont{C.}~\bibnamefont{{Giocoli}}},\ }%
  \bibfield{journal}{%
  \Doi{10.1093/mnras/stu1860}{\bibinfo {journal} {MNRAS}}\ }%
  \textbf{\bibinfo {volume} {445}},\ \bibinfo {pages} {1954} (\bibinfo {month}
  {Dec.}\ \bibinfo {year} {2014}),\
  \Eprint{http://arxiv.org/abs/1312.1536}{arXiv:1312.1536}%
  \bibAnnoteFile{NoStop}{2014MNRAS.445.1954P}%
\bibitem{2015PhRvD..91f3507L}%
  \BibitemOpen
  \bibfield{author}{%
  \bibinfo {author} {\bibfnamefont{J.}~\bibnamefont{{Liu}}}, \bibinfo {author}
  {\bibfnamefont{A.}~\bibnamefont{{Petri}}}, \bibinfo {author}
  {\bibfnamefont{Z.}~\bibnamefont{{Haiman}}}, \bibinfo {author}
  {\bibfnamefont{L.}~\bibnamefont{{Hui}}}, \bibinfo {author}
  {\bibfnamefont{J.~M.}\ \bibnamefont{{Kratochvil}}},\ and\ \bibinfo {author}
  {\bibfnamefont{M.}~\bibnamefont{{May}}},\ }%
  \bibfield{journal}{%
  \Doi{10.1103/PhysRevD.91.063507}{\bibinfo {journal} {PRD}}\ }%
  \textbf{\bibinfo {volume} {91}},\ \bibinfo {eid} {063507} (\bibinfo {month}
  {Mar.}\ \bibinfo {year} {2015}),\
  \Eprint{http://arxiv.org/abs/1412.0757}{arXiv:1412.0757}%
  \bibAnnoteFile{NoStop}{2015PhRvD..91f3507L}%
\bibitem{2015arXiv151108211G}%
  \BibitemOpen
  \bibfield{author}{%
  \bibinfo {author} {\bibfnamefont{C.}~\bibnamefont{{Giocoli}}}, \bibinfo
  {author} {\bibfnamefont{E.}~\bibnamefont{{Jullo}}}, \bibinfo {author}
  {\bibfnamefont{R.~B.}\ \bibnamefont{{Metcalf}}}, \bibinfo {author}
  {\bibfnamefont{S.}~\bibnamefont{{de la Torre}}}, \bibinfo {author}
  {\bibfnamefont{G.}~\bibnamefont{{Yepes}}}, \bibinfo {author}
  {\bibfnamefont{F.}~\bibnamefont{{Prada}}}, \bibinfo {author}
  {\bibfnamefont{J.}~\bibnamefont{{Comparat}}}, \bibinfo {author}
  {\bibfnamefont{S.}~\bibnamefont{{Goettlober}}}, \bibinfo {author}
  {\bibfnamefont{A.}~\bibnamefont{{Kyplin}}}, \bibinfo {author}
  {\bibfnamefont{J.-P.}\ \bibnamefont{{Kneib}}}, \bibinfo {author}
  {\bibfnamefont{M.}~\bibnamefont{{Petkova}}}, \bibinfo {author}
  {\bibfnamefont{H.}~\bibnamefont{{Shan}}},\ and\ \bibinfo {author}
  {\bibfnamefont{N.}~\bibnamefont{{Tessore}}},\ }%
  \bibfield{journal}{%
  \bibinfo {journal} {ArXiv e-prints}}%
   (\bibinfo {month} {Nov.}\ \bibinfo {year} {2015}),\
  \Eprint{http://arxiv.org/abs/1511.08211}{arXiv:1511.08211}%
  \bibAnnoteFile{NoStop}{2015arXiv151108211G}%
\bibitem{2009MNRAS.396..772C}%
  \BibitemOpen
  \bibfield{author}{%
  \bibinfo {author} {\bibfnamefont{Y.-C.}\ \bibnamefont{{Cai}}}, \bibinfo
  {author} {\bibfnamefont{S.}~\bibnamefont{{Cole}}}, \bibinfo {author}
  {\bibfnamefont{A.}~\bibnamefont{{Jenkins}}},\ and\ \bibinfo {author}
  {\bibfnamefont{C.}~\bibnamefont{{Frenk}}},\ }%
  \bibfield{journal}{%
  \Doi{10.1111/j.1365-2966.2009.14780.x}{\bibinfo {journal} {MNRAS}}\ }%
  \textbf{\bibinfo {volume} {396}},\ \bibinfo {pages} {772} (\bibinfo {month}
  {Jun.}\ \bibinfo {year} {2009}),\
  \Eprint{http://arxiv.org/abs/0809.4488}{arXiv:0809.4488}%
  \bibAnnoteFile{NoStop}{2009MNRAS.396..772C}%
\bibitem{whitehu2000}%
  \BibitemOpen
  \bibfield{author}{%
  \bibinfo {author} {\bibfnamefont{M.}~\bibnamefont{{White}}}\ and\ \bibinfo
  {author} {\bibfnamefont{W.}~\bibnamefont{{Hu}}},\ }%
  \bibfield{journal}{%
  \Doi{10.1086/309009}{\bibinfo {journal} {APJ}}\ }%
  \textbf{\bibinfo {volume} {537}},\ \bibinfo {pages} {1} (\bibinfo {month}
  {Jul.}\ \bibinfo {year} {2000}),\
  \Eprint{http://arxiv.org/abs/astro-ph/9909165}{astro-ph/9909165}%
  \bibAnnoteFile{NoStop}{whitehu2000}%
\bibitem{li2001}%
  \BibitemOpen
  \bibfield{author}{%
  \bibinfo {author} {\bibfnamefont{B.}~\bibnamefont{{Li}}}, \bibinfo {author}
  {\bibfnamefont{L.~J.}\ \bibnamefont{{King}}}, \bibinfo {author}
  {\bibfnamefont{G.-B.}\ \bibnamefont{{Zhao}}},\ and\ \bibinfo {author}
  {\bibfnamefont{H.}~\bibnamefont{{Zhao}}},\ }%
  \bibfield{journal}{%
  \Doi{10.1111/j.1365-2966.2011.18754.x}{\bibinfo {journal} {MNRAS}}\ }%
  \textbf{\bibinfo {volume} {415}},\ \bibinfo {pages} {881} (\bibinfo {month}
  {Jul.}\ \bibinfo {year} {2011}),\
  \Eprint{http://arxiv.org/abs/1012.1625}{arXiv:1012.1625}%
  \bibAnnoteFile{NoStop}{li2001}%
\bibitem{2002A&A...385..337T}%
  \BibitemOpen
  \bibfield{author}{%
  \bibinfo {author} {\bibfnamefont{R.}~\bibnamefont{{Teyssier}}},\ }%
  \bibfield{journal}{%
  \Doi{10.1051/0004-6361:20011817}{\bibinfo {journal} {AAP}}\ }%
  \textbf{\bibinfo {volume} {385}},\ \bibinfo {pages} {337} (\bibinfo {month}
  {Apr.}\ \bibinfo {year} {2002}),\
  \Eprint{http://arxiv.org/abs/astro-ph/0111367}{astro-ph/0111367}%
  \bibAnnoteFile{NoStop}{2002A&A...385..337T}%
\bibitem{2010MNRAS.407..201C}%
  \BibitemOpen
  \bibfield{author}{%
  \bibinfo {author} {\bibfnamefont{Y.-C.}\ \bibnamefont{{Cai}}}, \bibinfo
  {author} {\bibfnamefont{S.}~\bibnamefont{{Cole}}}, \bibinfo {author}
  {\bibfnamefont{A.}~\bibnamefont{{Jenkins}}},\ and\ \bibinfo {author}
  {\bibfnamefont{C.~S.}\ \bibnamefont{{Frenk}}},\ }%
  \bibfield{journal}{%
  \Doi{10.1111/j.1365-2966.2010.16946.x}{\bibinfo {journal} {MNRAS}}\ }%
  \textbf{\bibinfo {volume} {407}},\ \bibinfo {pages} {201} (\bibinfo {month}
  {Sep.}\ \bibinfo {year} {2010}),\
  \Eprint{http://arxiv.org/abs/1003.0974}{arXiv:1003.0974}%
  \bibAnnoteFile{NoStop}{2010MNRAS.407..201C}%
\bibitem{2014MNRAS.438..412W}%
  \BibitemOpen
  \bibfield{author}{%
  \bibinfo {author} {\bibfnamefont{W.~A.}\ \bibnamefont{{Watson}}}, \bibinfo
  {author} {\bibfnamefont{J.~M.}\ \bibnamefont{{Diego}}}, \bibinfo {author}
  {\bibfnamefont{S.}~\bibnamefont{{Gottl{\"o}ber}}}, \bibinfo {author}
  {\bibfnamefont{I.~T.}\ \bibnamefont{{Iliev}}}, \bibinfo {author}
  {\bibfnamefont{A.}~\bibnamefont{{Knebe}}}, \bibinfo {author}
  {\bibfnamefont{E.}~\bibnamefont{{Mart{\'{\i}}nez-Gonz{\'a}lez}}}, \bibinfo
  {author} {\bibfnamefont{G.}~\bibnamefont{{Yepes}}}, \bibinfo {author}
  {\bibfnamefont{R.~B.}\ \bibnamefont{{Barreiro}}}, \bibinfo {author}
  {\bibfnamefont{J.}~\bibnamefont{{Gonz{\'a}lez-Nuevo}}}, \bibinfo {author}
  {\bibfnamefont{S.}~\bibnamefont{{Hotchkiss}}}, \bibinfo {author}
  {\bibfnamefont{A.}~\bibnamefont{{Marcos-Caballero}}}, \bibinfo {author}
  {\bibfnamefont{S.}~\bibnamefont{{Nadathur}}},\ and\ \bibinfo {author}
  {\bibfnamefont{P.}~\bibnamefont{{Vielva}}},\ }%
  \bibfield{journal}{%
  \Doi{10.1093/mnras/stt2208}{\bibinfo {journal} {MNRAS}}\ }%
  \textbf{\bibinfo {volume} {438}},\ \bibinfo {pages} {412} (\bibinfo {month}
  {Feb.}\ \bibinfo {year} {2014}),\
  \Eprint{http://arxiv.org/abs/1307.1712}{arXiv:1307.1712}%
  \bibAnnoteFile{NoStop}{2014MNRAS.438..412W}%
\bibitem{2000MNRAS.317...37D}%
  \BibitemOpen
  \bibfield{author}{%
  \bibinfo {author} {\bibfnamefont{A.~C.}\ \bibnamefont{{da Silva}}}, \bibinfo
  {author} {\bibfnamefont{D.}~\bibnamefont{{Barbosa}}}, \bibinfo {author}
  {\bibfnamefont{A.~R.}\ \bibnamefont{{Liddle}}},\ and\ \bibinfo {author}
  {\bibfnamefont{P.~A.}\ \bibnamefont{{Thomas}}},\ }%
  \bibfield{journal}{%
  \Doi{10.1046/j.1365-8711.2000.03553.x}{\bibinfo {journal} {MNRAS}}\ }%
  \textbf{\bibinfo {volume} {317}},\ \bibinfo {pages} {37} (\bibinfo {month}
  {Sep.}\ \bibinfo {year} {2000}),\
  \Eprint{http://arxiv.org/abs/astro-ph/9907224}{astro-ph/9907224}%
  \bibAnnoteFile{NoStop}{2000MNRAS.317...37D}%
\bibitem{2001MNRAS.326..155D}%
  \BibitemOpen
  \bibfield{author}{%
  \bibinfo {author} {\bibfnamefont{A.~C.}\ \bibnamefont{{da Silva}}}, \bibinfo
  {author} {\bibfnamefont{D.}~\bibnamefont{{Barbosa}}}, \bibinfo {author}
  {\bibfnamefont{A.~R.}\ \bibnamefont{{Liddle}}},\ and\ \bibinfo {author}
  {\bibfnamefont{P.~A.}\ \bibnamefont{{Thomas}}},\ }%
  \bibfield{journal}{%
  \Doi{10.1046/j.1365-8711.2001.04580.x}{\bibinfo {journal} {MNRAS}}\ }%
  \textbf{\bibinfo {volume} {326}},\ \bibinfo {pages} {155} (\bibinfo {month}
  {Sep.}\ \bibinfo {year} {2001}),\
  \Eprint{http://arxiv.org/abs/astro-ph/0011187}{astro-ph/0011187}%
  \bibAnnoteFile{NoStop}{2001MNRAS.326..155D}%
\bibitem{2001ApJ...549..681S}%
  \BibitemOpen
  \bibfield{author}{%
  \bibinfo {author} {\bibfnamefont{V.}~\bibnamefont{{Springel}}}, \bibinfo
  {author} {\bibfnamefont{M.}~\bibnamefont{{White}}},\ and\ \bibinfo {author}
  {\bibfnamefont{L.}~\bibnamefont{{Hernquist}}},\ }%
  \bibfield{journal}{%
  \Doi{10.1086/319473}{\bibinfo {journal} {APJ}}\ }%
  \textbf{\bibinfo {volume} {549}},\ \bibinfo {pages} {681} (\bibinfo {month}
  {Mar.}\ \bibinfo {year} {2001}),\
  \Eprint{http://arxiv.org/abs/astro-ph/0008133}{astro-ph/0008133}%
  \bibAnnoteFile{NoStop}{2001ApJ...549..681S}%
\bibitem{2014MNRAS.440.3645M}%
  \BibitemOpen
  \bibfield{author}{%
  \bibinfo {author} {\bibfnamefont{I.~G.}\ \bibnamefont{{McCarthy}}}, \bibinfo
  {author} {\bibfnamefont{A.~M.~C.}\ \bibnamefont{{Le Brun}}}, \bibinfo
  {author} {\bibfnamefont{J.}~\bibnamefont{{Schaye}}},\ and\ \bibinfo {author}
  {\bibfnamefont{G.~P.}\ \bibnamefont{{Holder}}},\ }%
  \bibfield{journal}{%
  \Doi{10.1093/mnras/stu543}{\bibinfo {journal} {MNRAS}}\ }%
  \textbf{\bibinfo {volume} {440}},\ \bibinfo {pages} {3645} (\bibinfo {month}
  {Jun.}\ \bibinfo {year} {2014}),\
  \Eprint{http://arxiv.org/abs/1312.5341}{arXiv:1312.5341}%
  \bibAnnoteFile{NoStop}{2014MNRAS.440.3645M}%
\bibitem{2015arXiv150905134D}%
  \BibitemOpen
  \bibfield{author}{%
  \bibinfo {author} {\bibfnamefont{K.}~\bibnamefont{{Dolag}}}, \bibinfo
  {author} {\bibfnamefont{E.}~\bibnamefont{{Komatsu}}},\ and\ \bibinfo {author}
  {\bibfnamefont{R.}~\bibnamefont{{Sunyaev}}},\ }%
  \bibfield{journal}{%
  \bibinfo {journal} {ArXiv e-prints}}%
   (\bibinfo {month} {Sep.}\ \bibinfo {year} {2015}),\
  \Eprint{http://arxiv.org/abs/1509.05134}{arXiv:1509.05134}%
  \bibAnnoteFile{NoStop}{2015arXiv150905134D}%
\bibitem{2006MNRAS.371.1057I}%
  \BibitemOpen
  \bibfield{author}{%
  \bibinfo {author} {\bibfnamefont{I.~T.}\ \bibnamefont{{Iliev}}}, \bibinfo
  {author} {\bibfnamefont{B.}~\bibnamefont{{Ciardi}}}, \bibinfo {author}
  {\bibfnamefont{M.~A.}\ \bibnamefont{{Alvarez}}}, \bibinfo {author}
  {\bibfnamefont{A.}~\bibnamefont{{Maselli}}}, \bibinfo {author}
  {\bibfnamefont{A.}~\bibnamefont{{Ferrara}}}, \bibinfo {author}
  {\bibfnamefont{N.~Y.}\ \bibnamefont{{Gnedin}}}, \bibinfo {author}
  {\bibfnamefont{G.}~\bibnamefont{{Mellema}}}, \bibinfo {author}
  {\bibfnamefont{T.}~\bibnamefont{{Nakamoto}}}, \bibinfo {author}
  {\bibfnamefont{M.~L.}\ \bibnamefont{{Norman}}}, \bibinfo {author}
  {\bibfnamefont{A.~O.}\ \bibnamefont{{Razoumov}}}, \bibinfo {author}
  {\bibfnamefont{E.-J.}\ \bibnamefont{{Rijkhorst}}}, \bibinfo {author}
  {\bibfnamefont{J.}~\bibnamefont{{Ritzerveld}}}, \bibinfo {author}
  {\bibfnamefont{P.~R.}\ \bibnamefont{{Shapiro}}}, \bibinfo {author}
  {\bibfnamefont{H.}~\bibnamefont{{Susa}}}, \bibinfo {author}
  {\bibfnamefont{M.}~\bibnamefont{{Umemura}}},\ and\ \bibinfo {author}
  {\bibfnamefont{D.~J.}\ \bibnamefont{{Whalen}}},\ }%
  \bibfield{journal}{%
  \Doi{10.1111/j.1365-2966.2006.10775.x}{\bibinfo {journal} {MNRAS}}\ }%
  \textbf{\bibinfo {volume} {371}},\ \bibinfo {pages} {1057} (\bibinfo {month}
  {Sep.}\ \bibinfo {year} {2006}),\
  \Eprint{http://arxiv.org/abs/astro-ph/0603199}{astro-ph/0603199}%
  \bibAnnoteFile{NoStop}{2006MNRAS.371.1057I}%
\bibitem{2009MNRAS.400.1283I}%
  \BibitemOpen
  \bibfield{author}{%
  \bibinfo {author} {\bibfnamefont{I.~T.}\ \bibnamefont{{Iliev}}}, \bibinfo
  {author} {\bibfnamefont{D.}~\bibnamefont{{Whalen}}}, \bibinfo {author}
  {\bibfnamefont{G.}~\bibnamefont{{Mellema}}}, \bibinfo {author}
  {\bibfnamefont{K.}~\bibnamefont{{Ahn}}}, \bibinfo {author}
  {\bibfnamefont{S.}~\bibnamefont{{Baek}}}, \bibinfo {author}
  {\bibfnamefont{N.~Y.}\ \bibnamefont{{Gnedin}}}, \bibinfo {author}
  {\bibfnamefont{A.~V.}\ \bibnamefont{{Kravtsov}}}, \bibinfo {author}
  {\bibfnamefont{M.}~\bibnamefont{{Norman}}}, \bibinfo {author}
  {\bibfnamefont{M.}~\bibnamefont{{Raicevic}}}, \bibinfo {author}
  {\bibfnamefont{D.~R.}\ \bibnamefont{{Reynolds}}}, \bibinfo {author}
  {\bibfnamefont{D.}~\bibnamefont{{Sato}}}, \bibinfo {author}
  {\bibfnamefont{P.~R.}\ \bibnamefont{{Shapiro}}}, \bibinfo {author}
  {\bibfnamefont{B.}~\bibnamefont{{Semelin}}}, \bibinfo {author}
  {\bibfnamefont{J.}~\bibnamefont{{Smidt}}}, \bibinfo {author}
  {\bibfnamefont{H.}~\bibnamefont{{Susa}}}, \bibinfo {author}
  {\bibfnamefont{T.}~\bibnamefont{{Theuns}}},\ and\ \bibinfo {author}
  {\bibfnamefont{M.}~\bibnamefont{{Umemura}}},\ }%
  \bibfield{journal}{%
  \Doi{10.1111/j.1365-2966.2009.15558.x}{\bibinfo {journal} {MNRAS}}\ }%
  \textbf{\bibinfo {volume} {400}},\ \bibinfo {pages} {1283} (\bibinfo {month}
  {Dec.}\ \bibinfo {year} {2009}),\
  \Eprint{http://arxiv.org/abs/0905.2920}{arXiv:0905.2920}%
  \bibAnnoteFile{NoStop}{2009MNRAS.400.1283I}%
\bibitem{2011MNRAS.414.3458W}%
  \BibitemOpen
  \bibfield{author}{%
  \bibinfo {author} {\bibfnamefont{J.~H.}\ \bibnamefont{{Wise}}}\ and\ \bibinfo
  {author} {\bibfnamefont{T.}~\bibnamefont{{Abel}}},\ }%
  \bibfield{journal}{%
  \Doi{10.1111/j.1365-2966.2011.18646.x}{\bibinfo {journal} {MNRAS}}\ }%
  \textbf{\bibinfo {volume} {414}},\ \bibinfo {pages} {3458} (\bibinfo {month}
  {Jul.}\ \bibinfo {year} {2011}),\
  \Eprint{http://arxiv.org/abs/1012.2865}{arXiv:1012.2865 [astro-ph.IM]}%
  \bibAnnoteFile{NoStop}{2011MNRAS.414.3458W}%
\bibitem{2013MNRAS.436.2188R}%
  \BibitemOpen
  \bibfield{author}{%
  \bibinfo {author} {\bibfnamefont{J.}~\bibnamefont{{Rosdahl}}}, \bibinfo
  {author} {\bibfnamefont{J.}~\bibnamefont{{Blaizot}}}, \bibinfo {author}
  {\bibfnamefont{D.}~\bibnamefont{{Aubert}}}, \bibinfo {author}
  {\bibfnamefont{T.}~\bibnamefont{{Stranex}}},\ and\ \bibinfo {author}
  {\bibfnamefont{R.}~\bibnamefont{{Teyssier}}},\ }%
  \bibfield{journal}{%
  \Doi{10.1093/mnras/stt1722}{\bibinfo {journal} {MNRAS}}\ }%
  \textbf{\bibinfo {volume} {436}},\ \bibinfo {pages} {2188} (\bibinfo {month}
  {Dec.}\ \bibinfo {year} {2013}),\
  \Eprint{http://arxiv.org/abs/1304.7126}{arXiv:1304.7126}%
  \bibAnnoteFile{NoStop}{2013MNRAS.436.2188R}%
\bibitem{2013MNRAS.434..748A}%
  \BibitemOpen
  \bibfield{author}{%
  \bibinfo {author} {\bibfnamefont{G.}~\bibnamefont{{Altay}}}\ and\ \bibinfo
  {author} {\bibfnamefont{T.}~\bibnamefont{{Theuns}}},\ }%
  \bibfield{journal}{%
  \Doi{10.1093/mnras/stt1067}{\bibinfo {journal} {MNRAS}}\ }%
  \textbf{\bibinfo {volume} {434}},\ \bibinfo {pages} {748} (\bibinfo {month}
  {Sep.}\ \bibinfo {year} {2013}),\
  \Eprint{http://arxiv.org/abs/1304.4235}{arXiv:1304.4235 [astro-ph.CO]}%
  \bibAnnoteFile{NoStop}{2013MNRAS.434..748A}%
\bibitem{2005ApJ...622..759G}%
  \BibitemOpen
  \bibfield{author}{%
  \bibinfo {author} {\bibfnamefont{K.~M.}\ \bibnamefont{{G{\'o}rski}}},
  \bibinfo {author} {\bibfnamefont{E.}~\bibnamefont{{Hivon}}}, \bibinfo
  {author} {\bibfnamefont{A.~J.}\ \bibnamefont{{Banday}}}, \bibinfo {author}
  {\bibfnamefont{B.~D.}\ \bibnamefont{{Wandelt}}}, \bibinfo {author}
  {\bibfnamefont{F.~K.}\ \bibnamefont{{Hansen}}}, \bibinfo {author}
  {\bibfnamefont{M.}~\bibnamefont{{Reinecke}}},\ and\ \bibinfo {author}
  {\bibfnamefont{M.}~\bibnamefont{{Bartelmann}}},\ }%
  \bibfield{journal}{%
  \Doi{10.1086/427976}{\bibinfo {journal} {APJ}}\ }%
  \textbf{\bibinfo {volume} {622}},\ \bibinfo {pages} {759} (\bibinfo {month}
  {Apr.}\ \bibinfo {year} {2005}),\
  \Eprint{http://arxiv.org/abs/astro-ph/0409513}{astro-ph/0409513}%
  \bibAnnoteFile{NoStop}{2005ApJ...622..759G}%
\bibitem{2005PhRvD..72b3516C}%
  \BibitemOpen
  \bibfield{author}{%
  \bibinfo {author} {\bibfnamefont{P.~G.}\ \bibnamefont{{Castro}}}, \bibinfo
  {author} {\bibfnamefont{A.~F.}\ \bibnamefont{{Heavens}}},\ and\ \bibinfo
  {author} {\bibfnamefont{T.~D.}\ \bibnamefont{{Kitching}}},\ }%
  \bibfield{journal}{%
  \Doi{10.1103/PhysRevD.72.023516}{\bibinfo {journal} {PRD}}\ }%
  \textbf{\bibinfo {volume} {72}},\ \bibinfo {eid} {023516} (\bibinfo {month}
  {Jul.}\ \bibinfo {year} {2005}),\
  \Eprint{http://arxiv.org/abs/astro-ph/0503479}{astro-ph/0503479}%
  \bibAnnoteFile{NoStop}{2005PhRvD..72b3516C}%
\bibitem{2010PhRvD..81h3002B}%
  \BibitemOpen
  \bibfield{author}{%
  \bibinfo {author} {\bibfnamefont{F.}~\bibnamefont{{Bernardeau}}}, \bibinfo
  {author} {\bibfnamefont{C.}~\bibnamefont{{Bonvin}}},\ and\ \bibinfo {author}
  {\bibfnamefont{F.}~\bibnamefont{{Vernizzi}}},\ }%
  \bibfield{journal}{%
  \Doi{10.1103/PhysRevD.81.083002}{\bibinfo {journal} {PRD}}\ }%
  \textbf{\bibinfo {volume} {81}},\ \bibinfo {eid} {083002} (\bibinfo {month}
  {Apr.}\ \bibinfo {year} {2010}),\
  \Eprint{http://arxiv.org/abs/0911.2244}{arXiv:0911.2244 [astro-ph.CO]}%
  \bibAnnoteFile{NoStop}{2010PhRvD..81h3002B}%
\bibitem{1993ApJ...404..441K}%
  \BibitemOpen
  \bibfield{author}{%
  \bibinfo {author} {\bibfnamefont{N.}~\bibnamefont{{Kaiser}}}\ and\ \bibinfo
  {author} {\bibfnamefont{G.}~\bibnamefont{{Squires}}},\ }%
  \bibfield{journal}{%
  \Doi{10.1086/172297}{\bibinfo {journal} {APJ}}\ }%
  \textbf{\bibinfo {volume} {404}},\ \bibinfo {pages} {441} (\bibinfo {month}
  {Feb.}\ \bibinfo {year} {1993})%
  \bibAnnoteFile{NoStop}{1993ApJ...404..441K}%
\bibitem{Gough:2009:GSL:1538674}%
  \BibitemOpen
  \bibfield{author}{%
  \bibinfo {author} {\bibfnamefont{B.}~\bibnamefont{Gough}},\ }%
  \emph{\bibinfo {title} {GNU Scientific Library Reference Manual - Third
  Edition}},\ \bibinfo {edition} {3rd}\ ed.\ (\bibinfo {publisher} {Network
  Theory Ltd.},\ \bibinfo {year} {2009})\ ISBN \bibinfo {isbn} {0954612078,
  9780954612078}%
  \bibAnnoteFile{NoStop}{Gough:2009:GSL:1538674}%
\bibitem{2014A&A...571A..16P}%
  \BibitemOpen
  \bibfield{author}{%
  \bibinfo {author} {\bibnamefont{{Planck Collaboration}}}, \bibinfo {author}
  {\bibfnamefont{P.~A.~R.}\ \bibnamefont{{Ade}}}, \bibinfo {author}
  {\bibfnamefont{N.}~\bibnamefont{{Aghanim}}}, \bibinfo {author}
  {\bibfnamefont{C.}~\bibnamefont{{Armitage-Caplan}}}, \bibinfo {author}
  {\bibfnamefont{M.}~\bibnamefont{{Arnaud}}}, \bibinfo {author}
  {\bibfnamefont{M.}~\bibnamefont{{Ashdown}}}, \bibinfo {author}
  {\bibfnamefont{F.}~\bibnamefont{{Atrio-Barandela}}}, \bibinfo {author}
  {\bibfnamefont{J.}~\bibnamefont{{Aumont}}}, \bibinfo {author}
  {\bibfnamefont{C.}~\bibnamefont{{Baccigalupi}}}, \bibinfo {author}
  {\bibfnamefont{A.~J.}\ \bibnamefont{{Banday}}},\ and\ \bibinfo {author}
  {\bibnamefont{et~al.}},\ }%
  \bibfield{journal}{%
  \Doi{10.1051/0004-6361/201321591}{\bibinfo {journal} {AAP}}\ }%
  \textbf{\bibinfo {volume} {571}},\ \bibinfo {eid} {A16} (\bibinfo {month}
  {Nov.}\ \bibinfo {year} {2014}),\
  \Eprint{http://arxiv.org/abs/1303.5076}{arXiv:1303.5076}%
  \bibAnnoteFile{NoStop}{2014A&A...571A..16P}%
\bibitem{2015arXiv150201589P}%
  \BibitemOpen
  \bibfield{author}{%
  \bibinfo {author} {\bibnamefont{{Planck Collaboration}}}, \bibinfo {author}
  {\bibfnamefont{P.~A.~R.}\ \bibnamefont{{Ade}}}, \bibinfo {author}
  {\bibfnamefont{N.}~\bibnamefont{{Aghanim}}}, \bibinfo {author}
  {\bibfnamefont{M.}~\bibnamefont{{Arnaud}}}, \bibinfo {author}
  {\bibfnamefont{M.}~\bibnamefont{{Ashdown}}}, \bibinfo {author}
  {\bibfnamefont{J.}~\bibnamefont{{Aumont}}}, \bibinfo {author}
  {\bibfnamefont{C.}~\bibnamefont{{Baccigalupi}}}, \bibinfo {author}
  {\bibfnamefont{A.~J.}\ \bibnamefont{{Banday}}}, \bibinfo {author}
  {\bibfnamefont{R.~B.}\ \bibnamefont{{Barreiro}}}, \bibinfo {author}
  {\bibfnamefont{J.~G.}\ \bibnamefont{{Bartlett}}},\ and\ \bibinfo {author}
  {\bibnamefont{et~al.}},\ }%
  \bibfield{journal}{%
  \bibinfo {journal} {ArXiv e-prints}}%
   (\bibinfo {month} {Feb.}\ \bibinfo {year} {2015}),\
  \Eprint{http://arxiv.org/abs/1502.01589}{arXiv:1502.01589}%
  \bibAnnoteFile{NoStop}{2015arXiv150201589P}%
\bibitem{2008ApJS..178..179P}%
  \BibitemOpen
  \bibfield{author}{%
  \bibinfo {author} {\bibfnamefont{S.}~\bibnamefont{{Prunet}}}, \bibinfo
  {author} {\bibfnamefont{C.}~\bibnamefont{{Pichon}}}, \bibinfo {author}
  {\bibfnamefont{D.}~\bibnamefont{{Aubert}}}, \bibinfo {author}
  {\bibfnamefont{D.}~\bibnamefont{{Pogosyan}}}, \bibinfo {author}
  {\bibfnamefont{R.}~\bibnamefont{{Teyssier}}},\ and\ \bibinfo {author}
  {\bibfnamefont{S.}~\bibnamefont{{Gottloeber}}},\ }%
  \bibfield{journal}{%
  \Doi{10.1086/590370}{\bibinfo {journal} {APJS}}\ }%
  \textbf{\bibinfo {volume} {178}},\ \bibinfo {pages} {179} (\bibinfo {month}
  {Oct.}\ \bibinfo {year} {2008}),\
  \Eprint{http://arxiv.org/abs/0804.3536}{arXiv:0804.3536}%
  \bibAnnoteFile{NoStop}{2008ApJS..178..179P}%
\bibitem{2013MNRAS.433.3373V}%
  \BibitemOpen
  \bibfield{author}{%
  \bibinfo {author} {\bibfnamefont{L.}~\bibnamefont{{Van Waerbeke}}}, \bibinfo
  {author} {\bibfnamefont{J.}~\bibnamefont{{Benjamin}}}, \bibinfo {author}
  {\bibfnamefont{T.}~\bibnamefont{{Erben}}}, \bibinfo {author}
  {\bibfnamefont{C.}~\bibnamefont{{Heymans}}}, \bibinfo {author}
  {\bibfnamefont{H.}~\bibnamefont{{Hildebrandt}}}, \bibinfo {author}
  {\bibfnamefont{H.}~\bibnamefont{{Hoekstra}}}, \bibinfo {author}
  {\bibfnamefont{T.~D.}\ \bibnamefont{{Kitching}}}, \bibinfo {author}
  {\bibfnamefont{Y.}~\bibnamefont{{Mellier}}}, \bibinfo {author}
  {\bibfnamefont{L.}~\bibnamefont{{Miller}}}, \bibinfo {author}
  {\bibfnamefont{J.}~\bibnamefont{{Coupon}}}, \bibinfo {author}
  {\bibfnamefont{J.}~\bibnamefont{{Harnois-D{\'e}raps}}}, \bibinfo {author}
  {\bibfnamefont{L.}~\bibnamefont{{Fu}}}, \bibinfo {author}
  {\bibfnamefont{M.}~\bibnamefont{{Hudson}}}, \bibinfo {author}
  {\bibfnamefont{M.}~\bibnamefont{{Kilbinger}}}, \bibinfo {author}
  {\bibfnamefont{K.}~\bibnamefont{{Kuijken}}}, \bibinfo {author}
  {\bibfnamefont{B.}~\bibnamefont{{Rowe}}}, \bibinfo {author}
  {\bibfnamefont{T.}~\bibnamefont{{Schrabback}}}, \bibinfo {author}
  {\bibfnamefont{E.}~\bibnamefont{{Semboloni}}}, \bibinfo {author}
  {\bibfnamefont{S.}~\bibnamefont{{Vafaei}}}, \bibinfo {author}
  {\bibfnamefont{E.}~\bibnamefont{{van Uitert}}},\ and\ \bibinfo {author}
  {\bibfnamefont{M.}~\bibnamefont{{Velander}}},\ }%
  \bibfield{journal}{%
  \Doi{10.1093/mnras/stt971}{\bibinfo {journal} {MNRAS}}\ }%
  \textbf{\bibinfo {volume} {433}},\ \bibinfo {pages} {3373} (\bibinfo {month}
  {Aug.}\ \bibinfo {year} {2013}),\
  \Eprint{http://arxiv.org/abs/1303.1806}{arXiv:1303.1806}%
  \bibAnnoteFile{NoStop}{2013MNRAS.433.3373V}%
\bibitem{2015PhRvD..92b2006V}%
  \BibitemOpen
  \bibfield{author}{%
  \bibinfo {author} {\bibfnamefont{V.}~\bibnamefont{{Vikram et al}}},\ }%
  \bibfield{journal}{%
  \Doi{10.1103/PhysRevD.92.022006}{\bibinfo {journal} {PRD}}\ }%
  \textbf{\bibinfo {volume} {92}},\ \bibinfo {eid} {022006} (\bibinfo {month}
  {Jul.}\ \bibinfo {year} {2015}),\
  \Eprint{http://arxiv.org/abs/1504.03002}{arXiv:1504.03002}%
  \bibAnnoteFile{NoStop}{2015PhRvD..92b2006V}%
\bibitem{2003MNRAS.341.1311S}%
  \BibitemOpen
  \bibfield{author}{%
  \bibinfo {author} {\bibfnamefont{R.~E.}\ \bibnamefont{{Smith}}}, \bibinfo
  {author} {\bibfnamefont{J.~A.}\ \bibnamefont{{Peacock}}}, \bibinfo {author}
  {\bibfnamefont{A.}~\bibnamefont{{Jenkins}}}, \bibinfo {author}
  {\bibfnamefont{S.~D.~M.}\ \bibnamefont{{White}}}, \bibinfo {author}
  {\bibfnamefont{C.~S.}\ \bibnamefont{{Frenk}}}, \bibinfo {author}
  {\bibfnamefont{F.~R.}\ \bibnamefont{{Pearce}}}, \bibinfo {author}
  {\bibfnamefont{P.~A.}\ \bibnamefont{{Thomas}}}, \bibinfo {author}
  {\bibfnamefont{G.}~\bibnamefont{{Efstathiou}}},\ and\ \bibinfo {author}
  {\bibfnamefont{H.~M.~P.}\ \bibnamefont{{Couchman}}},\ }%
  \bibfield{journal}{%
  \Doi{10.1046/j.1365-8711.2003.06503.x}{\bibinfo {journal} {MNRAS}}\ }%
  \textbf{\bibinfo {volume} {341}},\ \bibinfo {pages} {1311} (\bibinfo {month}
  {Jun.}\ \bibinfo {year} {2003}),\
  \Eprint{http://arxiv.org/abs/astro-ph/0207664}{astro-ph/0207664}%
  \bibAnnoteFile{NoStop}{2003MNRAS.341.1311S}%
\bibitem{2012ApJ...761..152T}%
  \BibitemOpen
  \bibfield{author}{%
  \bibinfo {author} {\bibfnamefont{R.}~\bibnamefont{{Takahashi}}}, \bibinfo
  {author} {\bibfnamefont{M.}~\bibnamefont{{Sato}}}, \bibinfo {author}
  {\bibfnamefont{T.}~\bibnamefont{{Nishimichi}}}, \bibinfo {author}
  {\bibfnamefont{A.}~\bibnamefont{{Taruya}}},\ and\ \bibinfo {author}
  {\bibfnamefont{M.}~\bibnamefont{{Oguri}}},\ }%
  \bibfield{journal}{%
  \Doi{10.1088/0004-637X/761/2/152}{\bibinfo {journal} {APJ}}\ }%
  \textbf{\bibinfo {volume} {761}},\ \bibinfo {eid} {152} (\bibinfo {month}
  {Dec.}\ \bibinfo {year} {2012}),\
  \Eprint{http://arxiv.org/abs/1208.2701}{arXiv:1208.2701}%
  \bibAnnoteFile{NoStop}{2012ApJ...761..152T}%
\bibitem{2009A&A...499...31H}%
  \BibitemOpen
  \bibfield{author}{%
  \bibinfo {author} {\bibfnamefont{S.}~\bibnamefont{{Hilbert}}}, \bibinfo
  {author} {\bibfnamefont{J.}~\bibnamefont{{Hartlap}}}, \bibinfo {author}
  {\bibfnamefont{S.~D.~M.}\ \bibnamefont{{White}}},\ and\ \bibinfo {author}
  {\bibfnamefont{P.}~\bibnamefont{{Schneider}}},\ }%
  \bibfield{journal}{%
  \Doi{10.1051/0004-6361/200811054}{\bibinfo {journal} {AAP}}\ }%
  \textbf{\bibinfo {volume} {499}},\ \bibinfo {pages} {31} (\bibinfo {month}
  {May}\ \bibinfo {year} {2009}),\
  \Eprint{http://arxiv.org/abs/0809.5035}{arXiv:0809.5035}%
  \bibAnnoteFile{NoStop}{2009A&A...499...31H}%
\bibitem{2013ApJ...762..109B}%
  \BibitemOpen
  \bibfield{author}{%
  \bibinfo {author} {\bibfnamefont{P.~S.}\ \bibnamefont{{Behroozi}}}, \bibinfo
  {author} {\bibfnamefont{R.~H.}\ \bibnamefont{{Wechsler}}},\ and\ \bibinfo
  {author} {\bibfnamefont{H.-Y.}\ \bibnamefont{{Wu}}},\ }%
  \bibfield{journal}{%
  \Doi{10.1088/0004-637X/762/2/109}{\bibinfo {journal} {APJ}}\ }%
  \textbf{\bibinfo {volume} {762}},\ \bibinfo {eid} {109} (\bibinfo {month}
  {Jan.}\ \bibinfo {year} {2013}),\
  \Eprint{http://arxiv.org/abs/1110.4372}{arXiv:1110.4372 [astro-ph.CO]}%
  \bibAnnoteFile{NoStop}{2013ApJ...762..109B}%
\bibitem{Navarro:1996gj}%
  \BibitemOpen
  \bibfield{author}{%
  \bibinfo {author} {\bibfnamefont{J.~F.}\ \bibnamefont{Navarro}}, \bibinfo
  {author} {\bibfnamefont{C.~S.}\ \bibnamefont{Frenk}},\ and\ \bibinfo {author}
  {\bibfnamefont{S.~D.}\ \bibnamefont{White}},\ }%
  \bibfield{journal}{%
  \Doi{10.1086/304888}{\bibinfo {journal} {Astrophys.J.}}\ }%
  \textbf{\bibinfo {volume} {490}},\ \bibinfo {pages} {493} (\bibinfo {year}
  {1997}),\
  \Eprint{http://arxiv.org/abs/astro-ph/9611107}{arXiv:astro-ph/9611107
  [astro-ph]}%
  \bibAnnoteFile{NoStop}{Navarro:1996gj}%
\bibitem{1999astro.ph..8213O}%
  \BibitemOpen
  \bibfield{author}{%
  \bibinfo {author} {\bibfnamefont{C.}~\bibnamefont{{Oaxaca Wright}}}\ and\
  \bibinfo {author} {\bibfnamefont{T.~G.}\ \bibnamefont{{Brainerd}}},\ }%
  \bibfield{journal}{%
  \bibinfo {journal} {ArXiv Astrophysics e-prints}}%
   (\bibinfo {month} {Aug.}\ \bibinfo {year} {1999}),\
  \Eprint{http://arxiv.org/abs/astro-ph/9908213}{astro-ph/9908213}%
  \bibAnnoteFile{NoStop}{1999astro.ph..8213O}%
\bibitem{1996A&A...313..697B}%
  \BibitemOpen
  \bibfield{author}{%
  \bibinfo {author} {\bibfnamefont{M.}~\bibnamefont{{Bartelmann}}},\ }%
  \bibfield{journal}{%
  \bibinfo {journal} {AAP}\ }%
  \textbf{\bibinfo {volume} {313}},\ \bibinfo {pages} {697} (\bibinfo {month}
  {Sep.}\ \bibinfo {year} {1996}),\
  \Eprint{http://arxiv.org/abs/astro-ph/9602053}{astro-ph/9602053}%
  \bibAnnoteFile{NoStop}{1996A&A...313..697B}%
\bibitem{2010arXiv1002.3952U}%
  \BibitemOpen
  \bibfield{author}{%
  \bibinfo {author} {\bibfnamefont{K.}~\bibnamefont{{Umetsu}}},\ }%
  \bibfield{journal}{%
  \bibinfo {journal} {ArXiv e-prints}}%
   (\bibinfo {month} {Feb.}\ \bibinfo {year} {2010}),\
  \Eprint{http://arxiv.org/abs/1002.3952}{arXiv:1002.3952 [astro-ph.CO]}%
  \bibAnnoteFile{NoStop}{2010arXiv1002.3952U}%
\bibitem{2014ApJ...797...34M}%
  \BibitemOpen
  \bibfield{author}{%
  \bibinfo {author} {\bibnamefont{{Meneghetti et al}}},\ }%
  \bibfield{journal}{%
  \Doi{10.1088/0004-637X/797/1/34}{\bibinfo {journal} {APJ}}\ }%
  \textbf{\bibinfo {volume} {797}},\ \bibinfo {eid} {34} (\bibinfo {month}
  {Dec.}\ \bibinfo {year} {2014}),\
  \Eprint{http://arxiv.org/abs/1404.1384}{arXiv:1404.1384}%
  \bibAnnoteFile{NoStop}{2014ApJ...797...34M}%
\bibitem{2015ApJ...806....4M}%
  \BibitemOpen
  \bibfield{author}{%
  \bibinfo {author} {\bibfnamefont{J.}~\bibnamefont{{Merten et al}}},\ }%
  \bibfield{journal}{%
  \Doi{10.1088/0004-637X/806/1/4}{\bibinfo {journal} {APJ}}\ }%
  \textbf{\bibinfo {volume} {806}},\ \bibinfo {eid} {4} (\bibinfo {month}
  {Jun.}\ \bibinfo {year} {2015}),\
  \Eprint{http://arxiv.org/abs/1404.1376}{arXiv:1404.1376}%
  \bibAnnoteFile{NoStop}{2015ApJ...806....4M}%
\bibitem{2015ApJ...814..120D}%
  \BibitemOpen
  \bibfield{author}{%
  \bibinfo {author} {\bibfnamefont{W.}~\bibnamefont{{Du}}}, \bibinfo {author}
  {\bibfnamefont{Z.}~\bibnamefont{{Fan}}}, \bibinfo {author}
  {\bibfnamefont{H.}~\bibnamefont{{Shan}}}, \bibinfo {author}
  {\bibfnamefont{G.-B.}\ \bibnamefont{{Zhao}}}, \bibinfo {author}
  {\bibfnamefont{G.}~\bibnamefont{{Covone}}}, \bibinfo {author}
  {\bibfnamefont{L.}~\bibnamefont{{Fu}}},\ and\ \bibinfo {author}
  {\bibfnamefont{J.-P.}\ \bibnamefont{{Kneib}}},\ }%
  \bibfield{journal}{%
  \Doi{10.1088/0004-637X/814/2/120}{\bibinfo {journal} {APJ}}\ }%
  \textbf{\bibinfo {volume} {814}},\ \bibinfo {eid} {120} (\bibinfo {month}
  {Dec.}\ \bibinfo {year} {2015}),\
  \Eprint{http://arxiv.org/abs/1510.08193}{arXiv:1510.08193}%
  \bibAnnoteFile{NoStop}{2015ApJ...814..120D}%
\bibitem{2011PhRvD..84d3529Y}%
  \BibitemOpen
  \bibfield{author}{%
  \bibinfo {author} {\bibfnamefont{X.}~\bibnamefont{{Yang}}}, \bibinfo {author}
  {\bibfnamefont{J.~M.}\ \bibnamefont{{Kratochvil}}}, \bibinfo {author}
  {\bibfnamefont{S.}~\bibnamefont{{Wang}}}, \bibinfo {author}
  {\bibfnamefont{E.~A.}\ \bibnamefont{{Lim}}}, \bibinfo {author}
  {\bibfnamefont{Z.}~\bibnamefont{{Haiman}}},\ and\ \bibinfo {author}
  {\bibfnamefont{M.}~\bibnamefont{{May}}},\ }%
  \bibfield{journal}{%
  \Doi{10.1103/PhysRevD.84.043529}{\bibinfo {journal} {PRD}}\ }%
  \textbf{\bibinfo {volume} {84}},\ \bibinfo {eid} {043529} (\bibinfo {month}
  {Aug.}\ \bibinfo {year} {2011}),\
  \Eprint{http://arxiv.org/abs/1109.6333}{arXiv:1109.6333}%
  \bibAnnoteFile{NoStop}{2011PhRvD..84d3529Y}%
\bibitem{2004MNRAS.350..893H}%
  \BibitemOpen
  \bibfield{author}{%
  \bibinfo {author} {\bibfnamefont{T.}~\bibnamefont{{Hamana}}}, \bibinfo
  {author} {\bibfnamefont{M.}~\bibnamefont{{Takada}}},\ and\ \bibinfo {author}
  {\bibfnamefont{N.}~\bibnamefont{{Yoshida}}},\ }%
  \bibfield{journal}{%
  \Doi{10.1111/j.1365-2966.2004.07691.x}{\bibinfo {journal} {MNRAS}}\ }%
  \textbf{\bibinfo {volume} {350}},\ \bibinfo {pages} {893} (\bibinfo {month}
  {May}\ \bibinfo {year} {2004}),\
  \Eprint{http://arxiv.org/abs/astro-ph/0310607}{astro-ph/0310607}%
  \bibAnnoteFile{NoStop}{2004MNRAS.350..893H}%
\bibitem{2010ApJ...719.1408F}%
  \BibitemOpen
  \bibfield{author}{%
  \bibinfo {author} {\bibfnamefont{Z.}~\bibnamefont{{Fan}}}, \bibinfo {author}
  {\bibfnamefont{H.}~\bibnamefont{{Shan}}},\ and\ \bibinfo {author}
  {\bibfnamefont{J.}~\bibnamefont{{Liu}}},\ }%
  \bibfield{journal}{%
  \Doi{10.1088/0004-637X/719/2/1408}{\bibinfo {journal} {APJ}}\ }%
  \textbf{\bibinfo {volume} {719}},\ \bibinfo {pages} {1408} (\bibinfo {month}
  {Aug.}\ \bibinfo {year} {2010}),\
  \Eprint{http://arxiv.org/abs/1006.5121}{arXiv:1006.5121}%
  \bibAnnoteFile{NoStop}{2010ApJ...719.1408F}%
\bibitem{2012MNRAS.420.3213O}%
  \BibitemOpen
  \bibfield{author}{%
  \bibinfo {author} {\bibfnamefont{M.}~\bibnamefont{{Oguri}}}, \bibinfo
  {author} {\bibfnamefont{M.~B.}\ \bibnamefont{{Bayliss}}}, \bibinfo {author}
  {\bibfnamefont{H.}~\bibnamefont{{Dahle}}}, \bibinfo {author}
  {\bibfnamefont{K.}~\bibnamefont{{Sharon}}}, \bibinfo {author}
  {\bibfnamefont{M.~D.}\ \bibnamefont{{Gladders}}}, \bibinfo {author}
  {\bibfnamefont{P.}~\bibnamefont{{Natarajan}}}, \bibinfo {author}
  {\bibfnamefont{J.~F.}\ \bibnamefont{{Hennawi}}},\ and\ \bibinfo {author}
  {\bibfnamefont{B.~P.}\ \bibnamefont{{Koester}}},\ }%
  \bibfield{journal}{%
  \Doi{10.1111/j.1365-2966.2011.20248.x}{\bibinfo {journal} {MNRAS}}\ }%
  \textbf{\bibinfo {volume} {420}},\ \bibinfo {pages} {3213} (\bibinfo {month}
  {Mar.}\ \bibinfo {year} {2012}),\
  \Eprint{http://arxiv.org/abs/1109.2594}{arXiv:1109.2594}%
  \bibAnnoteFile{NoStop}{2012MNRAS.420.3213O}%
\bibitem{2011MNRAS.414.1851O}%
  \BibitemOpen
  \bibfield{author}{%
  \bibinfo {author} {\bibfnamefont{M.}~\bibnamefont{{Oguri}}}\ and\ \bibinfo
  {author} {\bibfnamefont{T.}~\bibnamefont{{Hamana}}},\ }%
  \bibfield{journal}{%
  \Doi{10.1111/j.1365-2966.2011.18481.x}{\bibinfo {journal} {MNRAS}}\ }%
  \textbf{\bibinfo {volume} {414}},\ \bibinfo {pages} {1851} (\bibinfo {month}
  {Jul.}\ \bibinfo {year} {2011}),\
  \Eprint{http://arxiv.org/abs/1101.0650}{arXiv:1101.0650}%
  \bibAnnoteFile{NoStop}{2011MNRAS.414.1851O}%
\bibitem{2014MNRAS.440.2922M}%
  \BibitemOpen
  \bibfield{author}{%
  \bibinfo {author} {\bibfnamefont{P.}~\bibnamefont{{Melchior}}}, \bibinfo
  {author} {\bibfnamefont{P.~M.}\ \bibnamefont{{Sutter}}}, \bibinfo {author}
  {\bibfnamefont{E.~S.}\ \bibnamefont{{Sheldon}}}, \bibinfo {author}
  {\bibfnamefont{E.}~\bibnamefont{{Krause}}},\ and\ \bibinfo {author}
  {\bibfnamefont{B.~D.}\ \bibnamefont{{Wandelt}}},\ }%
  \bibfield{journal}{%
  \Doi{10.1093/mnras/stu456}{\bibinfo {journal} {MNRAS}}\ }%
  \textbf{\bibinfo {volume} {440}},\ \bibinfo {pages} {2922} (\bibinfo {month}
  {Jun.}\ \bibinfo {year} {2014}),\
  \Eprint{http://arxiv.org/abs/1309.2045}{arXiv:1309.2045}%
  \bibAnnoteFile{NoStop}{2014MNRAS.440.2922M}%
\bibitem{2015MNRAS.454.3357C}%
  \BibitemOpen
  \bibfield{author}{%
  \bibinfo {author} {\bibfnamefont{J.}~\bibnamefont{{Clampitt}}}\ and\ \bibinfo
  {author} {\bibfnamefont{B.}~\bibnamefont{{Jain}}},\ }%
  \bibfield{journal}{%
  \Doi{10.1093/mnras/stv2215}{\bibinfo {journal} {MNRAS}}\ }%
  \textbf{\bibinfo {volume} {454}},\ \bibinfo {pages} {3357} (\bibinfo {month}
  {Dec.}\ \bibinfo {year} {2015}),\
  \Eprint{http://arxiv.org/abs/1404.1834}{arXiv:1404.1834}%
  \bibAnnoteFile{NoStop}{2015MNRAS.454.3357C}%
\bibitem{2013ApJ...762L..20K}%
  \BibitemOpen
  \bibfield{author}{%
  \bibinfo {author} {\bibfnamefont{E.}~\bibnamefont{{Krause}}}, \bibinfo
  {author} {\bibfnamefont{T.-C.}\ \bibnamefont{{Chang}}}, \bibinfo {author}
  {\bibfnamefont{O.}~\bibnamefont{{Dor{\'e}}}},\ and\ \bibinfo {author}
  {\bibfnamefont{K.}~\bibnamefont{{Umetsu}}},\ }%
  \bibfield{journal}{%
  \Doi{10.1088/2041-8205/762/2/L20}{\bibinfo {journal} {APJL}}\ }%
  \textbf{\bibinfo {volume} {762}},\ \bibinfo {eid} {L20} (\bibinfo {month}
  {Jan.}\ \bibinfo {year} {2013}),\
  \Eprint{http://arxiv.org/abs/1210.2446}{arXiv:1210.2446}%
  \bibAnnoteFile{NoStop}{2013ApJ...762L..20K}%
\bibitem{2013MNRAS.432.1021H}%
  \BibitemOpen
  \bibfield{author}{%
  \bibinfo {author} {\bibfnamefont{Y.}~\bibnamefont{{Higuchi}}}, \bibinfo
  {author} {\bibfnamefont{M.}~\bibnamefont{{Oguri}}},\ and\ \bibinfo {author}
  {\bibfnamefont{T.}~\bibnamefont{{Hamana}}},\ }%
  \bibfield{journal}{%
  \Doi{10.1093/mnras/stt521}{\bibinfo {journal} {MNRAS}}\ }%
  \textbf{\bibinfo {volume} {432}},\ \bibinfo {pages} {1021} (\bibinfo {month}
  {Jun.}\ \bibinfo {year} {2013}),\
  \Eprint{http://arxiv.org/abs/1211.5966}{arXiv:1211.5966}%
  \bibAnnoteFile{NoStop}{2013MNRAS.432.1021H}%
\bibitem{2014arXiv1410.8452H}%
  \BibitemOpen
  \bibfield{author}{%
  \bibinfo {author} {\bibfnamefont{S.}~\bibnamefont{{Hagstotz}}}, \bibinfo
  {author} {\bibfnamefont{B.~M.}\ \bibnamefont{{Sch{\"a}fer}}},\ and\ \bibinfo
  {author} {\bibfnamefont{P.~M.}\ \bibnamefont{{Merkel}}},\ }%
  \bibfield{journal}{%
  \bibinfo {journal} {ArXiv e-prints}}%
   (\bibinfo {month} {Oct.}\ \bibinfo {year} {2014}),\
  \Eprint{http://arxiv.org/abs/1410.8452}{arXiv:1410.8452}%
  \bibAnnoteFile{NoStop}{2014arXiv1410.8452H}%
\bibitem{2015JCAP...03..049C}%
  \BibitemOpen
  \bibfield{author}{%
  \bibinfo {author} {\bibfnamefont{M.}~\bibnamefont{{Calabrese}}}, \bibinfo
  {author} {\bibfnamefont{C.}~\bibnamefont{{Carbone}}}, \bibinfo {author}
  {\bibfnamefont{G.}~\bibnamefont{{Fabbian}}}, \bibinfo {author}
  {\bibfnamefont{M.}~\bibnamefont{{Baldi}}},\ and\ \bibinfo {author}
  {\bibfnamefont{C.}~\bibnamefont{{Baccigalupi}}},\ }%
  \bibfield{journal}{%
  \Doi{10.1088/1475-7516/2015/03/049}{\bibinfo {journal} {JCAP}}\ }%
  \textbf{\bibinfo {volume} {3}},\ \bibinfo {eid} {049} (\bibinfo {month}
  {Mar.}\ \bibinfo {year} {2015}),\
  \Eprint{http://arxiv.org/abs/1409.7680}{arXiv:1409.7680}%
  \bibAnnoteFile{NoStop}{2015JCAP...03..049C}%
\bibitem{2012arXiv1211.0310L}%
  \BibitemOpen
  \bibfield{author}{%
  \bibinfo {author} {\bibnamefont{{LSST Dark Energy Science Collaboration}}},\
  }%
  \bibfield{journal}{%
  \bibinfo {journal} {ArXiv e-prints}}%
   (\bibinfo {month} {Nov.}\ \bibinfo {year} {2012}),\
  \Eprint{http://arxiv.org/abs/1211.0310}{arXiv:1211.0310 [astro-ph.CO]}%
  \bibAnnoteFile{NoStop}{2012arXiv1211.0310L}%
\bibitem{2011arXiv1110.3193L}%
  \BibitemOpen
  \bibfield{author}{%
  \bibinfo {author} {\bibfnamefont{R.}~\bibnamefont{{Laureijs}}}, \bibinfo
  {author} {\bibfnamefont{J.}~\bibnamefont{{Amiaux}}}, \bibinfo {author}
  {\bibfnamefont{S.}~\bibnamefont{{Arduini}}}, \bibinfo {author}
  {\bibfnamefont{J.~.}\ \bibnamefont{{Augu{\`e}res}}}, \bibinfo {author}
  {\bibfnamefont{J.}~\bibnamefont{{Brinchmann}}}, \bibinfo {author}
  {\bibfnamefont{R.}~\bibnamefont{{Cole}}}, \bibinfo {author}
  {\bibfnamefont{M.}~\bibnamefont{{Cropper}}}, \bibinfo {author}
  {\bibfnamefont{C.}~\bibnamefont{{Dabin}}}, \bibinfo {author}
  {\bibfnamefont{L.}~\bibnamefont{{Duvet}}}, \bibinfo {author}
  {\bibfnamefont{A.}~\bibnamefont{{Ealet}}},\ and\ \bibinfo {author}
  {\bibnamefont{et~al.}},\ }%
  \bibfield{journal}{%
  \bibinfo {journal} {ArXiv e-prints}}%
   (\bibinfo {month} {Oct.}\ \bibinfo {year} {2011}),\
  \Eprint{http://arxiv.org/abs/1110.3193}{arXiv:1110.3193 [astro-ph.CO]}%
  \bibAnnoteFile{NoStop}{2011arXiv1110.3193L}%
\bibitem{2011MNRAS.417.2020S}%
  \BibitemOpen
  \bibfield{author}{%
  \bibinfo {author} {\bibfnamefont{E.}~\bibnamefont{{Semboloni}}}, \bibinfo
  {author} {\bibfnamefont{H.}~\bibnamefont{{Hoekstra}}}, \bibinfo {author}
  {\bibfnamefont{J.}~\bibnamefont{{Schaye}}}, \bibinfo {author}
  {\bibfnamefont{M.~P.}\ \bibnamefont{{van Daalen}}},\ and\ \bibinfo {author}
  {\bibfnamefont{I.~G.}\ \bibnamefont{{McCarthy}}},\ }%
  \bibfield{journal}{%
  \Doi{10.1111/j.1365-2966.2011.19385.x}{\bibinfo {journal} {MNRAS}}\ }%
  \textbf{\bibinfo {volume} {417}},\ \bibinfo {pages} {2020} (\bibinfo {month}
  {Nov.}\ \bibinfo {year} {2011}),\
  \Eprint{http://arxiv.org/abs/1105.1075}{arXiv:1105.1075}%
  \bibAnnoteFile{NoStop}{2011MNRAS.417.2020S}%
\bibitem{2014arXiv1410.6826M}%
  \BibitemOpen
  \bibfield{author}{%
  \bibinfo {author} {\bibfnamefont{I.}~\bibnamefont{{Mohammed}}}, \bibinfo
  {author} {\bibfnamefont{D.}~\bibnamefont{{Martizzi}}}, \bibinfo {author}
  {\bibfnamefont{R.}~\bibnamefont{{Teyssier}}},\ and\ \bibinfo {author}
  {\bibfnamefont{A.}~\bibnamefont{{Amara}}},\ }%
  \bibfield{journal}{%
  \bibinfo {journal} {ArXiv e-prints}}%
   (\bibinfo {month} {Oct.}\ \bibinfo {year} {2014}),\
  \Eprint{http://arxiv.org/abs/1410.6826}{arXiv:1410.6826}%
  \bibAnnoteFile{NoStop}{2014arXiv1410.6826M}%
\bibitem{2015ApJ...806..186O}%
  \BibitemOpen
  \bibfield{author}{%
  \bibinfo {author} {\bibfnamefont{K.}~\bibnamefont{{Osato}}}, \bibinfo
  {author} {\bibfnamefont{M.}~\bibnamefont{{Shirasaki}}},\ and\ \bibinfo
  {author} {\bibfnamefont{N.}~\bibnamefont{{Yoshida}}},\ }%
  \bibfield{journal}{%
  \Doi{10.1088/0004-637X/806/2/186}{\bibinfo {journal} {APJ}}\ }%
  \textbf{\bibinfo {volume} {806}},\ \bibinfo {eid} {186} (\bibinfo {month}
  {Jun.}\ \bibinfo {year} {2015}),\
  \Eprint{http://arxiv.org/abs/1501.02055}{arXiv:1501.02055}%
  \bibAnnoteFile{NoStop}{2015ApJ...806..186O}%
\bibitem{1968Natur.217..511R}%
  \BibitemOpen
  \bibfield{author}{%
  \bibinfo {author} {\bibfnamefont{M.~J.}\ \bibnamefont{{Rees}}}\ and\ \bibinfo
  {author} {\bibfnamefont{D.~W.}\ \bibnamefont{{Sciama}}},\ }%
  \bibfield{journal}{%
  \Doi{10.1038/217511a0}{\bibinfo {journal} {NAT}}\ }%
  \textbf{\bibinfo {volume} {217}},\ \bibinfo {pages} {511} (\bibinfo {month}
  {Feb.}\ \bibinfo {year} {1968})%
  \bibAnnoteFile{NoStop}{1968Natur.217..511R}%
\bibitem{2012PhR...513....1C}%
  \BibitemOpen
  \bibfield{author}{%
  \bibinfo {author} {\bibfnamefont{T.}~\bibnamefont{{Clifton}}}, \bibinfo
  {author} {\bibfnamefont{P.~G.}\ \bibnamefont{{Ferreira}}}, \bibinfo {author}
  {\bibfnamefont{A.}~\bibnamefont{{Padilla}}},\ and\ \bibinfo {author}
  {\bibfnamefont{C.}~\bibnamefont{{Skordis}}},\ }%
  \bibfield{journal}{%
  \Doi{10.1016/j.physrep.2012.01.001}{\bibinfo {journal} {PHYSREP}}\ }%
  \textbf{\bibinfo {volume} {513}},\ \bibinfo {pages} {1} (\bibinfo {month}
  {Mar.}\ \bibinfo {year} {2012}),\
  \Eprint{http://arxiv.org/abs/1106.2476}{arXiv:1106.2476 [astro-ph.CO]}%
  \bibAnnoteFile{NoStop}{2012PhR...513....1C}%
\bibitem{Joyce:2014kja}%
  \BibitemOpen
  \bibfield{author}{%
  \bibinfo {author} {\bibfnamefont{A.}~\bibnamefont{Joyce}}, \bibinfo {author}
  {\bibfnamefont{B.}~\bibnamefont{Jain}}, \bibinfo {author}
  {\bibfnamefont{J.}~\bibnamefont{Khoury}},\ and\ \bibinfo {author}
  {\bibfnamefont{M.}~\bibnamefont{Trodden}}}%
   (\bibinfo {year} {2014}),\
  \Eprint{http://arxiv.org/abs/1407.0059}{arXiv:1407.0059 [astro-ph.CO]}%
  \bibAnnoteFile{NoStop}{Joyce:2014kja}%
\bibitem{2015arXiv150404623K}%
  \BibitemOpen
  \bibfield{author}{%
  \bibinfo {author} {\bibfnamefont{K.}~\bibnamefont{{Koyama}}},\ }%
  \bibfield{journal}{%
  \bibinfo {journal} {ArXiv e-prints}}%
   (\bibinfo {month} {Apr.}\ \bibinfo {year} {2015}),\
  \Eprint{http://arxiv.org/abs/1504.04623}{arXiv:1504.04623}%
  \bibAnnoteFile{NoStop}{2015arXiv150404623K}%
\bibitem{2012JCAP...01..051L}%
  \BibitemOpen
  \bibfield{author}{%
  \bibinfo {author} {\bibfnamefont{B.}~\bibnamefont{{Li}}}, \bibinfo {author}
  {\bibfnamefont{G.-B.}\ \bibnamefont{{Zhao}}}, \bibinfo {author}
  {\bibfnamefont{R.}~\bibnamefont{{Teyssier}}},\ and\ \bibinfo {author}
  {\bibfnamefont{K.}~\bibnamefont{{Koyama}}},\ }%
  \bibfield{journal}{%
  \Doi{10.1088/1475-7516/2012/01/051}{\bibinfo {journal} {JCAP}}\ }%
  \textbf{\bibinfo {volume} {1}},\ \bibinfo {eid} {051} (\bibinfo {month}
  {Jan.}\ \bibinfo {year} {2012}),\
  \Eprint{http://arxiv.org/abs/1110.1379}{arXiv:1110.1379 [astro-ph.CO]}%
  \bibAnnoteFile{NoStop}{2012JCAP...01..051L}%
\bibitem{baojiudgp}%
  \BibitemOpen
  \bibfield{author}{%
  \bibinfo {author} {\bibfnamefont{B.}~\bibnamefont{{Li}}}, \bibinfo {author}
  {\bibfnamefont{G.-B.}\ \bibnamefont{{Zhao}}},\ and\ \bibinfo {author}
  {\bibfnamefont{K.}~\bibnamefont{{Koyama}}},\ }%
  \bibfield{journal}{%
  \Doi{10.1088/1475-7516/2013/05/023}{\bibinfo {journal} {JCAP}}\ }%
  \textbf{\bibinfo {volume} {5}},\ \bibinfo {eid} {023} (\bibinfo {month}
  {May}\ \bibinfo {year} {2013}),\
  \Eprint{http://arxiv.org/abs/1303.0008}{arXiv:1303.0008 [astro-ph.CO]}%
  \bibAnnoteFile{NoStop}{baojiudgp}%
\bibitem{2013JCAP...11..012L}%
  \BibitemOpen
  \bibfield{author}{%
  \bibinfo {author} {\bibfnamefont{B.}~\bibnamefont{{Li}}}, \bibinfo {author}
  {\bibfnamefont{A.}~\bibnamefont{{Barreira}}}, \bibinfo {author}
  {\bibfnamefont{C.~M.}\ \bibnamefont{{Baugh}}}, \bibinfo {author}
  {\bibfnamefont{W.~A.}\ \bibnamefont{{Hellwing}}}, \bibinfo {author}
  {\bibfnamefont{K.}~\bibnamefont{{Koyama}}}, \bibinfo {author}
  {\bibfnamefont{S.}~\bibnamefont{{Pascoli}}},\ and\ \bibinfo {author}
  {\bibfnamefont{G.-B.}\ \bibnamefont{{Zhao}}},\ }%
  \bibfield{journal}{%
  \Doi{10.1088/1475-7516/2013/11/012}{\bibinfo {journal} {JCAP}}\ }%
  \textbf{\bibinfo {volume} {11}},\ \bibinfo {eid} {012} (\bibinfo {month}
  {Nov.}\ \bibinfo {year} {2013}),\
  \Eprint{http://arxiv.org/abs/1308.3491}{arXiv:1308.3491 [astro-ph.CO]}%
  \bibAnnoteFile{NoStop}{2013JCAP...11..012L}%
\bibitem{2015arXiv151108200B}%
  \BibitemOpen
  \bibfield{author}{%
  \bibinfo {author} {\bibfnamefont{A.}~\bibnamefont{{Barreira}}}, \bibinfo
  {author} {\bibfnamefont{S.}~\bibnamefont{{Bose}}},\ and\ \bibinfo {author}
  {\bibfnamefont{B.}~\bibnamefont{{Li}}},\ }%
  \bibfield{journal}{%
  \bibinfo {journal} {ArXiv e-prints}}%
   (\bibinfo {month} {Nov.}\ \bibinfo {year} {2015}),\
  \Eprint{http://arxiv.org/abs/1511.08200}{arXiv:1511.08200}%
  \bibAnnoteFile{NoStop}{2015arXiv151108200B}%
\bibitem{2014AA...562A..78L}%
  \BibitemOpen
  \bibfield{author}{%
  \bibinfo {author} {\bibfnamefont{C.}~\bibnamefont{{Llinares}}}, \bibinfo
  {author} {\bibfnamefont{D.~F.}\ \bibnamefont{{Mota}}},\ and\ \bibinfo
  {author} {\bibfnamefont{H.~A.}\ \bibnamefont{{Winther}}},\ }%
  \bibfield{journal}{%
  \Doi{10.1051/0004-6361/201322412}{\bibinfo {journal} {AAP}}\ }%
  \textbf{\bibinfo {volume} {562}},\ \bibinfo {eid} {A78} (\bibinfo {month}
  {Feb.}\ \bibinfo {year} {2014}),\
  \Eprint{http://arxiv.org/abs/1307.6748}{arXiv:1307.6748}%
  \bibAnnoteFile{NoStop}{2014AA...562A..78L}%
\bibitem{2009PhRvD..79f4036N}%
  \BibitemOpen
  \bibfield{author}{%
  \bibinfo {author} {\bibfnamefont{A.}~\bibnamefont{{Nicolis}}}, \bibinfo
  {author} {\bibfnamefont{R.}~\bibnamefont{{Rattazzi}}},\ and\ \bibinfo
  {author} {\bibfnamefont{E.}~\bibnamefont{{Trincherini}}},\ }%
  \bibfield{journal}{%
  \Doi{10.1103/PhysRevD.79.064036}{\bibinfo {journal} {PRD}}\ }%
  \textbf{\bibinfo {volume} {79}},\ \bibinfo {eid} {064036} (\bibinfo {month}
  {Mar.}\ \bibinfo {year} {2009}),\
  \Eprint{http://arxiv.org/abs/0811.2197}{arXiv:0811.2197 [hep-th]}%
  \bibAnnoteFile{NoStop}{2009PhRvD..79f4036N}%
\bibitem{2009PhRvD..79h4003D}%
  \BibitemOpen
  \bibfield{author}{%
  \bibinfo {author} {\bibfnamefont{C.}~\bibnamefont{{Deffayet}}}, \bibinfo
  {author} {\bibfnamefont{G.}~\bibnamefont{{Esposito-Far{\`e}se}}},\ and\
  \bibinfo {author} {\bibfnamefont{A.}~\bibnamefont{{Vikman}}},\ }%
  \bibfield{journal}{%
  \Doi{10.1103/PhysRevD.79.084003}{\bibinfo {journal} {PRD}}\ }%
  \textbf{\bibinfo {volume} {79}},\ \bibinfo {eid} {084003} (\bibinfo {month}
  {Apr.}\ \bibinfo {year} {2009}),\
  \Eprint{http://arxiv.org/abs/0901.1314}{arXiv:0901.1314 [hep-th]}%
  \bibAnnoteFile{NoStop}{2009PhRvD..79h4003D}%
\bibitem{2015PhRvD..91f4012P}%
  \BibitemOpen
  \bibfield{author}{%
  \bibinfo {author} {\bibfnamefont{Y.}~\bibnamefont{{Park}}}\ and\ \bibinfo
  {author} {\bibfnamefont{M.}~\bibnamefont{{Wyman}}},\ }%
  \bibfield{journal}{%
  \Doi{10.1103/PhysRevD.91.064012}{\bibinfo {journal} {PRD}}\ }%
  \textbf{\bibinfo {volume} {91}},\ \bibinfo {eid} {064012} (\bibinfo {month}
  {Mar.}\ \bibinfo {year} {2015}),\
  \Eprint{http://arxiv.org/abs/1408.4773}{arXiv:1408.4773}%
  \bibAnnoteFile{NoStop}{2015PhRvD..91f4012P}%
\bibitem{2015MNRAS.454.4085B}%
  \BibitemOpen
  \bibfield{author}{%
  \bibinfo {author} {\bibfnamefont{A.}~\bibnamefont{{Barreira}}}, \bibinfo
  {author} {\bibfnamefont{B.}~\bibnamefont{{Li}}}, \bibinfo {author}
  {\bibfnamefont{E.}~\bibnamefont{{Jennings}}}, \bibinfo {author}
  {\bibfnamefont{J.}~\bibnamefont{{Merten}}}, \bibinfo {author}
  {\bibfnamefont{L.}~\bibnamefont{{King}}}, \bibinfo {author}
  {\bibfnamefont{C.~M.}\ \bibnamefont{{Baugh}}},\ and\ \bibinfo {author}
  {\bibfnamefont{S.}~\bibnamefont{{Pascoli}}},\ }%
  \bibfield{journal}{%
  \Doi{10.1093/mnras/stv2211}{\bibinfo {journal} {MNRAS}}\ }%
  \textbf{\bibinfo {volume} {454}},\ \bibinfo {pages} {4085} (\bibinfo {month}
  {Dec.}\ \bibinfo {year} {2015}),\
  \Eprint{http://arxiv.org/abs/1505.03468}{arXiv:1505.03468}%
  \bibAnnoteFile{NoStop}{2015MNRAS.454.4085B}%
\bibitem{2015JCAP...08..028B}%
  \BibitemOpen
  \bibfield{author}{%
  \bibinfo {author} {\bibfnamefont{A.}~\bibnamefont{{Barreira}}}, \bibinfo
  {author} {\bibfnamefont{M.}~\bibnamefont{{Cautun}}}, \bibinfo {author}
  {\bibfnamefont{B.}~\bibnamefont{{Li}}}, \bibinfo {author}
  {\bibfnamefont{C.~M.}\ \bibnamefont{{Baugh}}},\ and\ \bibinfo {author}
  {\bibfnamefont{S.}~\bibnamefont{{Pascoli}}},\ }%
  \bibfield{journal}{%
  \Doi{10.1088/1475-7516/2015/08/028}{\bibinfo {journal} {JCAP}}\ }%
  \textbf{\bibinfo {volume} {8}},\ \bibinfo {eid} {028} (\bibinfo {month}
  {Aug.}\ \bibinfo {year} {2015}),\
  \Eprint{http://arxiv.org/abs/1505.05809}{arXiv:1505.05809}%
  \bibAnnoteFile{NoStop}{2015JCAP...08..028B}%
\bibitem{2014JCAP...08..059B}%
  \BibitemOpen
  \bibfield{author}{%
  \bibinfo {author} {\bibfnamefont{A.}~\bibnamefont{{Barreira}}}, \bibinfo
  {author} {\bibfnamefont{B.}~\bibnamefont{{Li}}}, \bibinfo {author}
  {\bibfnamefont{C.~M.}\ \bibnamefont{{Baugh}}},\ and\ \bibinfo {author}
  {\bibfnamefont{S.}~\bibnamefont{{Pascoli}}},\ }%
  \bibfield{journal}{%
  \Doi{10.1088/1475-7516/2014/08/059}{\bibinfo {journal} {JCAP}}\ }%
  \textbf{\bibinfo {volume} {8}},\ \bibinfo {eid} {059} (\bibinfo {month}
  {Aug.}\ \bibinfo {year} {2014}),\
  \Eprint{http://arxiv.org/abs/1406.0485}{arXiv:1406.0485}%
  \bibAnnoteFile{NoStop}{2014JCAP...08..059B}%
\bibitem{2013JCAP...10..027B}%
  \BibitemOpen
  \bibfield{author}{%
  \bibinfo {author} {\bibfnamefont{A.}~\bibnamefont{{Barreira}}}, \bibinfo
  {author} {\bibfnamefont{B.}~\bibnamefont{{Li}}}, \bibinfo {author}
  {\bibfnamefont{W.~A.}\ \bibnamefont{{Hellwing}}}, \bibinfo {author}
  {\bibfnamefont{C.~M.}\ \bibnamefont{{Baugh}}},\ and\ \bibinfo {author}
  {\bibfnamefont{S.}~\bibnamefont{{Pascoli}}},\ }%
  \bibfield{journal}{%
  \Doi{10.1088/1475-7516/2013/10/027}{\bibinfo {journal} {JCAP}}\ }%
  \textbf{\bibinfo {volume} {10}},\ \bibinfo {eid} {027} (\bibinfo {month}
  {Oct.}\ \bibinfo {year} {2013}),\
  \Eprint{http://arxiv.org/abs/1306.3219}{arXiv:1306.3219}%
  \bibAnnoteFile{NoStop}{2013JCAP...10..027B}%
\bibitem{2012PhRvD..86l4016B}%
  \BibitemOpen
  \bibfield{author}{%
  \bibinfo {author} {\bibfnamefont{A.}~\bibnamefont{{Barreira}}}, \bibinfo
  {author} {\bibfnamefont{B.}~\bibnamefont{{Li}}}, \bibinfo {author}
  {\bibfnamefont{C.~M.}\ \bibnamefont{{Baugh}}},\ and\ \bibinfo {author}
  {\bibfnamefont{S.}~\bibnamefont{{Pascoli}}},\ }%
  \bibfield{journal}{%
  \Doi{10.1103/PhysRevD.86.124016}{\bibinfo {journal} {PRD}}\ }%
  \textbf{\bibinfo {volume} {86}},\ \bibinfo {eid} {124016} (\bibinfo {month}
  {Dec.}\ \bibinfo {year} {2012}),\
  \Eprint{http://arxiv.org/abs/1208.0600}{arXiv:1208.0600 [astro-ph.CO]}%
  \bibAnnoteFile{NoStop}{2012PhRvD..86l4016B}%
\bibitem{2007PhRvL..99k1301D}%
  \BibitemOpen
  \bibfield{author}{%
  \bibinfo {author} {\bibfnamefont{S.}~\bibnamefont{{Deser}}}\ and\ \bibinfo
  {author} {\bibfnamefont{R.~P.}\ \bibnamefont{{Woodard}}},\ }%
  \bibfield{journal}{%
  \Doi{10.1103/PhysRevLett.99.111301}{\bibinfo {journal} {Physical Review
  Letters}}\ }%
  \textbf{\bibinfo {volume} {99}},\ \bibinfo {eid} {111301} (\bibinfo {month}
  {Sep.}\ \bibinfo {year} {2007}),\
  \Eprint{http://arxiv.org/abs/0706.2151}{arXiv:0706.2151}%
  \bibAnnoteFile{NoStop}{2007PhRvL..99k1301D}%
\bibitem{2014IJMPA..2950116F}%
  \BibitemOpen
  \bibfield{author}{%
  \bibinfo {author} {\bibfnamefont{S.}~\bibnamefont{{Foffa}}}, \bibinfo
  {author} {\bibfnamefont{M.}~\bibnamefont{{Maggiore}}},\ and\ \bibinfo
  {author} {\bibfnamefont{E.}~\bibnamefont{{Mitsou}}},\ }%
  \bibfield{journal}{%
  \Doi{10.1142/S0217751X14501164}{\bibinfo {journal} {International Journal of
  Modern Physics A}}\ }%
  \textbf{\bibinfo {volume} {29}},\ \bibinfo {eid} {1450116} (\bibinfo {month}
  {Aug.}\ \bibinfo {year} {2014}),\
  \Eprint{http://arxiv.org/abs/1311.3435}{arXiv:1311.3435 [hep-th]}%
  \bibAnnoteFile{NoStop}{2014IJMPA..2950116F}%
\bibitem{2014JCAP...09..031B}%
  \BibitemOpen
  \bibfield{author}{%
  \bibinfo {author} {\bibfnamefont{A.}~\bibnamefont{{Barreira}}}, \bibinfo
  {author} {\bibfnamefont{B.}~\bibnamefont{{Li}}}, \bibinfo {author}
  {\bibfnamefont{W.~A.}\ \bibnamefont{{Hellwing}}}, \bibinfo {author}
  {\bibfnamefont{C.~M.}\ \bibnamefont{{Baugh}}},\ and\ \bibinfo {author}
  {\bibfnamefont{S.}~\bibnamefont{{Pascoli}}},\ }%
  \bibfield{journal}{%
  \Doi{10.1088/1475-7516/2014/09/031}{\bibinfo {journal} {JCAP}}\ }%
  \textbf{\bibinfo {volume} {9}},\ \bibinfo {eid} {031} (\bibinfo {month}
  {Sep.}\ \bibinfo {year} {2014}),\
  \Eprint{http://arxiv.org/abs/1408.1084}{arXiv:1408.1084}%
  \bibAnnoteFile{NoStop}{2014JCAP...09..031B}%
\bibitem{2015PhRvD..91f3528B}%
  \BibitemOpen
  \bibfield{author}{%
  \bibinfo {author} {\bibfnamefont{A.}~\bibnamefont{{Barreira}}}, \bibinfo
  {author} {\bibfnamefont{P.}~\bibnamefont{{Brax}}}, \bibinfo {author}
  {\bibfnamefont{S.}~\bibnamefont{{Clesse}}}, \bibinfo {author}
  {\bibfnamefont{B.}~\bibnamefont{{Li}}},\ and\ \bibinfo {author}
  {\bibfnamefont{P.}~\bibnamefont{{Valageas}}},\ }%
  \bibfield{journal}{%
  \Doi{10.1103/PhysRevD.91.063528}{\bibinfo {journal} {PRD}}\ }%
  \textbf{\bibinfo {volume} {91}},\ \bibinfo {eid} {063528} (\bibinfo {month}
  {Mar.}\ \bibinfo {year} {2015}),\
  \Eprint{http://arxiv.org/abs/1411.5965}{arXiv:1411.5965}%
  \bibAnnoteFile{NoStop}{2015PhRvD..91f3528B}%
\bibitem{2014PhRvD..90b3508B}%
  \BibitemOpen
  \bibfield{author}{%
  \bibinfo {author} {\bibfnamefont{P.}~\bibnamefont{{Brax}}}\ and\ \bibinfo
  {author} {\bibfnamefont{P.}~\bibnamefont{{Valageas}}},\ }%
  \bibfield{journal}{%
  \Doi{10.1103/PhysRevD.90.023508}{\bibinfo {journal} {PRD}}\ }%
  \textbf{\bibinfo {volume} {90}},\ \bibinfo {eid} {023508} (\bibinfo {month}
  {Jul.}\ \bibinfo {year} {2014}),\
  \Eprint{http://arxiv.org/abs/1403.5424}{arXiv:1403.5424}%
  \bibAnnoteFile{NoStop}{2014PhRvD..90b3508B}%
\bibitem{2015JCAP...10..036T}%
  \BibitemOpen
  \bibfield{author}{%
  \bibinfo {author} {\bibfnamefont{N.}~\bibnamefont{{Tessore}}}, \bibinfo
  {author} {\bibfnamefont{H.~A.}\ \bibnamefont{{Winther}}}, \bibinfo {author}
  {\bibfnamefont{R.~B.}\ \bibnamefont{{Metcalf}}}, \bibinfo {author}
  {\bibfnamefont{P.~G.}\ \bibnamefont{{Ferreira}}},\ and\ \bibinfo {author}
  {\bibfnamefont{C.}~\bibnamefont{{Giocoli}}},\ }%
  \bibfield{journal}{%
  \Doi{10.1088/1475-7516/2015/10/036}{\bibinfo {journal} {JCAP}}\ }%
  \textbf{\bibinfo {volume} {10}},\ \bibinfo {eid} {036} (\bibinfo {month}
  {Oct.}\ \bibinfo {year} {2015}),\
  \Eprint{http://arxiv.org/abs/1508.04011}{arXiv:1508.04011}%
  \bibAnnoteFile{NoStop}{2015JCAP...10..036T}%
\bibitem{2005ApJ...619..741G}%
  \BibitemOpen
  \bibfield{author}{%
  \bibinfo {author} {\bibfnamefont{D.~M.}\ \bibnamefont{{Goldberg}}}\ and\
  \bibinfo {author} {\bibfnamefont{D.~J.}\ \bibnamefont{{Bacon}}},\ }%
  \bibfield{journal}{%
  \Doi{10.1086/426782}{\bibinfo {journal} {APJ}}\ }%
  \textbf{\bibinfo {volume} {619}},\ \bibinfo {pages} {741} (\bibinfo {month}
  {Feb.}\ \bibinfo {year} {2005}),\
  \Eprint{http://arxiv.org/abs/astro-ph/0406376}{astro-ph/0406376}%
  \bibAnnoteFile{NoStop}{2005ApJ...619..741G}%
\bibitem{2006MNRAS.365..414B}%
  \BibitemOpen
  \bibfield{author}{%
  \bibinfo {author} {\bibfnamefont{D.~J.}\ \bibnamefont{{Bacon}}}, \bibinfo
  {author} {\bibfnamefont{D.~M.}\ \bibnamefont{{Goldberg}}}, \bibinfo {author}
  {\bibfnamefont{B.~T.~P.}\ \bibnamefont{{Rowe}}},\ and\ \bibinfo {author}
  {\bibfnamefont{A.~N.}\ \bibnamefont{{Taylor}}},\ }%
  \bibfield{journal}{%
  \Doi{10.1111/j.1365-2966.2005.09624.x}{\bibinfo {journal} {MNRAS}}\ }%
  \textbf{\bibinfo {volume} {365}},\ \bibinfo {pages} {414} (\bibinfo {month}
  {Jan.}\ \bibinfo {year} {2006}),\
  \Eprint{http://arxiv.org/abs/astro-ph/0504478}{astro-ph/0504478}%
  \bibAnnoteFile{NoStop}{2006MNRAS.365..414B}%
\bibitem{2007ApJ...660..995O}%
  \BibitemOpen
  \bibfield{author}{%
  \bibinfo {author} {\bibfnamefont{Y.}~\bibnamefont{{Okura}}}, \bibinfo
  {author} {\bibfnamefont{K.}~\bibnamefont{{Umetsu}}},\ and\ \bibinfo {author}
  {\bibfnamefont{T.}~\bibnamefont{{Futamase}}},\ }%
  \bibfield{journal}{%
  \Doi{10.1086/513135}{\bibinfo {journal} {APJ}}\ }%
  \textbf{\bibinfo {volume} {660}},\ \bibinfo {pages} {995} (\bibinfo {month}
  {May}\ \bibinfo {year} {2007}),\
  \Eprint{http://arxiv.org/abs/astro-ph/0607288}{astro-ph/0607288}%
  \bibAnnoteFile{NoStop}{2007ApJ...660..995O}%
\bibitem{2008A&A...485..363S}%
  \BibitemOpen
  \bibfield{author}{%
  \bibinfo {author} {\bibfnamefont{P.}~\bibnamefont{{Schneider}}}\ and\
  \bibinfo {author} {\bibfnamefont{X.}~\bibnamefont{{Er}}},\ }%
  \bibfield{journal}{%
  \Doi{10.1051/0004-6361:20078631}{\bibinfo {journal} {AAP}}\ }%
  \textbf{\bibinfo {volume} {485}},\ \bibinfo {pages} {363} (\bibinfo {month}
  {Jul.}\ \bibinfo {year} {2008}),\
  \Eprint{http://arxiv.org/abs/0709.1003}{arXiv:0709.1003}%
  \bibAnnoteFile{NoStop}{2008A&A...485..363S}%
\bibitem{2013MNRAS.435..822R}%
  \BibitemOpen
  \bibfield{author}{%
  \bibinfo {author} {\bibfnamefont{B.}~\bibnamefont{{Rowe}}}, \bibinfo {author}
  {\bibfnamefont{D.}~\bibnamefont{{Bacon}}}, \bibinfo {author}
  {\bibfnamefont{R.}~\bibnamefont{{Massey}}}, \bibinfo {author}
  {\bibfnamefont{C.}~\bibnamefont{{Heymans}}}, \bibinfo {author}
  {\bibfnamefont{B.}~\bibnamefont{{H{\"a}u{\ss}ler}}}, \bibinfo {author}
  {\bibfnamefont{A.}~\bibnamefont{{Taylor}}}, \bibinfo {author}
  {\bibfnamefont{J.}~\bibnamefont{{Rhodes}}},\ and\ \bibinfo {author}
  {\bibfnamefont{Y.}~\bibnamefont{{Mellier}}},\ }%
  \bibfield{journal}{%
  \Doi{10.1093/mnras/stt1353}{\bibinfo {journal} {MNRAS}}\ }%
  \textbf{\bibinfo {volume} {435}},\ \bibinfo {pages} {822} (\bibinfo {month}
  {Oct.}\ \bibinfo {year} {2013}),\
  \Eprint{http://arxiv.org/abs/1211.0966}{arXiv:1211.0966}%
  \bibAnnoteFile{NoStop}{2013MNRAS.435..822R}%
\bibitem{2014PhRvD..89d4010B}%
  \BibitemOpen
  \bibfield{author}{%
  \bibinfo {author} {\bibfnamefont{M.}~\bibnamefont{{Bruni}}}, \bibinfo
  {author} {\bibfnamefont{D.~B.}\ \bibnamefont{{Thomas}}},\ and\ \bibinfo
  {author} {\bibfnamefont{D.}~\bibnamefont{{Wands}}},\ }%
  \bibfield{journal}{%
  \Doi{10.1103/PhysRevD.89.044010}{\bibinfo {journal} {PRD}}\ }%
  \textbf{\bibinfo {volume} {89}},\ \bibinfo {eid} {044010} (\bibinfo {month}
  {Feb.}\ \bibinfo {year} {2014}),\
  \Eprint{http://arxiv.org/abs/1306.1562}{arXiv:1306.1562 [astro-ph.CO]}%
  \bibAnnoteFile{NoStop}{2014PhRvD..89d4010B}%
\bibitem{2015JCAP...09..021T}%
  \BibitemOpen
  \bibfield{author}{%
  \bibinfo {author} {\bibfnamefont{D.~B.}\ \bibnamefont{{Thomas}}}, \bibinfo
  {author} {\bibfnamefont{M.}~\bibnamefont{{Bruni}}},\ and\ \bibinfo {author}
  {\bibfnamefont{D.}~\bibnamefont{{Wands}}},\ }%
  \bibfield{journal}{%
  \Doi{10.1088/1475-7516/2015/09/021}{\bibinfo {journal} {JCAP}}\ }%
  \textbf{\bibinfo {volume} {9}},\ \bibinfo {eid} {021} (\bibinfo {month}
  {Sep.}\ \bibinfo {year} {2015}),\
  \Eprint{http://arxiv.org/abs/1403.4947}{arXiv:1403.4947}%
  \bibAnnoteFile{NoStop}{2015JCAP...09..021T}%
\end{thebibliography}%

\end{document}